\documentclass[11pt,a4paper]{article}
\pdfoutput=1

\usepackage{pgf}
\usepackage{tikz}
\usepackage{graphicx}
\usetikzlibrary{arrows,automata}
\usepackage{amssymb,amsmath,amsfonts}
\usepackage{longtable}
\usepackage{verbatim}
\usepackage{color}
\usepackage[mathscr]{euscript}
\usepackage{framed}
\usepackage{mdframed}
\usepackage{amsmath}
\usepackage{amsfonts}
\usepackage{mathrsfs}
\usepackage{amssymb}
\usepackage{mathrsfs}  
\usepackage{subcaption}
\usepackage{caption}
\usepackage[many]{tcolorbox}

\usepackage[left=26mm,top=20mm,right=26mm,bottom=20mm]{geometry}

\usepackage{color}
\usepackage{cite}
\usepackage{hyperref}
\hypersetup{colorlinks, citecolor=bluscuro, linkcolor=black, urlcolor=bluscuro}
\definecolor{rossos}{cmyk}{0,1,1,0.55}
\definecolor{bluscuro}{rgb}{0.15, 0.2, .85}
\definecolor{bluchiaro}{cmyk}{1,.3,0.,0.1}

\graphicspath{{./Figures/}}

\newcommand{\be}{\begin{equation}}
\newcommand{\ee}{\end{equation}}
\newcommand{\bea}{\begin{eqnarray}}
\newcommand{\eea}{\end{eqnarray}}
\newcommand{\beq}{\begin{equation}}
\newcommand{\eeq}{\end{equation}}

\def\beqa{\begin{eqnarray}}

\def\eeqa{\end{eqnarray}}

\def\lsim{\mathrel{\rlap{\lower4pt\hbox{\hskip0.5pt$\sim$}}
    \raise1pt\hbox{$<$}}}         
\def\gsim{\mathrel{\rlap{\lower4pt\hbox{\hskip0.5pt$\sim$}}
    \raise1pt\hbox{$>$}}}         

\usepackage[normalem]{ulem}

\newcommand{\arXiv}[2]{\href{http://arxiv.org/pdf/#1}{{\tt [#2/#1]}}}
\newcommand{\arXivold}[1]{\href{http://arxiv.org/pdf/#1}{{\tt [#1]}}}

\begin{document}

\vspace{0.1in}

\begin{center}
{\Large\bf\color{black}
Constraints on the 
curvature power spectrum 
from primordial black hole evaporation 
}\\
\bigskip\color{black}
\vspace{.5cm}
{ {\large  {Ioannis Dalianis}$^a$ }
\vspace{0.3cm}
} \\[5mm]
{\it {$^a$ Physics Division, National Technical University of Athens\\ 15780 Zografou Campus, Athens, Greece}}
\\[2mm]
$\text{dalianis@mail.ntua.gr}$
\\[2mm]

\end{center}

\vskip.2in

\noindent
\rule{16.6cm}{0.4pt}

\vspace{.3cm}
\noindent
{\bf \large {Abstract}}
\vskip.15in 
We estimate the maximum allowed amplitude for the power spectrum of the  primordial curvature perturbations, ${\cal P_R}(k)$,  on all scales
 from the absence of any detection signal of sub-solar mass black holes. 
In particular we analyze the constraints on the PBHs and we focus on the low mass limit where the Hawking radiation is expected to significantly influence the big bang observables, considering  also different early cosmic histories.
 We derive the upper bounds for the variance of density perturbations, $\sigma(M)$, for any possible reheating temperature  as well as for the cosmological scenario of a scalar condensate domination.
 We expect our results to have considerable implications for models designed to generate PBHs, especially in the low mass range, and provide additional constraints to a large class of  inflationary models.
\noindent

\bigskip

\noindent
\rule{16.6cm}{0.4pt}

\vskip.4in

\section{Introduction}  


The primordial power spectrum of the comoving curvature perturbations ${\cal P_R}(k)$ has been precisely measured by the CMB probes in the range of scales  between $k\sim 10^{-3}$ Mpc$^{-1}$ and 1 Mpc$^{-1}$.  In smaller scales the ${\cal P_R}(k)$ is poorly and indirectly constrained by observations of nonlinear structures. The relevant limits are very weak coming mainly from the mass fraction  of the universe, $\beta(M)$, that collapsed and formed primordial black holes (PBH) of mass $M$. 
Black holes affect dynamical systems and cause microlensing events and so a bound on the $\beta(M)$ value is obtained.  The 
Hawking prediction \cite{Hawking:1974sw, Hawking:1974rv}
  that black holes radiate thermally with temperature
\begin{align}
T_\text{BH}=1.06 \, \frac{M^2_\text{Pl}}{M_\text{}} 
\simeq \, \left( \frac{10^{13}\text{g}}{M_\text{}} \right) \, \text{GeV}\,,
\end{align}
and evaporate after a timescale  
\begin{align}
\tau(M_\text{}) \simeq 4 \times 10^{11} \left( \frac{M_\text{}}{10^{13} \, \text{g}} \right)^3\, \text{s}\,,
\end{align}
provides us with additional bounds on $\beta(M)$ for  small mass PBHs from  
 the absence of any evidence for black hole evaporation. 
Consequently, ${\cal P_R}(k)$ bounds in the smallest range of scales  
can be obtained.
 In the inflationary framework the measurement of the ${\cal P_R}(k)$ can be regarded as an insight into the microscopic dynamics of the field(s) that dominated the energy density of the early universe and generated the primordial perturbations.
The purpose of  this paper is to  make use of the limits on $\beta(M)$ 
coming primarily from the CMB and BBN observables to constrain 
the variance of the density perturbations 
and therefore, the cosmological scenarios, such as inflation, designed to trigger PBH formation. 

PBHs form from the collapse of large-amplitude inhomogeneities  \cite{PBH1, PBH2, Carr:1975qj, PBH4, Lindley:1980bu}.
In order to decouple from the background expansion it has to be $GM/R\sim 1$, for a region of mass $M$ over a scale $R$.  This can be achieved if the power spectrum  ${\cal P_R}(k)$ is enhanced at a scale $R^{-1}\sim k$, characteristic of the PBH mass, by many orders of magnitude. 
Large wavenumbers yield light PBH which if they have mass  $M \lesssim 10^{15}$ g evaporate at timescales less than the age of the universe.  
 PBHs with $M > 10^{15}$ g  would still survive today and would be dynamically cold component of the dark matter in galactic structures.
 To distinguish between the nonevaporated and the evaporated PBH we label the mass and the characteristic wavenumber of the former with a dark dot subscript, i.e. $M_\bullet$, $k_\bullet$ respectively. 

The formation of  a primordial  black hole of mass similar to the black holes detected by LIGO\cite{ligo}, $M_\bullet\sim 30 M_\odot$  requires ${\cal P_R}(10^{10}\, \text{Mpc}^{-1}) \sim 10^{-2}$. 
 Similar values for the  ${\cal P_R}(k)$ are required for the formation of lighter primordial black holes that, although lack observational support, are 
 well motivated  dark matter candidates.  
Actually, it is the low mass window, $M_\bullet \ll M_\odot$, the most promising one for explaining the dark matter in the galaxies,  according to the current  observational constraints. 
 Several inflation models that achieve the required  ${\cal P_R}(k)$ enhancement have been proposed the last years \cite{Kawaguchi:2007fz, Alabidi:2009bk, Drees:2011hb, Drees:2011yz,  Kawasaki2013, Kohri:2012yw, Kawasaki:2016pql, Inomata:2016rbd, gauge, sone1, ssm1, sone0, sone2, ssm2, Hertzberg:2017dkh, Ando:2017veq, sone3, Ozsoy:2018flq, Dalianis:2018frf, Ballesteros:2018wlw, Cai:2018tuh}  putting forward new ideas and elaborating further 
 earlier works \cite{s1, Yokoyama:1995ex, s2, s3, Kawasaki:1997ju, Yokoyama:1999xi}. 
The shape of the ${\cal P_R}(k)$ at small scales is mainly constrained at a scale $k_\bullet$ where the abundance of the nonevaporated PBHs maximizes. 
The aim of this work is to stress that, depending on the postinflationary expansion history, the shape of the ${\cal P_R}(k)$ at smaller scales $k\gg k_\bullet$ is crucial to affirm the viability of a model designed to generate dark matter PBHs.

Large values for the ${\cal P_R}(k)$ at small scales may generate short-lived PBH that, although absent from the cosmic structures today, evaporate in  the early universe leaving potentially observable signatures. 
The thermal emission of black holes affects the BBN \cite{Vainer, PBH4, Miyama:1978mp, Vainer2, Lindley, Kohri:1999ex, Kawasaki:2004qu, Carr:2009jm
}
in the mass range $10^9-10^{13}$ g and  bounds on the fraction of the universe mass that collapses into black holes, $\beta(M)$ are induced.  
In addition, the diffuse extragalactic $\gamma$-ray background 
put constraints on the mass range  $10^{14}-10^{17}$ g \cite{Page:1976wx, MacGibbon:1991vc, Carr:1998fw, Barrau:2003nj,  Carr:2016hva}. The most stringent constraint on the mean number density of the short-lived PBHs  comes from  
the CMB anisotropy damping \cite{PBH4} which limits the $\beta(M) \lesssim 10^{-29}$ \cite{Zhang:2007zzh}
 in the mass range  $10^{13} - 10^{14}$ g.  
 In Ref.\cite{Carr:2009jm} these constraints are outlined and further references can be found therein.
 
Our investigation is focused on the $\beta(M)$ bounds for $M\ll 10^{15}$ g it will be shown that extra important constraints can be put on the  ${\cal P_R}(k)$. 
From a different perspective these constraints can be viewed also as an insight into the unknown cosmic history of the early universe if a measurement of the  ${\cal P_R}(k)$ on small scales is made possible. Apparently, the key relation is the one that connects the power spectrum ${\cal P_R}(k)$ and the $\beta(M)$. The knowledge of the $\beta(M)$ can constrain the ${\cal P_R}(k)$ only if one assumes a model for the PBH formation. 
In the following analysis we assume 
spherical symmetric Gaussian primordial perturbations and that the PBHs form on the high $\sigma$-tail according to the Press-Schechter formalism \cite{Press}. 
We consider gravitational collapse during radiation era as well as during presureless matter era taking into account spin effects. 
This is actually a distinct ingredient of this work.
We follow the monochromatic mass spectrum approximation and assume a one-to-one correspondence between the scale of
perturbation and the mass of PBHs. We do not consider possible impacts on the power spectrum from non-Gaussianities \cite{Franciolini:2018vbk, Atal:2018neu} and quantum diffusion effects \cite{Pattison:2017mbe, Biagetti:2018pjj, Ezquiaga:2018gbw, Cruces:2018cvq}.

The main result of this work is the derivation of upper bounds  for the variance of  comoving density contrast at horizon entry $\sigma(M)$ on all scales for different reheating temperatures and cosmological scenarios and translate these bounds onto $P_{\cal R}(k)$ bounds.
In particular, we derive the upper bound for the $\sigma(M)$ in order that  the CMB and BBN observables remain intact  for any possible reheating temperature.
These bounds are the most stringent ones for promptly evaporating PBHs.
Most of our analysis is general regardless the mechanism that generates the perturbation spectrum. We implicitly assume that it is inflation  behind the $P_{\cal R}(k)$ generation, however, we do not specify the inflaton dynamics apart from the energy scale that inflation ends.

 Large-amplitude inhomogeneities are necessary for the PBH formation however, the formation rate may significantly increase \cite{Harada:2016mhb, Harada:2017fjm} if the equation of state of the background energy density, $w$, becomes soft or zero. This is a rather plausible scenario for PBHs that form not long after the end of the inflationary phase where the inflaton coherent oscillations result in an early matter domination era. Other scenarios, such a modulus domination that is natural in several extensions of the Standard Model of particle physics,  also result in an early non-thermal phase.  

The most  important implications for the PBH formation models of such an early matter domination era is that the variance of the density perturbations, which determines the $\beta(M)$, can be smaller for a fixed PBH abundance.   
To be explicit let us express in a qualitative level how the relic abundance of the PBHs depends on the cosmic era that the collapse takes place.  It is
\begin{align}
f^\text{(RD)}_\text{PBH}\,  \propto \, \sqrt{{\cal P_R}} \,\,
e^{-{\delta^2_c}/{{\cal P_R}} }  \,, \quad\quad\quad
f^\text{(MD)}_\text{PBH} \, \propto \, {\cal P_R}^{5/2} \,.
\end{align}
The  expression on the left hand side is for collapse during radiation and on the right hand side for spinless collapse during matter era.
Thus, an overdensity is more probable  to collapse  during matter era for ${\cal P_R}\ll 1$. 
Since the power spectrum is expressed in terms of the wavenumber $k$ whereas the fractional PBH abundance, $f_\text{PBH}$, is expressed in terms of the PBH mass $M$, the $k=k(M)$ relation is required in order to connect the $f_\text{PBH}(M)$ and the ${\cal P_R}(k)$. 
 During matter era the horizon size grows with a different rate and
the relation between the mass  $M$ contained in the comoving Hubble radius of size $k^{-1}=(aH)^{-1}$ is different from that of a radiation era. In particular the $k(M)$ relation has the following scaling during radiation and matter eras respectively,
\begin{align}
k_\text{RD}(M)\,  \propto \, M^{-1/2} \,\,
  \,, \quad\quad\quad
k_\text{MD}(M, T_\text{rh}) \, \propto  M^{-1/3}\, T_\text{rh}^{1/3}\, .
\end{align}
For the matter era there is the extra dependence on the reheating temperature.

Given these relations, in this work, we will transform the observational upper bounds on the 
 PBH yield, $Y_\text{PBH}$, that is the PBH number density-over-entropy with mass  $M$ into upper bounds on the power spectrum ${\cal P_R}(k)$. This requires two computational steps. The first step is to specify the maximum value for the $\beta(M)$ from the yield, something that is possible  only after the reheating temperature is known. 
The second step is to go from the $\beta(M)$ to the variance $\sigma(M)$. This is a more elaborated step since the size of the $\beta$ depends on the variance of the density perturbations in a different way if the collapse takes place in a background with or without  pressure. Also, if there is thermal pressure the collapse is effectively instantaneous whereas, if the there is no pressure  the collapse has a duration determined by a critical $\sigma$ value. Moreover, in matter era the amplitude of the variance determines whether the PBH formation-rate is affected by angular momentum effects. Taking all these into account, we derive the maximum value allowed by observations for the ${\cal P_R}(k)$ on all scales.

The recent works \cite{Carr:2017edp, Cole:2017gle}  have also examined the  ${\cal P_R}(k)$ constraints on all scales in a similar context. 
In this work we present new results and elucidate different questions. 
In particular, we complement part of their analysis 
 by including the spin effects for gravitational collapse, that are crucial and change considerably the corresponding bounds on the variance $\sigma(M)$ and the power spectrum ${\cal P_R}(k)$. 
 We also derive the constraints on the PBH production scenarios for any reheating temperature and in addition we examine the scenario of a non-thermal phase due to a modulus field. 
  Finally, we estimate constraints for the spectral index value of the tail of the power spectrum with respect to the reheating temperature and 
 estimate the maximum allowed value for the power spectrum taking into account the BBN and CMB constraints along with the $f_\text{PBH}$  bounds.

The structure of the paper is the following. 
In section \ref{2} we discuss the observational bounds on the mass fraction of the universe that collapses into PBH, $\beta(M)$, introducing the $\beta(M)$ constraints for the early matter domination (eMD)  era  in addition to those for the radiation domination (RD) era.
In section \ref{3}  we derive the expressions that relate the PBH mass and the comoving horizon scale for different cosmic histories, that we generalize in section \ref{7}. In section \ref{4} we derive the principle results of this work, that is the upper bound for the variance of the density perturbations for any reheating temperature. 
In section \ref{5} we estimate the maximum possible amplitude for the power spectrum, ${\cal P_R}_\text{max}$, with respect to the reheating temperature  considering  constraints both on the  PBH dark matter abundance and the Hawking radiation. In section \ref{6} we examine the  cosmological scenario of an intermediate non-thermal phase due to modulus condensate domination. In section \ref{7} we present the full power spectrum constraints and  briefly discuss additional constraints that apply at larger scales. We conclude in section \ref{8} where we outline and  discuss the implications of the constraints derived in this work for the inflationary models. 
In Appendix we 
assume a particular morphology for the ${\cal P_R}(k)$ for large wavenumbers and illustrate the tension with the big bang observables that a wide power spectrum peak may generate.

\begin{figure}[!htbp]
  \begin{subfigure}{.5\textwidth}
  \centering
  \includegraphics[width=.95 \linewidth]{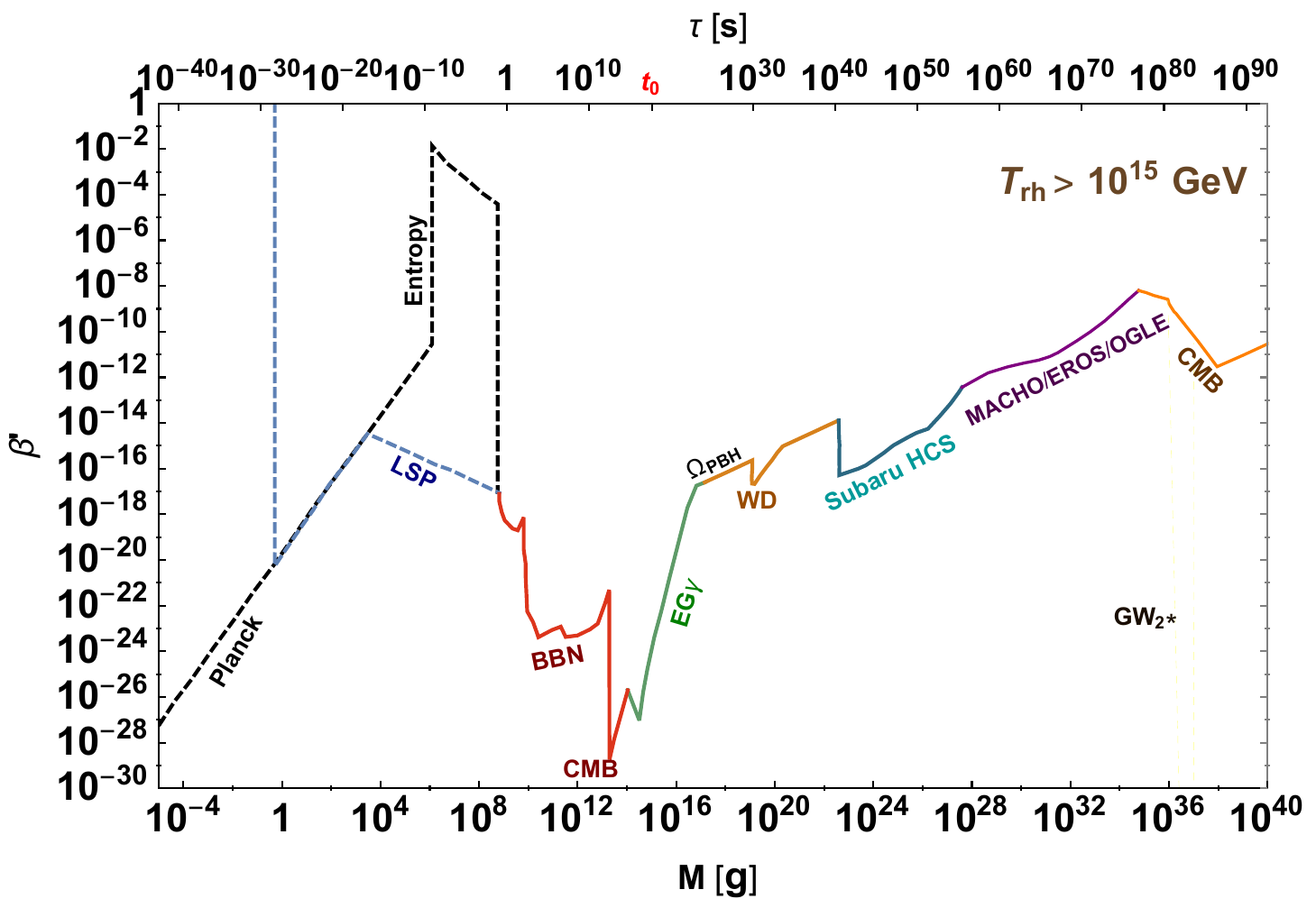}
\end{subfigure}  
  \begin{subfigure}{.5\textwidth}
  \centering
  \includegraphics[width=.95\linewidth]{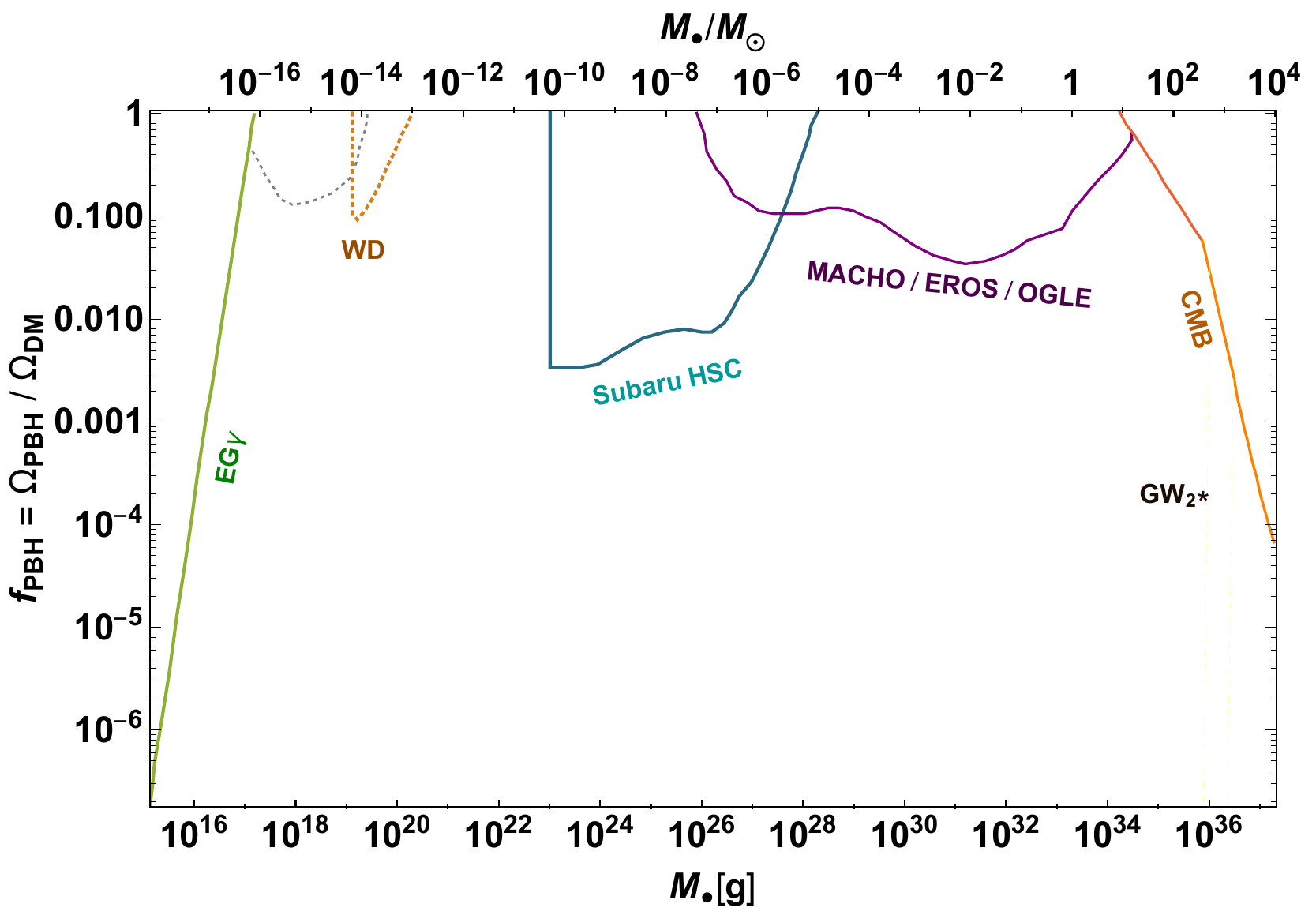}
\end{subfigure}
 \caption{\label{YokoR}~~  
{\it Left panel:} The figure depicts updated upper bounds on $\beta'(M)$ (and  $\beta'(\tau_\text{})$ on the upper axis with $t_0$ indicating the age of the universe) from Planck and LSP relics, BBN, CMB anisotropy, extragalactic gamma rays, due to evaporating black holes and density bounds on nonevaporated black holes for arbitrarily large reheating temperature, from Ref.\cite{Carr:2009jm}. 
The DM constraint is depicted with dotted-dashed line because it is model dependent and subject to much uncertainty; the constraint associated with GWs is partial and included only as reference following Ref.  \cite{Carr:2009jm}. The focus of the present work is mostly on the red line, i.e. the BBN and CMB constraints. 
 {\it Right panel:} The fractional abundance of the nonevaporated PBHs.
  }
\end{figure}

\section{The $\beta$ bound for PBH formation during radiation and matter domination  eras} \label{2}

\subsection{Radiation domination era}

A PBH with mass $M$ forms during a  radiation dominated (RD) era  if a preexisting overdensity with wavelength $k^{-1}$  enters the horizon after the reheating of the universe. 
The PBH mass is equal to $\gamma M_\text{hor}$ where $M_\text{hor}$ is the horizon mass and $\gamma$  a numerical factor which depends on the details of gravitational collapse. 
This consideration is regarded as the conventional one for PBH formation, 
for a different recent suggestion  see Ref. \cite{Germani:2018jgr}.
The present ratio of the abundance of PBHs  with mass $M$ over the total dark matter (DM) abundance,  $f_\text{PBH}(M)\equiv {\Omega_{\text{PBH}}(M)}/{\Omega_{\text{DM}}}$,  can be expressed as 
 \begin{equation}
f_\text{PBH} (M)\, =\,\left(\frac{\beta_\text{}(M)}{7.3 \times 10^{-15}}\right) \, 
\left(   \frac{\Omega_{\text{DM}}h^2}{0.12}   \right)^{-1}  
 \Big(\frac{\gamma_\text{R}}{0.2}\Big)^{\frac{3}{2}} 
\left(\frac{g_*}{106.75}\right)^{-\frac{1}{4}}  
\left(\frac{M}{10^{20}\text{g}}\right)^{-1/2}\,,
\end{equation}
where we took again the effective degrees of freedom $g_*$ and $g_s$ approximately equal.  
We also wrote the numerical factor $\gamma$ during RD as $\gamma_\text{R}$ in order to distinguish it from that during eMD, labeled $\gamma_\text{M}$, since their values are expected to be different. 
The observational constraints put upon the PBH dark matter fraction $f_\text{PBH}(M)$ an upper bound $f_\text{max}$.  
However for evaporating PBH it is more meaningful to use the $\beta(M)$ instead,  
\begin{align} \label{b-f}
\beta'(M) =6.5 \times 10^{-21} \, f_\text{PBH} \left( \frac{M}{10^{10} \text{g}}\right)^{1/2}
\end{align}
where 
we defined
\begin{align} \label{betapR}
\beta'(M)\equiv \gamma_\text{R}^{3/2}\left(\frac{g_*}{106.75}\right)^{-1/4}\, \beta(M)
\end{align}
and took $\Omega_{\text{DM}}h^2=0.12$\footnote{We do not assume that the dark matter is necessarily composed of PBHs.}.
The observational upper bound on $f_\text{PBH}(M)$ which from Eq. (\ref{b-f}) is translated into an upper bound on $\beta(M)$ that we call $\boldsymbol{C_M}$, 
\begin{align}  \label{gencon}
\beta'(M)  \,< \,  \boldsymbol{C_M}\, \equiv \, \beta'_\text{RD, max}(M)\,. 
\end{align}
The evaporation of the PBH formed will not affect the cosmological observables   if the constraint (\ref{gencon}) is satisfied.  The constraint line $\boldsymbol{C_M}$ is depicted in Fig. \ref{YokoR}, following the results of Ref. \cite{Carr:2009jm}.

The inequality (\ref{b-f}) is written in terms of the black holes lifetime and yield as $\beta'(\tau) <5.3\times10^{21} (\tau/s)^{1/2} Y_\text{PBH}$. 
For lifetime $\tau<10^2$ seconds 
there are limits on the amount of the thermal radiation from the PBHs evaporation due to the  production of entropy, that may be in conflict with the cosmological photon-to-baryon ratio, \cite{Zel'dovich3},  dark matter, e.g. the lightest supersymmetric particle, or Planck-mass remnants \cite{MacGibbon:1987my, Barrow:1992hq, Carr:1994ar, Dalianis:2019asr}, labeled {\it entropy, DM} (with dotted-dashed line due to model dependence of this constraint) and {\it Planck} respectively in our figures. 
These ultra light PBHs give interesting constraints or might explain the entire dark matter in the universe, see \cite{Dalianis:2019asr} for a recent work.  In the following analysis we will focus on PBHs with larger lifetimes since the presence of such PBHs might be in conflict with the {\it BBN} and {\it CMB} observables implying that  the $\beta(M)$ has to be particularly suppressed. 
For $\tau=10^{2}-10^{7}s$, that corresponds to $M=10^{10}-10^{12}$ g, hadrodissociation processes become important and the debris deuterons and nonthermally produced $^6$Li constrain the $\beta(M)$;  for 
$\tau= 10^7 -10^{12}$  s, that corresponds to $M=10^{12} - 10^{13}$ g, 
 photodissociation processes 
  overproduce  $^3$He and D  and put strong constraints on $\beta(M)$ \cite{Page:1976wx, MacGibbon:1991vc, Carr:1998fw, Barrau:2003nj,  Carr:2016hva}.
In addition, the heat produced by PBHs evaporation after the time of recombination may 
damp small-scale CMB anisotropies contrary to observations.
The electrons and positron scatter off the the CMB photons and heat the surrounding matter. 
The small scale CMB anisotropies  will remain intact by the PBHs  evaporation 
if $\beta'(M)\lesssim 3\times 10^{-30}(M/10^{13}\text{g})^{3.1}$ 
for $ 2.5 \times 10^{13}\lesssim M \lesssim 2.4 \times 10^{14} $  \cite{Chen:2003gz, Zhang:2007zzh, Carr:2009jm}.  This is stronger than all the other available limits on the $\beta'(M)$.  
In the next sections we will utilize the BBN bound and CMB (monochromatic) bounds  $\beta'(5 \times 10^{10} \text{g})<10^{-24}$ and $\beta'(2.5 \times 10^{13} \text{g})<5\times 10^{-29}$  respectively, to derive the stringent constraints on the variance of the density perturbations on small scales.

Apart from the constraints on evaporated PBHs, that is the primary interest of this work, there are numerous constraints on PBHs present in the late universe, which are the most commonly applied.  
In the late universe, the PBH evaporation rate is constrained from the extra galactic gamma-ray background \cite{Page:1976wx, MacGibbon:1991vc, Carr:1998fw, Barrau:2003nj,  Carr:2016hva}.
  Black holes of mass above $10^{17}$g are subject to gravitational lensing constraints \cite{Barnacka:2012bm, Tisserand:2006zx, Niikura:2017zjd}, labeled with {\it Subaru HSC, MACHOS, EROS OGLE} in the plots.
The recent results of Ref. \cite{Katz:2018zrn} remove the femtolensing constraints and we accordingly updated the plots.  
Also, black holes influence the trajectory and the  dynamics of other astrophysial objects such as neutron stars and white dwarfs  \cite{Capela:2012jz, Capela:2013yf, Brandt:2016aco, Graham:2015apa, Gaggero:2016dpq}  that constrain the abundance of light black holes, labeled {\it WD}.
 The CMB constrains the PBH with mass above $10^{33}$g because the accretion of gas and the associated emission of radiation during the recombination epoch could affect the CMB anisotropies \cite{Ricotti:2007au}. Recently it has been claimed that the CMB bounds on massive PBHs may be relaxed due to uncertainties in the modeling of the relevant physical processes   \cite{Carr:2016drx, Clesse:2016vqa, kam}. Finally, there are indirect constraints from the  pulsar timing array experiments on the gravitational waves ({\it GW$_2$}*) associated with the formation of relatively massive PBHs at the epoch of horizon entry\footnote{The mass range that the GW$_2$* constraint applies has been {\it revised} over the last years. Here, the analysis and plots were done following Ref.\cite{Carr:2009jm}.}.
Notably,  a very severe constraint, $\beta(M) \lesssim 10^{-52}$, on the mass band $10^2 -10^4 M_\odot$ comes from pulsar timing data since the large scalar perturbations which are necessary to  produce the PBHs also generate second-order tensor perturbations \cite{Saito:2008jc}. 
GW can constrain a larger window of PBHs mass. Bounds from GW are very interesting but we do not examine them further since they apply on the large mass window of PBHs,  beyond the scope of this work. Only for comparison with the other observational constraints we include only the  GW$_2$ bound of Ref. \cite{Saito:2008jc} ({\it before} erratum) and add a brief discussion on secondary GW in section \ref{8}. 
We note finally that 21 cm observations \cite{Mack:2008nv} could potentially provide a stronger constraint in the  mass range around $10^{14}$ g,  with $\beta'(M)<3\times10^{-29}(M/10^{14}\text{g})^{7/2}$  for $M>10^{14}\text{g}$ but such limits do not exist at present.
The combined upper bounds on $\beta'(M)$ and $f_\text{PBH}$ are collectively  depicted in Fig. \ref{YokoR}.

\subsection{Matter domination era}

If the PBH form during the stage of oscillation of the inflaton field, i.e.  matter domination (eMD) era,  the corresponding wavelength $k^{-1}$ enters the horizon before the complete decay of the inflaton and it is
\begin{align}\label{fmdreM} 
  f^{(\text{MD})}_\text{PBH}(M)  = \,\left(\frac{\beta_\text{MD}(M)}{2.1 \times 10^{-14}}\right) \, 
\left(   \frac{\Omega_{\text{DM}}h^2}{0.12}   \right)^{-1}  
 \Big(\frac{\gamma_\text{M}}{0.1}\Big)^{\frac{3}{2}} 
\left(\frac{g_*(T_\text{rh})}{106.75}\right)^{-\frac{1}{4}}  
\left(\frac{M}{10^{20}\text{g}}\right)^{-1/2}\,
         \left(  \frac{k}{k_\text{rh}} \right)^{-3/2} 
 \end{align}
The extra factor $\left( {k}/{k_\text{rh}}\right)^{-3/2}$ accounts for the different redshift of the  energy density of the the matter dominated universe compared to the radiation dominated universe. This scenario in the realistic framework of the inflationary $\alpha$-attractors has been examined in Ref. \cite{Dalianis:2018frf}.
In this case we find that 
\begin{align} \label{b-f2}
\gamma_\text{M}^{3/2}\left(\frac{g_*}{106.75}\right)^{-1/4}\, \beta_\text{MD}(M) \, =\, 6.5 \times 10^{-21} \, f^\text{(MD)}_\text{PBH} \left( \frac{M}{10^{10} \text{g}}  \right)^{1/2} \,
         \left(  \frac{k}{k_\text{rh}} \right)^{-3/2} \,.
\end{align}
As we will show in the Section 3, it is $k/k_\text{rh}\propto  \gamma^{1/3}_\text{M} g^{-1/6}_* T_\text{rh}^{-2/3} M^{-1/3}$, and the PBH dark matter fraction is written as,
\begin{align}
f_\text{PBH}^\text{(MD)}\, = \, 1.6 \times 10^{19} \, \gamma_\text{M}\,  \beta_\text{MD}(M)\,  \left(\frac{T_\text{rh}}{10^{10}\, \text{GeV}}\right)\,.
\end{align}
The $f_\text{PBH}^\text{(MD)}$ value has to be below the observational bound $f_\text{max}$ and, on the same footing with the definition  (\ref{betapR}), we define for the matter domination era, 
\begin{align}
\beta'_\text{MD}(M) \equiv \gamma_\text{M} \, \beta_\text{MD}(M)
\end{align}
which is independent of thermal degrees of freedom $g_*$ as it should. 
The corresponding observational bound on $\beta'(M)$ in terms of the $\boldsymbol{C_M}$ reads, 
\begin{align}
\beta'_\text{MD}(M) \,  <  \, 9.8 \left(\frac{T_\text{rh}}{10^{10}\, \text{GeV}}\right)^{-1}\left( \frac{M}{10^{10}\, \text{g}}\right)^{-1/2} \boldsymbol{C_M}\,
\equiv\, \beta'_\text{MD, max}(M, T_\text{rh})\,. 
\end{align}
Hence,  for $k_\text{rh}< k $,  or equivalently $M/\gamma_\text{R}<M_\text{rh}$, where $M_\text{rh}$ the horizon mass at the end of reheating, the maximum mass fraction of the universe allowed to collapsed into PBH is temperature dependent, see Fig. \ref{betaT}.

In the inflationary framework the upper bound on $\beta'(M)$ 
is effective only if the formation of PBH with mass $M$ is possible.
The horizon mass right after inflation is $M_\text{end}=4\pi M^2_\text{Pl}/H_\text{end} $ 
and the bounds are meaningful for PBH with masses $M>\gamma_\text{}\, M_\text{end}$, that is for 
\begin{align} \label{Hcon}
H_\text{end}\,>\, 1.33 \times 10^4 \,  \text{GeV} \left( \frac{M_\text{}/\gamma_\text{R}}{10^{10}\, \text{g}}\right)^{-1}\, 
\end{align} 
The above inequality yields a lower bound for the inflation energy scale.
A  PBH with mass $M$ will form due to superhorizon perturbation if the  corresponding wavelength $k^{-1}$ is larger than the horizon distance at the end of inflation. Thus a different way to write the condition (\ref{Hcon}) is 
\begin{align}
M>M_\text{end} \quad \text{or}  \quad k_\text{end}>k\,.
\end{align}

\begin{figure}[hbt!]
\ContinuedFloat
\begin{subfigure}{.5\textwidth}
  \centering
  \includegraphics[width=1.\linewidth]{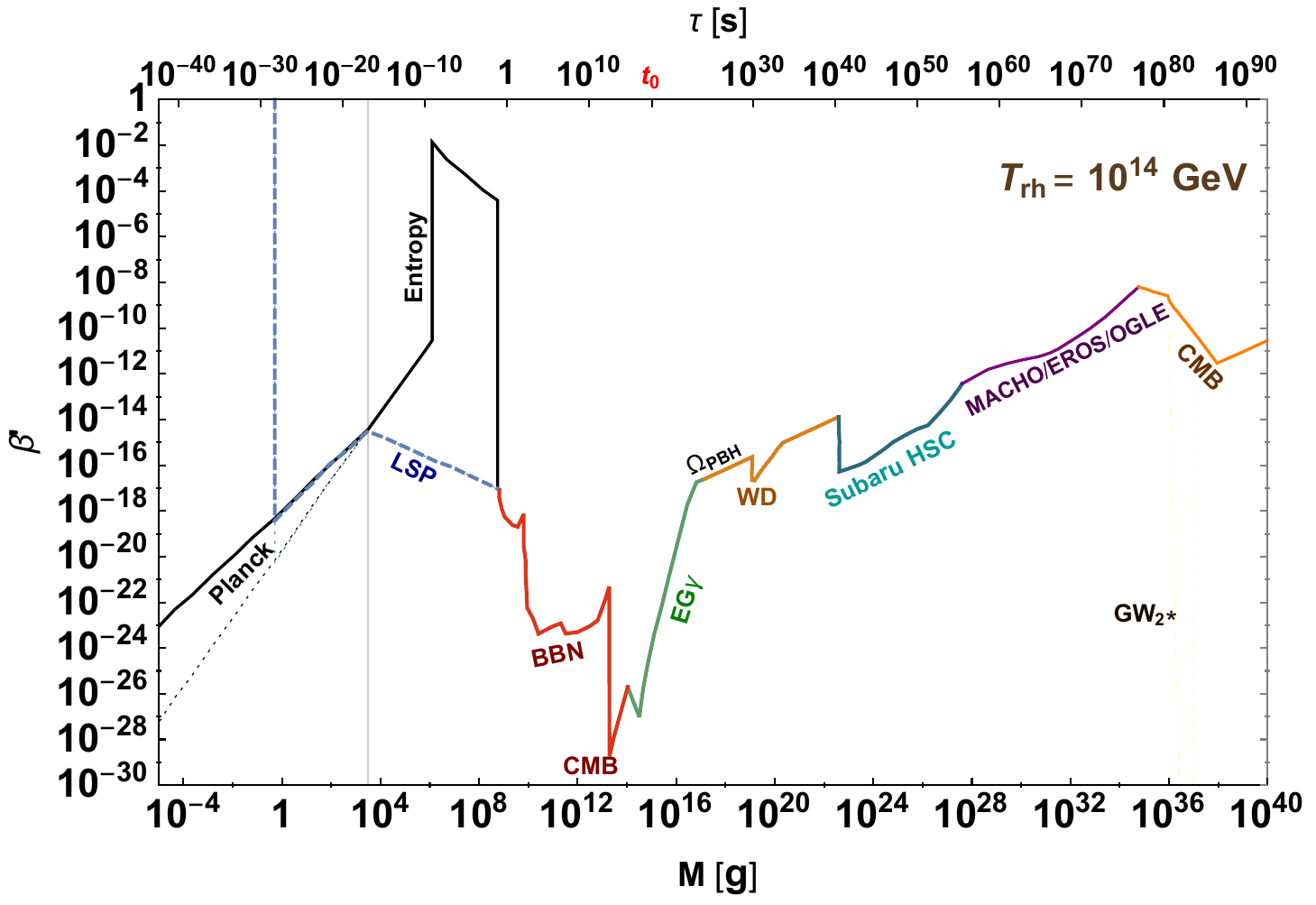}
\end{subfigure}%
\begin{subfigure}{.5\textwidth}
  \centering
  \includegraphics[width=1.\linewidth]{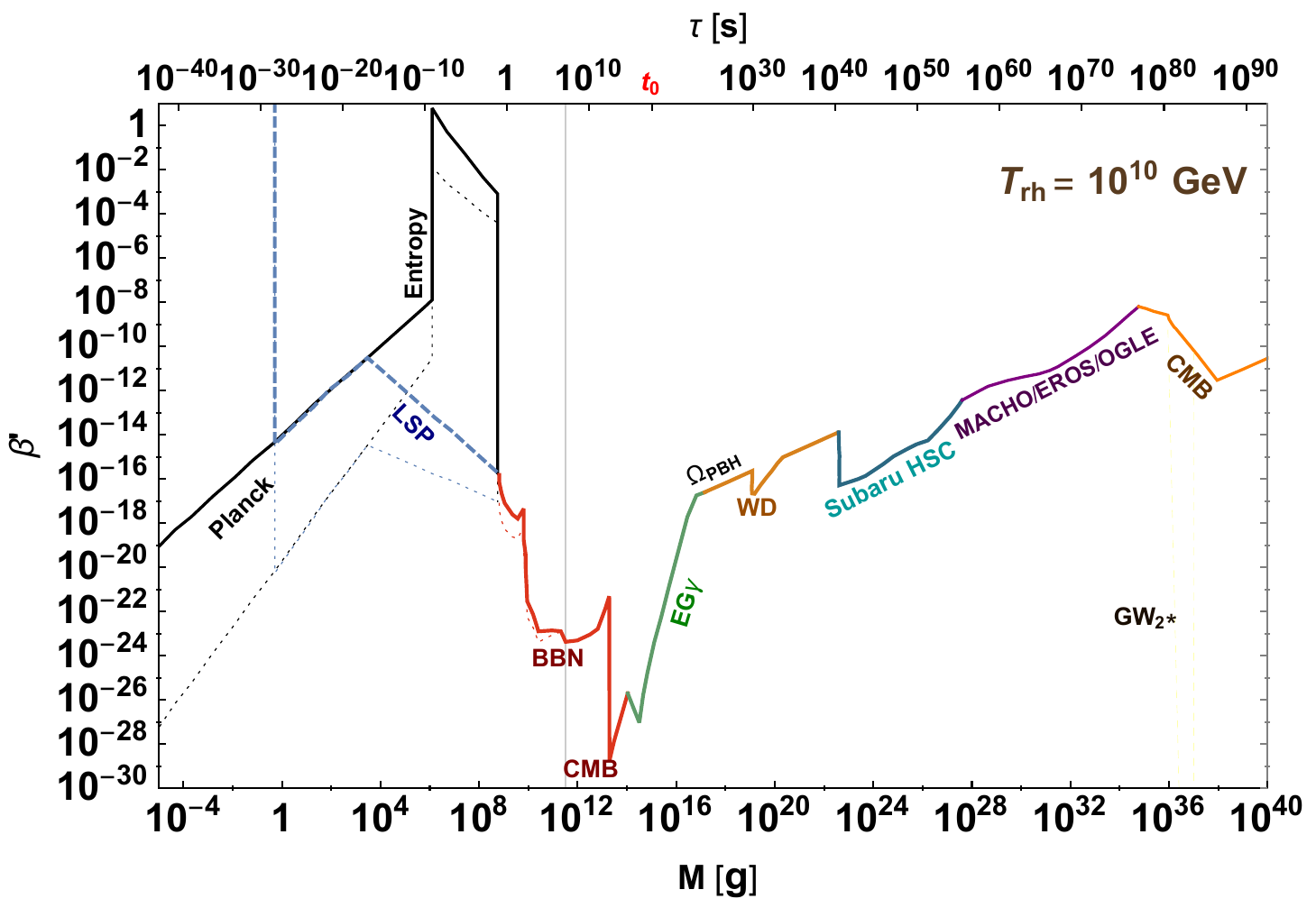}
\end{subfigure}
\\
\\
\\
\begin{subfigure}{.5\textwidth}
  \centering
  \includegraphics[width=1.\linewidth]{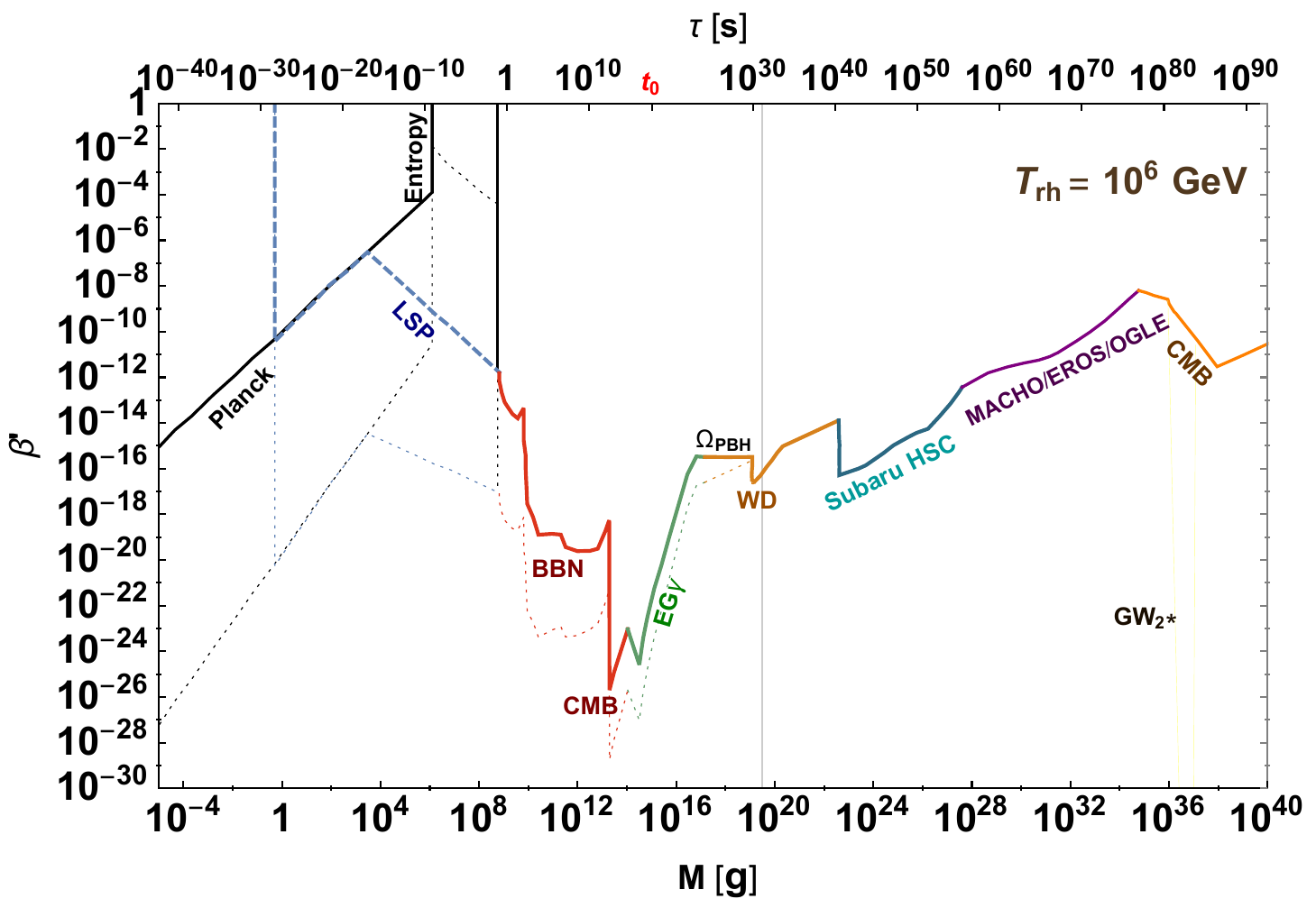}
\end{subfigure}%
\begin{subfigure}{.5\textwidth}
  \centering
  \includegraphics[width=1.\linewidth]{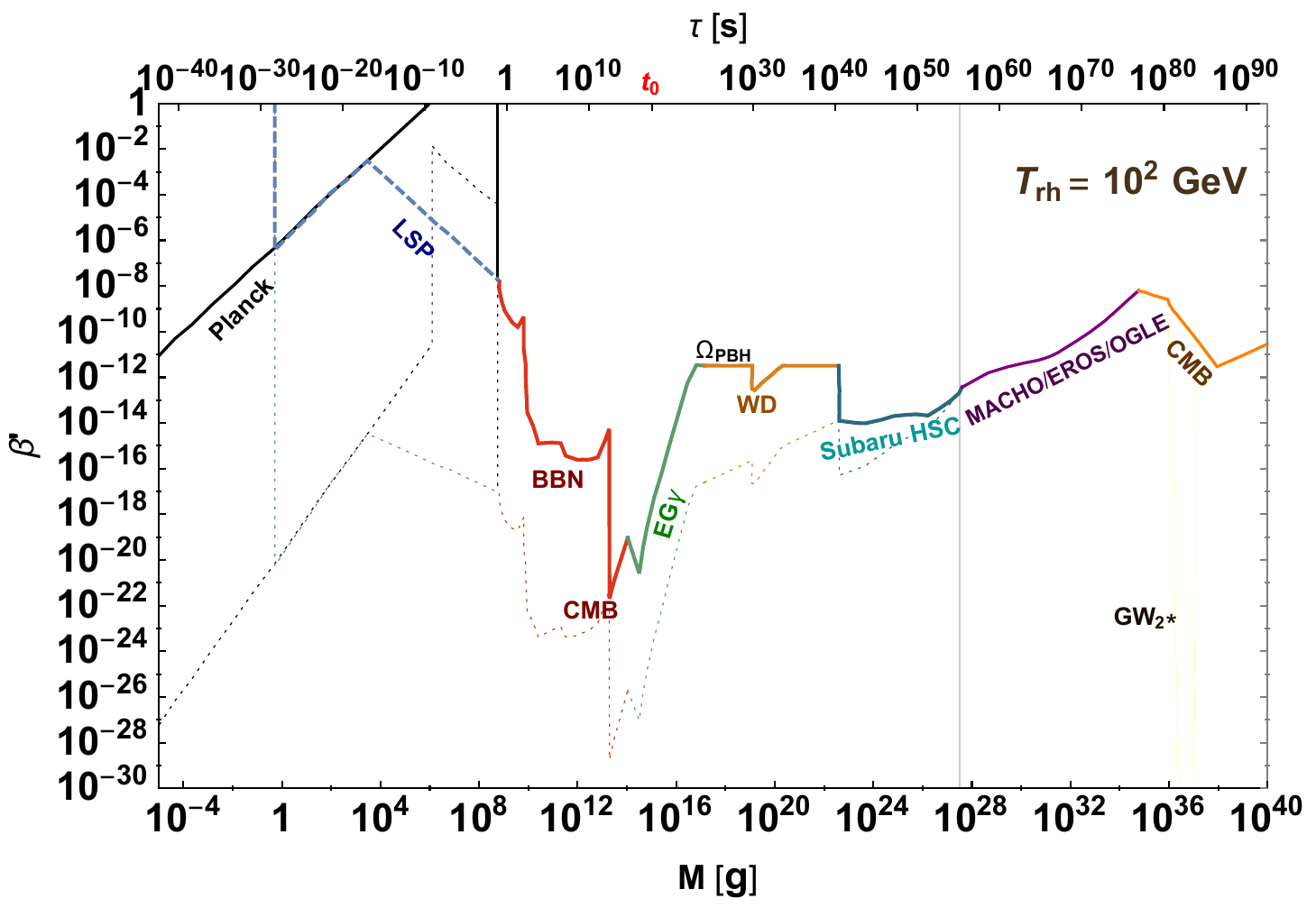}
\end{subfigure}
\caption{\label{betaT}~  
The plots depict combined upper bounds on $\beta'(M)$ for reheating temperatures $T_\text{rh}=10^{14}, 10^{10}, 10^{6}, 10^{2} $ GeV.  It is $\beta'(M)=\gamma^{3/2}_\text{R}(g_*/106.75)^{-1/4}\beta(M)$ for $M>\gamma_\text{R} M_\text{rh}$ and $\beta'(M)=\gamma_\text{M}\beta(M)$ for $M<\gamma_\text{R} M_\text{rh}$.
 The vertical line indicates the PBH mass forming at the epoch of reheating. The dotted lines depict the $\beta'(M)$ constraints for arbitrarily large reheating  temperature, as in Fig. \ref{YokoR}.
}
\end{figure} 
In the following we will find the expression $k(M)$ 
 in order to recast the $\beta'(M)$  bound into $\sigma(k)$ and put the constraints onto the power spectrum. This requires to determine the evolution of the cosmic horizon during RD and eMD eras.

\section{The relation between the PBH mass and the horizon scale} \label{3}

The relation between the mass  $M$ contained in the comoving horizon of size $k^{-1}$ is required in order to specify the PBH mass generated at that scale. This relation is also necessary in order to make contact of the $\beta(M)$ bounds with the power spectrum of the comoving curvature perturbation ${\cal P_R}(k)$.
In a radiation dominated universe the mass contained in the horizon of size $k^{-1}$ is   $M_\text{hor}(k) \simeq  M_\text{rh} \left( k/k_\text{rh} \right)^{-2}$, where $ M_\text{rh} $  the horizon mass  at the time of the reheating of the universe.
In the following we will determine the scales $k_\text{end}$ and $k_\text{rh}$. 
It is 
\begin{align} \label{end1M}
k_\text{end}=k_\text{rh} \, e^{\tilde{N}_\text{rh}/2}\,, \quad\quad  k_\text{rh}=k_\text{0.05} \, e^{\Delta\tilde{N}_{0.05}^\text{(RD)}} \left(\frac{g_*(T_\text{rh})}{g_*(T_{0.05})}\right)^{-1/6}
\end{align}
where,  for clarity,  we labeled  $k_{0.05}$  the Planck pivot scale $k_*=0.05$ Mpc$^{-1}\equiv k_{0.05}$ on the CMB sky \cite{Akrami:2018odb}.
$\tilde{N}_\text{rh}$ are the efolds that take place after inflation until the onset of the RD phase  and $\Delta \tilde{N}_\text{RD}$ are the efolds that take place from the onset of the continuous RD phase until the reentry epoch of the scale $k^{-1}_{0.05}$. At that time the thermal degrees of freedom are $g_*(T_{0.05})$.
In the case that an additional pressureless  non-thermal phase that last for $\tilde{N}_X$ efolds follows the reheating of the universe the  $k_\text{rh}$, given by Eq. (\ref{end1M}), is further displaced   $e^{\tilde{N}_X/2}$ times.

 The $\tilde{N}_\text{rh}$ and $\Delta \tilde{N}_{0.05}^\text{(RD)}$  are related with the inflationary dynamics by the expression
\begin{align} \label{twoNM}
 N_{0.05}+\ln\left( \frac{H_\text{end}}{H_{0.05}}\right)_\text{INF} = \frac12 \tilde{N}_\text{rh}+\Delta \tilde{N}_{0.05}^\text{(RD)}-\ln\left(\frac{g_*(T_\text{rh})}{g_*(T_{0.05})}\right)^{1/6}
\end{align}
In following sections we will consider the scenario of an early cosmic era dominated by a scalar-condensate with zero effective equation of state and the extra $\tilde{N}_X$ the e-folds have to be taken into account.

The number of e-folds during inflation $N_{0.05}$ are analyzed as \cite{Liddle:2003as, Dalianis:2018afb, Akrami:2018odb},
 \begin{align}
 N_{0.05} =67- \ln\left( \frac{k_{0.05}}{a_0 H_0}\right) +\frac14 \ln \left(\frac{V^2_{0.05}}{M^4_\text{Pl}\, \rho_\text{end}} \right) +\frac{1-3\bar{w}_\text{rh}}{12(1+\bar{w}_\text{rh})} \ln \frac{\rho_\text{rh}}{\rho_\text{end}} -
 \frac{1}{12}\ln \left({g_*}\right)\,.
 \end{align}
The measured value ${\cal P_R}=V_{
 0.05}/(24\pi^2 \epsilon_{0.05} M^4_\text{Pl}) =2.2 \times 10^{-9}$ gives that $\ln (V^{1/4}_{0.05} M^{-1}_\text{Pl}/\sqrt{3})=-4.2+1/4\ln(\epsilon_{0.05})$,
and $\ln(k_{0.05}/(a_0 H_0)) \simeq 5.4$.
For $\bar{w}_\text{rh}=0$ and recalling that  $H_{0.05}$ is written in terms of the inflationary observable $r_{0.05}$ as 
$H_{0.05}(r)=8.38\times 10^{13} \, \text{GeV}\times $ $\sqrt{ {r_{0.05}}/{0.1}}$
we attain
\begin{align} \label{radNM}
N_{0.05}+\ln\left( \frac{H_\text{end}}{H_{0.05}}\right)_\text{INF} 
=40.1+\frac12 \ln \left(\frac{H_\text{end}}{\text{GeV}}\right)-\frac14  \tilde{N}_\text{rh} - \frac{1}{12}\ln \left({g_*}\right)
\end{align}
Now, the $\tilde{N}_\text{rh}$ is related to the reheating temperature as $\rho_\text{rh}\, =\, \rho_\text{end}\,e^{-3\tilde{N}_\text{rh}} \,=\,\pi^2g_*T^4_\text{rh}/30\,$
thus
\begin{align}
\tilde{N}_\text{rh}(T_\text{rh},H_\text{end}, g_*)=-\frac43 \ln\left[\left(\frac{\pi^2 g_*}{90}  \right)^{1/4} \frac{T_\text{rh}}{(H_\text{end}M_\text{Pl})^{1/2}} \right]\,.
\end{align}
One sees that  $T_\text{rh}$ maximizes when $\tilde{N}_\text{rh}=0$ and $H_\text{end}$ is maximum. 
From Eq. (\ref{end1M}) and (\ref{twoNM}) we obtain the wavenumber that corresponds to the horizon mass at the end of inflation,
\begin{align} \label{kend}
 k_\text{end}\,(T_\text{rh},H_\text{end})\, \simeq \,
k_{0.05} 
\, e^{40.1} \,\left( \frac{H_\text{end}}{\text{GeV}}\right)^{1/3} \,\left[\left(\frac{\pi^2}{90}  \right)^{1/4} \frac{T_\text{rh}}{(\text{GeV}\,M_\text{Pl})^{1/2}} \right]^{1/3} 
\end{align}
Accordingly we obtain the horizon wavenumber $k_\text{rh}=k_\text{end}e^{-\frac12 \tilde{N}_\text{rh}}$  at the moment of reheating, $\Gamma_\text{inf}=H$,
\begin{align} \label{krhM}
 k_\text{rh}\,(T_\text{rh}, g_*)\, \simeq \,
k_{0.05} 
\, e^{40.1} \,\left(\frac{\pi^2}{90}  \right)^{1/4} \frac{T_\text{rh}}{(\text{GeV}\,M_\text{Pl})^{1/2}}  \, g_*^{1/6}(T_\text{rh})
\end{align} 
During the matter domination, the relation between the the scale $k^{-1}_\text{MD}$ and the horizon mass $M/\gamma_{M}$ is 
\begin{align} \label{kmd1}
k_\text{MD}=k_\text{end}\,  \left( \frac{4\pi M^2_\text{Pl}}{H_\text{end}}\right)^{1/3} \left(\frac{M}{\gamma_\text{M}}\right)^{-1/3}\,, \quad \text{for} \,\quad
k_\text{MD}>k_\text{rh}\,.
\end{align}
Utilizing the relation (\ref{kend}) and after normalizing the PBH mass, the reheating temperature and the relativistic degrees of freedom  we obtain for $N_X=0$,
\begin{align}\label{kmdM}  
k_\text{MD} (M, T_\text{rh}) \, =\,  7.1   \times 10^{17}  \, \text{Mpc}^{-1}
\gamma^{1/3}_\text{M} \,
 \left(\frac{M}{10^{10}\, \text{g}} \right)^{-1/3} 
   \left(\frac{T_\text{rh}}{10^{10}\, \text{GeV}} \right)^{1/3}
  \,.
\end{align}
After the completion of reheating the universe is in a thermal equilibrium state with temperature $T_\text{rh}$ and the radiation domination phase commends.  The horizon mass at that stage is $M_\text{rh}=M_{\text{hor}}(T_\text{rh}, g_*)$,
\begin{align}
M_\text{rh}=4\pi \left( \frac{\pi^2 g_*}{90}\right)^{-1/2} \frac{M^3_\text{Pl}}{T^2_\text{rh}}\,=\, 9.5 \times10^{11} \, \text{g}\, \left( \frac{T_\text{rh}}{10^{10}\, \text{GeV}}\right)^{-2} \left(\frac{g_*(T_\text{rh})}{106.75} \right)^{-1/2}\,.
\end{align}
During the RD era the relation between the  scale $k^{-1}_\text{RD}$  and the horizon mass $M/\gamma_\text{R}$ is 
\begin{align}
k_\text{RD}=k_\text{rh}\,  \left( \frac{M/\gamma_\text{R}}{M_{\text{rh}}}\right)^{-1/2}  \left(\frac{g_*}{g_*(T_\text{rh})} \right)^{-1/12}
\,, \, \quad\quad\quad \text{for} \quad k_\text{RD} <k_\text{rh} \,,
\end{align} 
where we substituted the expression for $k_\text{rh}$.  Plugging in numbers we obtain
\begin{align} \label{kRDm} 
k_\text{RD}(M)\, = \, 1.8 \times 10^{18}\, \text{Mpc}^{-1}\,  \gamma^{1/2}_\text{R} \,  \left(\frac{M}{10^{10}\, \text{g}} \right)^{-1/2}   \left(\frac{g_*}{106.75} \right)^{-1/12} \,.
\end{align}
Collectively we write the PBH mass $M$ and the horizon scale relation,
\begin{equation} \label{gen}
  k(M, T_\text{rh}, g_*) \, =
  \begin{cases} 
 \,   k_\text{MD}(M, T_\text{rh})
       \,,  \quad\quad\quad\quad\quad\quad\quad\quad\quad\quad\quad \text{for} \quad\quad k >k_\text{rh}\,  \\      \\ 
       
    \,  k_\text{RD}(M, g_*),
        \quad \quad\quad \quad\quad\quad\quad\quad\quad\quad\quad\quad \text{for} \quad\quad k <k_\text{rh}\,
  \end{cases}
   \end{equation}

\begin{figure}[!htbp]
\begin{subfigure}{.5\textwidth}
  \centering
  \includegraphics[width=1.\linewidth]{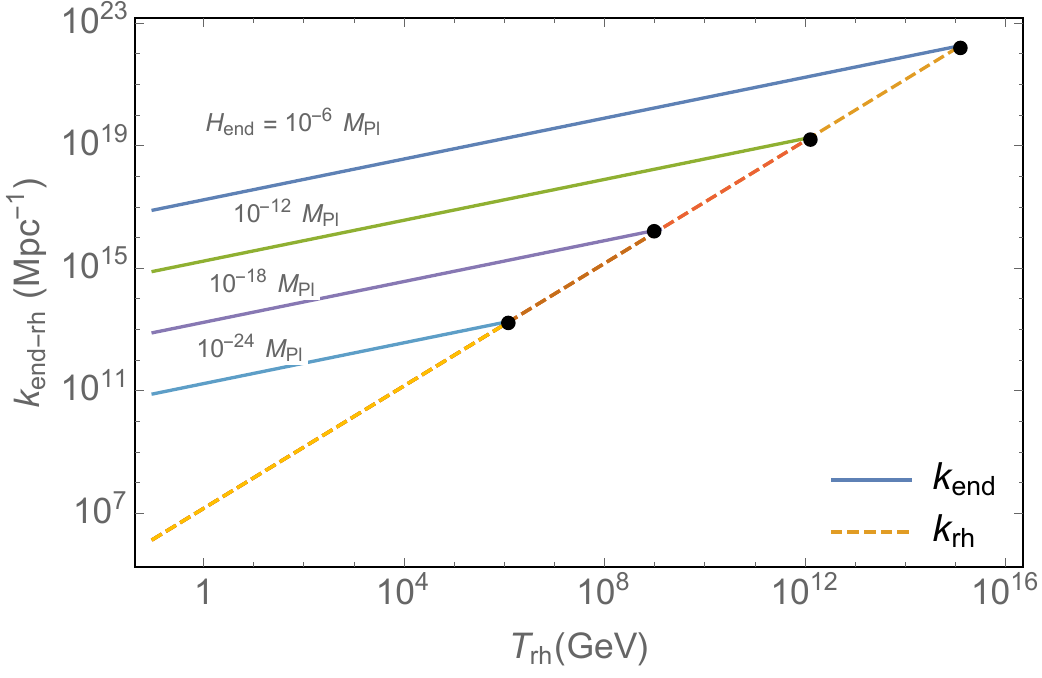}
\end{subfigure}%
\begin{subfigure}{.5\textwidth}
  \centering
  \includegraphics[width=1.\linewidth]{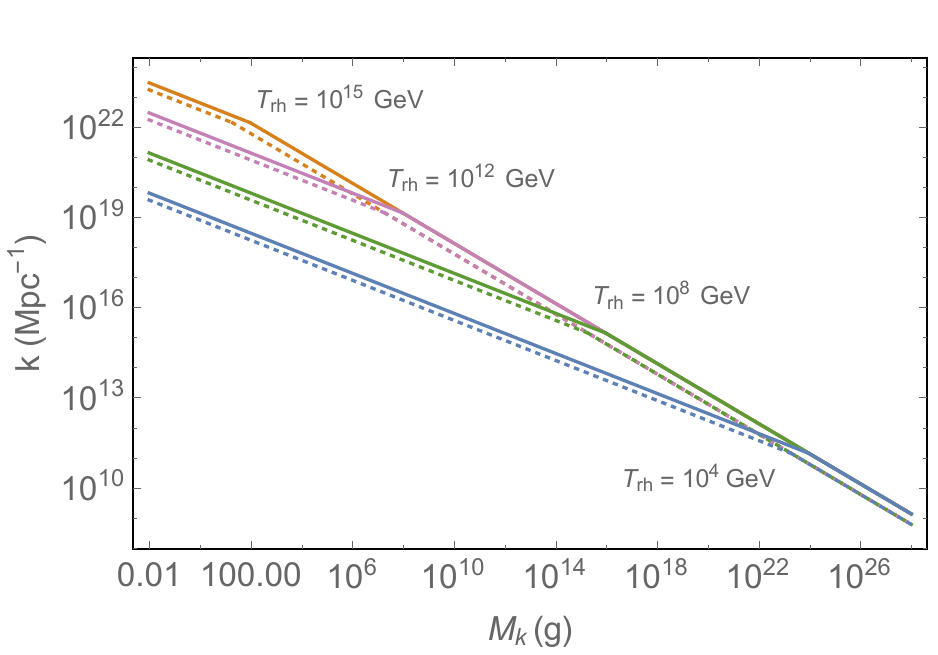}
\end{subfigure}
\caption{\label{NT}~ {\it Left panel}: 
The dependence  of the wavenumber at  the end of inflation, $k^{}_\text{end}$,  (solid lines)
and at the beginning of the radiation dominated era, $k^{}_\text{rh}$ (dashed line),  on the reheating temperature are depicted.  Four different values for the $H_\text{end}$ were considered. 
The black dots show the  $k^{}_\text{end}$, $k^{}_\text{rh}$ when the reheating temperature is maximum,  i.e. $N_\text{rh}= 0$.
 {\it Right panel}: 
The $k=k(M)$ relation (\ref{gen})  for four different reheating temperatures $T_\text{rh}$. The change of the slope happens at $k=k_\text{rh}$. For smaller $T_\text{rh}$ the smaller the $k_\text{end}$ at the end of inflation is. 
 The solid (dotted) line is for $\gamma_\text{R}=\gamma_\text{M}=1$ (0.2). 
}
\end{figure}

For a range of reheating temperatures the $k=k(M)$ relation is depicted in Fig.  \ref{NT}.
This is a necessary  relation  in order  someone to apply the $\beta(M)$ constraints onto an inflationary  model that yields a particular ${\cal P_ R}(k)$. 
Throughout this paper we assume  a one-to-one correspondence between the scale of perturbation and the mass of the PBHs.
Our analysis is supported by the findings of the Refs. \cite{Niemeyer:1997mt, Yokoyama:1998xd} that the typical mass of the PBHs is about the horizon mass at the moment of formation. Nevertheless, the work of \cite{Yokoyama:1998xd} points out that a tiny amount of black holes are created at the low-mass tail of the near-critical collapse. This finding is rather interesting, nevertheless in the current analysis we omit possible effects from PBHs in the low-mass tail.

\subsection*{The transition from matter to radiation}

A key quantity is the moment that the transition from the matter domination to radiation  takes place. If it happens instantaneously then the horizon mass at the transition epoch is equal to $M_\text{rh}$. But, the decay of the inflaton condensate is 
not an instantaneous process. It happens with a decay rate $\Gamma_\text{inf}$ and the completion is usually defined at the moment that $H=\Gamma_\text{inf}$. However 
radiation is gradually generated by the partial inflaton decay implies that the transition from eMD to RD may take place either before or after the moment $H=\Gamma_\text{inf}$. 

Let us define the moment of the transition as $H_\text{tr} =\Gamma_\text{inf}/\alpha $.  Then the efolds that take place from the end of inflation until the transition epoch are $N_\text{tr}=N_\text{rh}+\frac23 \, \ln\alpha$.
Also, the wavenumber at the transition is $k_\text{tr}=\alpha^{-1/3} k_\text{rh} $ and the horizon mass, $M_\text{tr}=\alpha^{2/3} \, M_\text{rh}$.
The moment of the transition might be when  the energy density of the universe is equally partitioned between the inflaton condensate and the entropy produced by the inflaton decay, or when the probability for PBH formation coincides for the two production mechanisms \cite{Carr:2018nkm}. In the first case it is $\alpha>1$
 since about 21$\%$ of the (comoving) energy density of the inflaton condensate has been extracted at reheat time \cite{Giudice:2000ex}, thus 
$N_\text{tr} > N_\text{rh}$.  In the second case it is $\alpha<1$ and $N_\text{tr} \simeq 90\%\, N_\text{rh}<N_\text{rh}$ \cite{Carr:2018nkm}. 
For clarity and simplicity, in the following analysis we approximate  $N_\text{tr} = N_\text{rh}$, $k_\text{tr} = k_\text{rh}$ and $M_\text{tr} = M_\text{rh}$.

\section{The upper bounds for the variance of the density perturbation} \label{4}

PBH formation is possible during the early stages of the Universe when superhorizon fluctuations in the curvature of spacetime cross into the horizon  and collapse under their own self-gravitation.  We will assume the approximation that the mass distribution of the PBH formed is contracted about the horizon mass.
In the heated universe the PBHs are expected to form with mass $M=\gamma_\text{R} M_\text{hor}$ when the cosmic temperature is
\begin{align} \label{TM}
T(M)= 9.7  \times 10^{10} \, \text{GeV}\, \gamma^{1/2}_\text{R}\, \left(\frac{M}{10^{10}\,\text{g}}\right)^{-1/2} 
\left( \frac{g_*}{106.75}\right)^{-1/4}\,\,.
 \end{align}   
For $M=5\times 10^{10}$ g we define the BBN critical temperature where PBHs of that mass form,
\begin{align}
T_\text{bbn}\, \equiv\, 4.3  \times 10^{10} \, \text{GeV}\ \gamma^{1/2}_\text{R}\,
\left( \frac{g_*}{106.75}\right)^{-1/4}\,.
\end{align}
PBHs with such a mass emit a flux of Hawking thermal radiation in a timescale $\tau(M)$ which might alter the BBN observables. 
Respectively, for $M=2.5 \times 10^{13}$ g we define the CMB critical temperature,
\begin{align}
T_\text{cmb}\, \equiv\, 1.9  \times 10^{9} \, \text{GeV}\ \gamma^{1/2}_\text{R}\,
\left( \frac{g_*}{106.75}\right)^{-1/4}\,,
\end{align}
 that PBHs with lifetime $\tau(M)$, characteristic of the CMB physics timescale, form.

\subsection{Radiation domination era}
The $\beta'(M)< \boldsymbol{C_M} $ bound applies for
PBH with mass  $ M >\gamma_\text{R} M_\text{hor}(T_\text{rh} )$, or  equivalently for scales $k^{-1}$  
that enter after the completion of the reheating phase, i.e. $k < k_\text{rh}$.
From Eq. 
(\ref{krhM}) this is recast into the condition for the reheating temperature,
\begin{align} \label{kcrh}
T_\text{rh}> 9.7  \times 10^{10} \, \text{GeV}\, \gamma^{1/2}_\text{R}\, \left(\frac{M}{10^{10}\,\text{g}}\right)^{-1/2} 
\left( \frac{g_*}{106.75}\right)^{-1/4}\,
 \end{align}  

 For that large reheating temperatures the mass fraction $\beta'$ of the universe that collapses into PBH has to be smaller than $ \boldsymbol{C_M}$,
 \begin{align} \label{cbr1}
 \beta_\text{RD}(M)\, \simeq  \, \frac{1}{\sqrt{2\pi}}\frac{\sigma(M)}{\delta_c} \,e^{-\delta_c^2/2\sigma^2(M)}\,<\, \gamma_\text{R}^{-3/2}\left( \frac{g_*}{106.75}\right)^{1/4} \,  \boldsymbol{C_M}\,\equiv\, \beta_\text{RD, max}(M, \gamma_\text{R}, g_*)\,.
 \end{align}
We have assumed that the fluctuations at horizon crossing are Gaussian with variance $\sigma(M)$. 
A black hole forms if the density contrast at horizon crossing $k=aH$ exceeds a critical value $\delta_c$. The value of $\delta_c$ varies in the literature, e.g. $\delta_c=1/3$ or 0.45 are used which means that black holes form only from the tail of the density fluctuation distribution and the overdense regions are likely to be spherical.  In the comoving gauge Ref. \cite{Harada:2013epa} finds that 
\begin{equation} \label{dc}
\delta_c = \frac{3(1+w)}{5+3w}\sin^2 \frac{\pi \, \sqrt{w}}{1+3w}\,.
\end{equation}
For $w=1/3$ it is $\delta_c=0.41$.
We note that different values for $\delta_c$ are cited in the literature, see also \cite{Niemeyer:1997mt,  Shibata:1999zs, Musco:2008hv,  Musco:2012au}.
The numerical value of the $\gamma_\text{M}$ is unknown.
It depends on the details of gravitational collapse. Simple analytical calculation
suggests that it is $\gamma_\text{R}\sim 0.2$ \cite{Carr:1975qj}.
Critical phenomena, such as the reduction of pressure could reduce the $\gamma_\text{M}$ value \cite{Khlopov:1980mg, Polnarev:1986bi}. 
In our analysis we leave the $\gamma$ numerical values unspecified 
and for clarity in some expressions we will normalize the $\gamma_\text{R}$ with $0.2$ and the $\gamma_\text{M}$ with 0.1.

\subsubsection*{BBN and CMB constraints}
The $\beta(M)$ depends very mildly on the degrees of freedom thus for simplicity and without cost in accuracy we assume below that $g_*=106.75$. For $|\ln(\boldsymbol{C_M}/\sqrt{\gamma_\text{R}})/|\gg |\ln (\sigma/\delta_c)|$ the (\ref{cbr1})  rewrites
\begin{align} \label{sRD}
\sigma(M)\, < \, \frac{\delta_c}{\sqrt{2}} \left[\ln\left(\frac{\sqrt{\gamma_\text{R}}}{\sqrt{2\pi}\, \boldsymbol{C_M} } \right)\right]^{-1/2}\, \equiv \sigma_\text{RD, max}(M, \delta_c, \gamma_\text{R})\,.
\end{align}
For $M\simeq 5\times 10^{10}$ g the BBN constraint is $\boldsymbol{C_M} \simeq 10^{-24}$ and for $M\simeq 2.5\times 10^{13}$ g the CMB constraint is $\boldsymbol{C_M}  \simeq 5\times 10^{-29}$, hence we can obtain the  $\sigma(5\times10^{10}\text{g})$ and  the $\sigma(2.5 \times10^{13}\text{g})$ bound respectively.  
Using the expression (\ref{kRDm}) the constraints read in the momentum space,
\begin{align}
&  \label{kspacBBN}
 \sigma\,(7.4 \times 10^{18}\, k_{0.05})\, \lesssim \,  0.038 \, \left(\frac{\delta_c}{0.41}\right) \left[1+0.028\ln\left(\frac{\gamma_\text{R}}{0.2} \right)\right]^{-1/2}\,,\,\quad\quad\quad\quad \text{(BBN)}
\end{align}
for $T_\text{rh}>4.3 \times 10^{10} \,\gamma_\text{R}^{1/2}$ GeV, and
 \begin{align}
\label{kspacCMB}
\sigma\,(3.7 \times 10^{17}\, k_{0.05}) \, \lesssim \, 0.035 \, \left(\frac{\delta_c}{0.41}\right) \left[1+ 0.023\ln\left(\frac{\gamma_\text{R}}{0.2} \right)\right]^{-1/2}
\,,\quad\quad\quad\quad \text{(CMB)}
\end{align}
for  $T_\text{rh}>2.2 \times 10^{9}\, \gamma_\text{R}^{1/2}$ GeV respectively.

Numerically we find that 
the approximation (\ref{sRD}) differs from the exact only about 0.1$\%$. 
In the next section, where the gravitational collapse during eMD era will be examined, we will see that the constraints (\ref{kspacBBN}) and (\ref{kspacCMB}) apply also for smaller reheating temperatures -about two orders of magnitude smaller-  due to the finite time required  for the gravitational collapse.

The above constraints on the variance of perturbations can be applied on the power spectrum of the primordial comoving curvature perturbations.  An explicit constraint on ${\cal P_R}$ can be found only if the ${\cal P_R}$ is known in a range of momenta $k$. Also, one has to  consider a window function to smooth the density contrast.
During RD the relation between the variance of the comoving density contrast and the ${\cal P_R}$ reads
\begin{equation}
\sigma^2(k)= \left( \frac{4}{9} \right)^2  \int \frac{dq}{q}\,W^2\left(\frac{q}{k}\right)\left(\frac{q}{k}\right)^4{\cal P_R}(q)\,,
\end{equation}
where $W(z)$ represents the Fourier transformed function of the Gaussian window,  $W(z)=e^{-z^2/2}$. 
For an order of magnitude estimation we can approximate $\sigma \sim (4/9) {\cal P_R}^{1/2}$ and the constraints on the power spectrum read in momentum space
${\cal P_R}(k)\lesssim {\cal O}(10^{-3})$.
 Increasing the $\delta_c$ value the bounds become weaker, for example for $\delta_c=0.5$  the bounds on
${\cal P_R}$ are relaxed  1.5 times.
\begin{figure}[!htbp]
\begin{subfigure}{.5\textwidth}
  \centering
  \includegraphics[width=1.\linewidth]{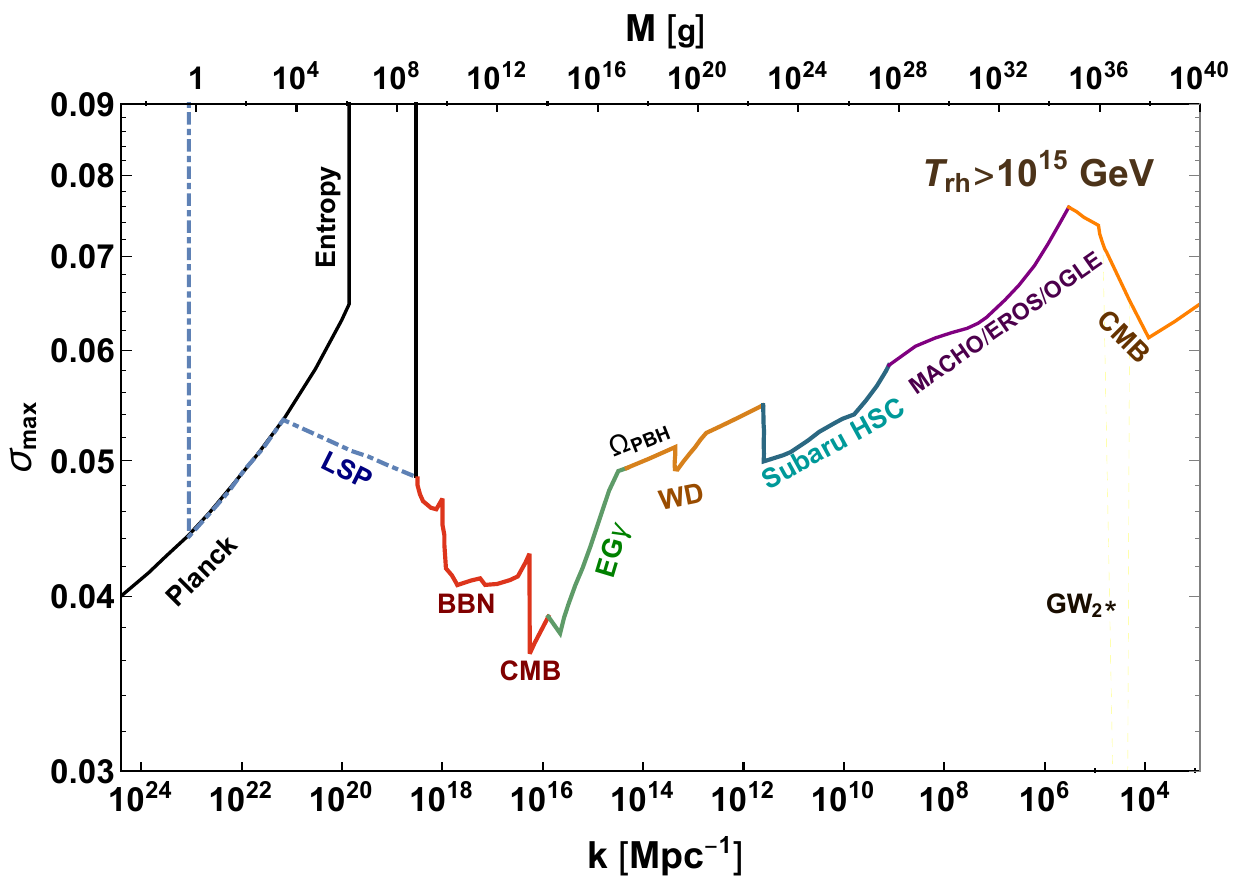}
\end{subfigure}%
\begin{subfigure}{.5\textwidth}
  \centering
  \includegraphics[width=1.\linewidth]{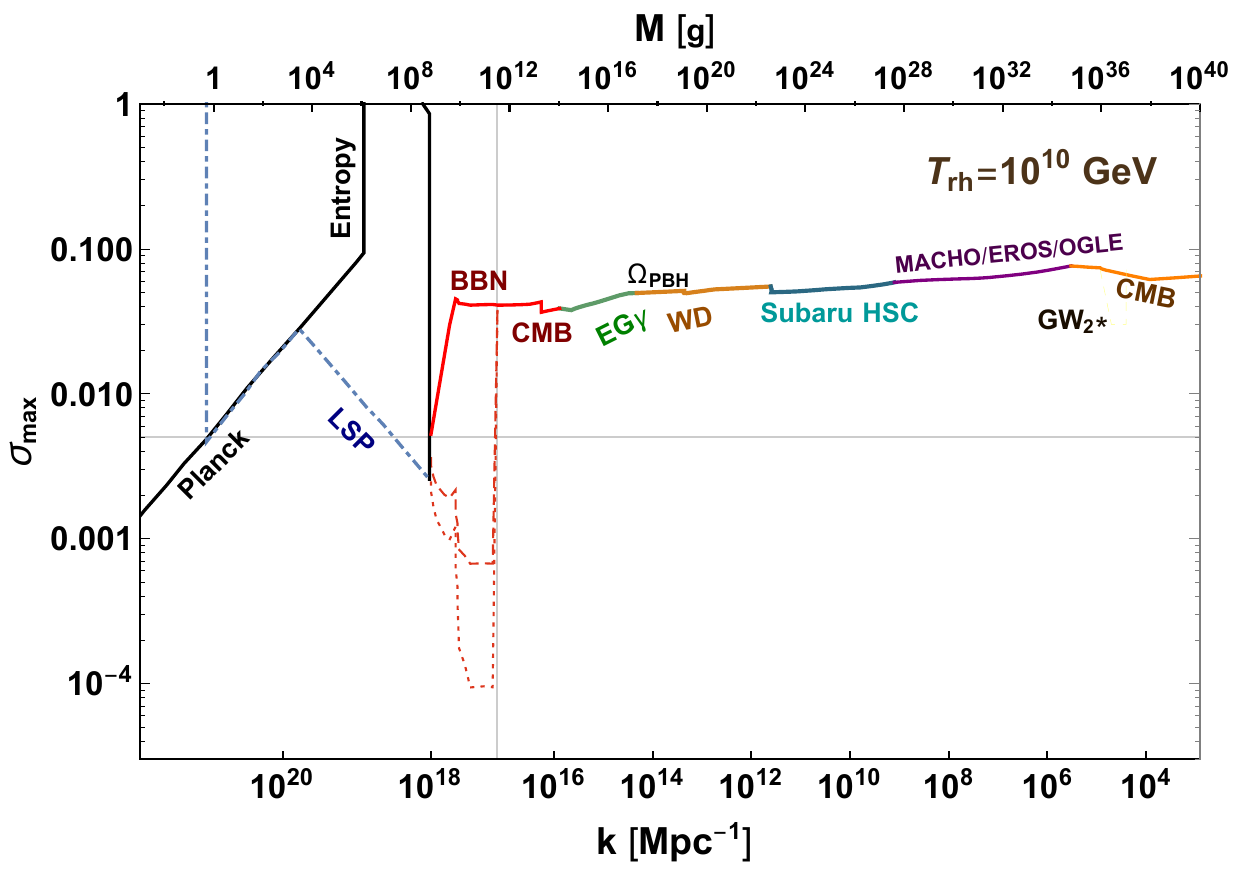}
\end{subfigure}
\\
\\
\begin{subfigure}{.5\textwidth}
  \centering
  \includegraphics[width=1.\linewidth]{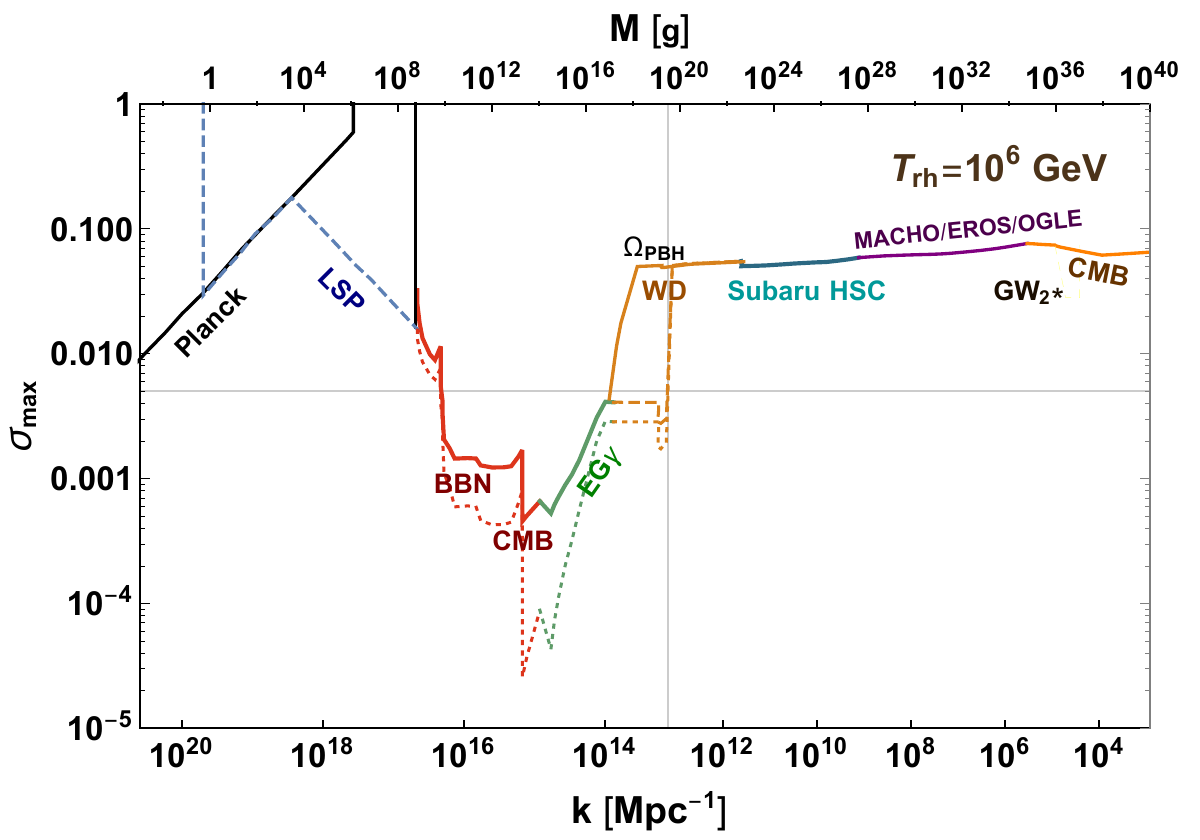}
\end{subfigure}%
\begin{subfigure}{.5\textwidth}
  \centering
  \includegraphics[width=1.\linewidth]{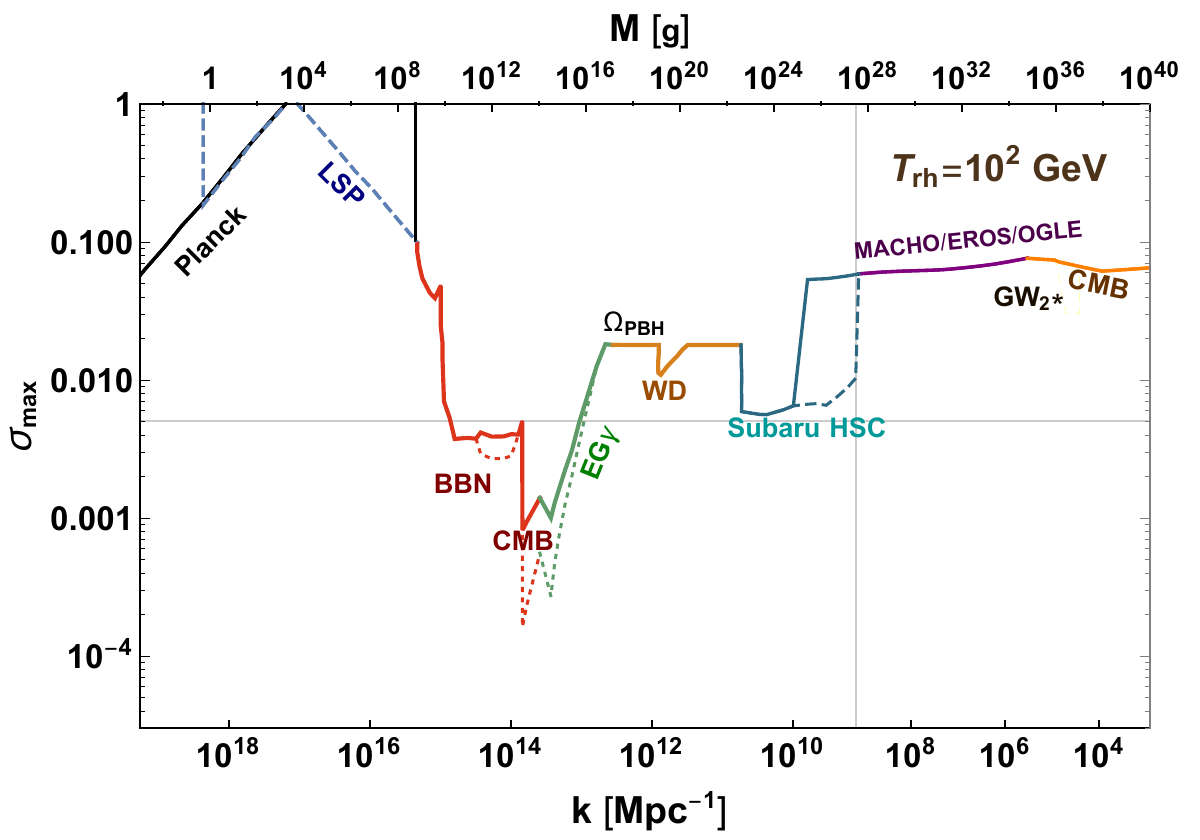}
\end{subfigure}
\caption{\label{sigmaT}~ The plots depict combined upper bounds on $\sigma(k)$ and  $\sigma(M)$ for reheating temperatures arbitrary large (upper left panel) and $T_\text{rh}= 10^{10}, 10^{6}, 10^{2} $ GeV. The vertical line indicates the PBH mass forming at the epoch of reheating and the horizontal line the threshold value  $\sigma_\text{thr}= 0.005$. The dotted lines below $\sigma_\text{thr}$  depict the $\sigma(M)$ upper bound for vanishing spin.  The dashed lines next to the reheating scale show the maximum $\sigma$ if one neglects the finite time for the gravitational collapse. 
In the plots benchmark values, $\gamma_\text{R}=0.2$, $\gamma_\text{M}=0.1$, $\delta_c=0.41$ have been used.
}
\end{figure}

\subsection{Matter domination era}
A presureless matter domination era is naturally realized in the early universe due to the coherent oscillations of the inflaton or other scalar fields.
 During matter era  the Jeans pressure is negligible and scalar perturbations, that would be minor in  the radiation domination era, can grow linearly with  the scalar factor and lead to PBH formation.
If the reheating temperature is 
\begin{align} \label{kcrh2}
T_\text{rh} < 9.7 \times 10^{10} \, \text{GeV}\, \gamma^{1/2}_\text{R}\, \left(\frac{M}{10^{10}\,\text{g}}\right)^{-1/2} 
\left( \frac{g_*}{106.75}\right)^{-1/4}\,,
 \end{align}  
 then the formation rate of PBH with mass  less than $\gamma_\text{R} M_\text{rh}$  might change drastically.  In addition the PBH abundance formed during eMD era scales differently with time. 
For scales $k$ that enter during early matter domination  the bound is  $\beta'_\text{MD}(M_\text{})   \, < \, \gamma_\text{M}^{-1/2} (g_*/106.75)^{1/4}  
\left( {k}/{k_\text{rh}}\right)^{3/2}
 \, \boldsymbol{C_M} $, 
and in terms of $M$ and $T_\text{rh}$  is recast into
\begin{align} \label{betaMDbound}
\beta_\text{MD}(M) < \boldsymbol{{C}_M}\, (4.58)^{3/2}\, \gamma_\text{M}^{-1}\left( \frac{T_\text{rh}}{10^{10}\text{GeV}} \right)^{-1}
 \left( \frac{M_\text{}}{10^{10}\,\text{g}} \right)^{-1/2} \equiv \beta_\text{MD, max}( \boldsymbol{C_M}, M, T_\text{rh}, \gamma_\text{M}) 
\end{align}
The 
 $\beta'_\text{MD}(M)$ 
constraint is depicted in Fig. \ref{betaT}.

PBHs formed during a pressureless matter dominated (MD) era has been considered in Ref. \cite{Khlopov:1980mg, Polnarev:1986bi, Cotner:2017tir, Georg:2017mqk,  Carr:2017jsz}.  
Employing the results of Ref. \cite{Harada:2016mhb} for spinless gravitational collapse during eMD era the formation rate, which depends on the fraction of the regions which are sufficiently spherically symmetric,  is given by
\begin{align} \label{betaMD}
\beta_\text{MD}(M_\text{})   \, \simeq  \, 0.056\,  \sigma^5(M_\text{})\,.
\end{align}
We comment that in Ref. \cite{Khlopov:1980mg, Polnarev:1986bi} an additional suppression factor $\sigma^{3/2}$, refined also in Ref. \cite{Kokubu:2018fxy},  is included to take into account inhomogeneity effects that we do not consider here, following Ref. \cite{Harada:2016mhb}.
The PBH production rate in the eMD era is larger than that in the RD era for $\sigma \lesssim 0.05$ whereas for larger variance, due to the absence of relativistic pressure, the nonspherical effects  suppress the PBH formation rate in eMD eras\cite{Harada:2016mhb}.
From Eq. (\ref{betaMD}) we  attain a relation that relates the variance of the comoving density contrast $\sigma(M)$ to the observational bound $\boldsymbol{C_M}$,  
\begin{align} \label{Cs3}
\left. \sigma_\text{MD}(M)\right|_\text{spinless}  \, < \, 2.8  \,\, \gamma^{-1/5}_\text{M} \left( \frac{T_\text{rh}}{10^{10} \text{GeV}} \right)^{-1/5} 
\left( \frac{M}{10^{10} \text{g}} \right)^{-1/10} 
 \,\boldsymbol{{C}_M}^{1/5} 
\end{align}

The PBH production rate is modified when the collapsing region has spin. The angular momentum suppresses  the formation rate which now  reads\footnote{The erratum of \cite{Harada:2017fjm} has been taken into account.} \cite{Harada:2017fjm},
\begin{equation} \label{spin}
\beta_\text{MD}(M)=2\times 10^{-7}f_q(q_c) {\cal I}^6 \sigma(M)^2 e^{-0.147\frac{ \, {\cal I}^{4/3}}{ \sigma(M)^{2/3}}}\,.
\end{equation}
Benchmark values are $q_c=\sqrt{2}$,  $f_q \sim 1$ and ${\cal I}$ is a parameter of order unity \cite{Harada:2017fjm}. According to \cite{Harada:2017fjm} this expression applies for $\sigma(M) \lesssim 0.005 \equiv \sigma_\text{thr}$, whereas the equation $\beta_\text{MD}(M)   \, \simeq  \, 0.056\,  \sigma^5(M_\text{})$ applies for $0.005 \lesssim \sigma(M) \lesssim 0.2$.

\subsubsection*{The finite duration of the PBH formation}
An additional  critical parameter is the duration of the gravitational collapse.
PBH formation is strongly suppressed by a centrifugal force and it completes, that is to enter into the nonlinear regime, only if the eMD era lasts sufficiently long. According to  \cite{Harada:2017fjm}
the finite duration of the PBH formation can be neglected if the reheating time $t_\text{rh}$ satisfies
\begin{align}
t_\text{rh} >\left( \frac{2}{5} {\cal I} \, \sigma\right)^{-1} t\,,
\end{align}
where $t$ is the time of the horizon entry of the scale $k^{-1}$. In terms of wavenumbers and temperatures the above condition rewrites respectively,
\begin{align}
k_\text{rh} < \left( \frac{2}{5} {\cal I} \, \sigma\right)^{1/3} k \quad\quad \text{or} \quad\quad T_\text{rh}< \left( \frac{2}{5} {\cal I} \, \sigma\right)^{1/2} T\,,
\end{align}
where $T$ the temperature that the scale $k^{-1}$ enters the horizon. If these conditions are not fulfilled then the time duration for the overdensity to grow and enter the nonlinear regime is not adequate.
Due to the fact that the collapse does not happen instantaneously after the horizon crossing  the formation rate (\ref{spin}) applies only for the scales $k$
 that experience a variance of the  comoving density contrast at horizon entry that is larger than 
 \begin{align}
 \sigma > \sigma_\text{cr} \equiv \frac52 {\cal I}^{-1} \left( \frac{k_\text{rh}}{k}\right)^3\,.
 \end{align}
In terms of temperature this translates into $\sigma>5/2 \, {\cal I}^{-1} (T_\text{rh}/T)^2$. If $\sigma<\sigma_\text{cr}$ we will consider that the formation rate is that of the radiation era and in our numerics we will choose ${\cal I}=1$.

\subsubsection*{BBN and CMB constraints}

The constraints on the variance during eMD  apply only for those perturbations that have enough time to  gravitationally collapse during the reheating era, since the collapsing process is not instantaneous. 
For a given scale $k^{-1}$ and variance $\sigma(k)$ there is a maximum reheating temperature that the collapse is realized during the eMD era. 
For PBHs  to form during eMD  with masses, $M_\text{bbn}, M_\text{cmb}$,  associated with the BBN and CMB constraints, respectively,  it has to be 
\begin{align}\label{Tform}
T_\text{rh} < T_\text{bbn}^\text{(MD)} \equiv \left( \frac25\,{\cal I}\right)^{1/2} \,T_\text{bbn} \, \sigma^{1/2}(M_\text{bbn})\,,  \quad\quad
T_\text{rh} < T_\text{cmb}^\text{(MD)} \equiv \left( \frac25\,{\cal I}\right)^{1/2} \, T_\text{cmb} \, \sigma^{1/2}(M_\text{cmb})
\end{align}
Otherwise, the upper bound on the variance should be determined by  the RD era dynamics since we expect the relativistic pressure at times $\Gamma^{-1}_\text{inf}$ to cause a bounce on the ongoing collapsing process. 
Thus, the constraints on the variance of the density perturbations  given by Eq. (\ref{kspacBBN}) and (\ref{kspacCMB}) apply respectively for $T_\text{rh}> T_\text{bbn}^\text{(MD)}$ and $T_\text{rh}> T_\text{cmb}^\text{(MD)}$.  
The CMB constraint, as will be discussed below,  is the stringent one  except if the reheating temperature is in the window
\begin{align}
T_\text{cmb}^\text{(MD)} \,\,\,
 \lesssim T_\text{rh} \,\, 
 \lesssim \;\;
 T_\text{bbn}^\text{(MD)}\,,
\end{align}
For such reheating temperatures the PBH that form during eMD influence the BBN but not the CMB observables.

The eMD variance of the comoving density contrast at horizon entry is constrained by the BBN and CMB observables for $T_\text{rh}<T_\text{bbn}^\text{(MD)}$ and $T_\text{rh}<T_\text{cmb}^\text{(MD)}$ respectively at the scales, 
\begin{align} \label{kspacM}
& \sigma(5\times 10^{10}\text{g}) \,=\, \sigma\left(10^{19}\, \gamma_\text{M}^{1/3}\left( \frac{T_\text{rh}}{10^{10}\, \text{GeV}}\right)^{1/3} k_{0.05}\right)\\
& \sigma(2.5\times 10^{13}\text{g}) \,= \,\sigma\left(3\times 10^{18}\, \gamma_\text{M}^{1/3}\left( \frac{T_\text{rh}}{10^{10}\, \text{GeV}}\right)^{1/3} k_{0.05}\right)\,.
\end{align}
In order to recast the mass fraction $\beta(M)$ constraints  into constraints on the variance $\sigma(M)$ we have to solve the inequality 
\begin{align}
\beta_\text{MD}(\sigma(M)) \, < \, \beta_\text{MD, max}( \boldsymbol{C_M}, M, T_\text{rh}, \gamma_\text{M}) \,.
\end{align}
However, an analytic solution can be found only for the case of spinless gravitational collapse. For the spinning case any analytic approximation is not accurate enough and numerical solutions have to be pursued. 
It is actually the spin effects that determine the maximum value for the variance $\sigma$ and cannot be ignored.

\begin{figure}[!htbp]
\begin{subfigure}{.5\textwidth}
  \centering
  \includegraphics[width=1.\linewidth]{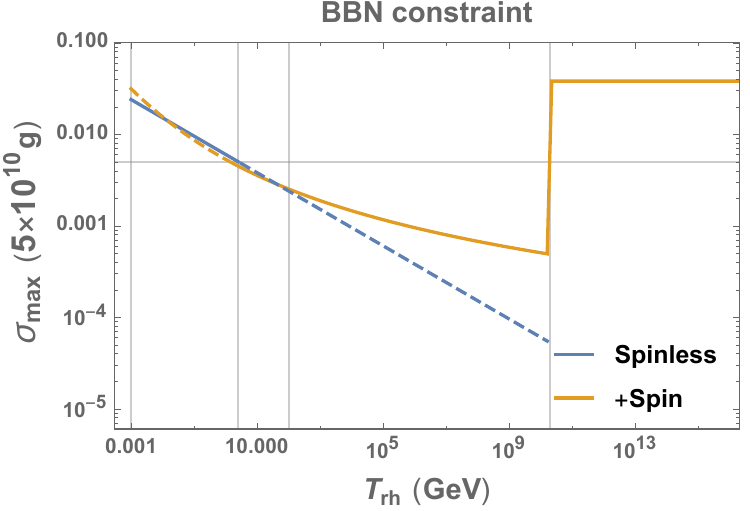}
\end{subfigure}%
\;
\begin{subfigure}{.5\textwidth}
  \centering
  \includegraphics[width=1.\linewidth]{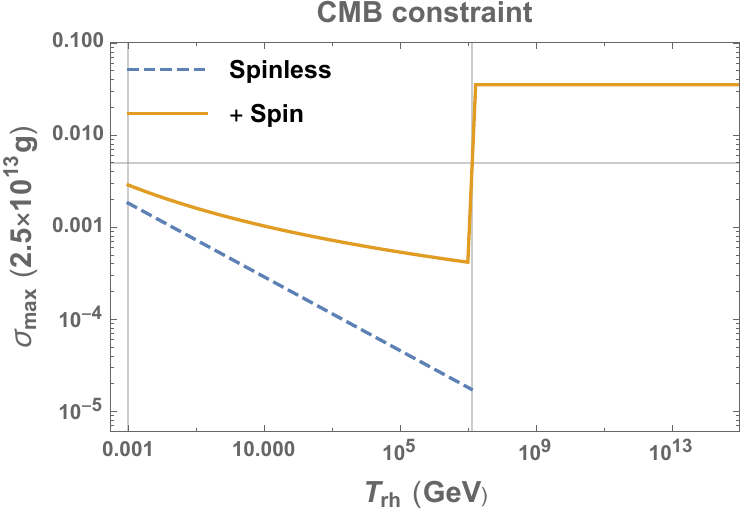}
\end{subfigure}
\caption{\label{sigmabbncmb}~  {\it Left panel}: 
The maximum $\sigma$ for $M=5\times 10^{10}$ g.  The blue line depicts the $\sigma$ for spinless collapse and the orange when spin effects are included. The solid line gives the correct upper bound if spin is considered. 
{\it Right panel}: The maximum $\sigma$ for $M=2.5\times 10^{13}$ g. 
Benchmark values $\gamma_\text{M}=0.1$ and  $\gamma_\text{R}=0.2$ have been assumed. The horizontal gridline depicts the $\sigma_\text{thr}=0.005$ threshold.  The change of the formation probability is determined  respectively by the temperatures $T^\text{(MD)}_\text{bbn}$ and  $T^\text{(MD)}_\text{cmb}$, see Eq. (\ref{Tformmax}).
} 
\end{figure}

Let us first calculate the constraints on the variance  of the comoving density contrast for the spinless collapse approximation, i.e. for  $\sigma>0.005$. It is
\begin{align} \label{sigma1BBN}
&
\left. \sigma(5\times 10^{10}\text{g})\right|_\text{spinless}   \, \lesssim \,  3.8 \times 10^{-5} \, \gamma^{-1/5}_\text{M} \left( \frac{T_\text{rh}}{10^{10}\, \text{GeV}}\right)^{-1/5}
\;\;\;\quad\quad\quad\quad \text{(BBN)}
 \\
&   \label{MDcmb}
\left. \sigma(2.5\times 10^{13}\text{g})\right|_\text{spinless} \, \lesssim \,  2.9 \times 10^{-6} \, \gamma^{-1/5}_\text{M} \left( \frac{T_\text{rh}}{10^{10}\, \text{GeV}}\right)^{-1/5}
 \quad\quad\quad\quad \text{(CMB)}
\end{align}
One sees that for the BBN constraint it is $\sigma>0.005$ for $T_\text{rh}\lesssim 3$ GeV. 
In fact,  for reheating temperatures $T_\text{rh}  <T_\text{cmb}^\text{(MD)}$ the CMB constraint is always the stringent one and there the spin effects  (that we discuss right after) determine the maximum allowed variance, $\sigma_\text{max}$, see the right panel of Fig. \ref{sigmabbncmb}.
   Hence, the bounds (\ref{sigma1BBN}) and (\ref{MDcmb}) should be seen only as indicative ones. 

Turning now to the  PBH formation rate  considering spin effects, dictated by Eq. (\ref{spin}), one has to solve numerically the inequality (\ref{betaMDbound}) in order to derive upper bounds for the variance of the density perturbations.  
After fitting the numerical solution we find the BBN and CMB constraints for the variance,
\begin{tcolorbox}[enhanced, ams align, drop fuzzy shadow,
  colback=yellow!10!white,colframe=yellow!50!black]
 \label{spinBBN}
\left.   \sigma(5\times 10^{10}\text{g})\right|_\text{+spin}   \, \lesssim \, 
\text{Exp}\left[ \right.
&
-4.74-0.23\, \ln \frac{T_\text{rh}}{\text{GeV}} 
+ 8.4 \times10^{-3}\,\left(\ln\frac{T_\text{rh}}{\text{GeV}}\right)^2 
\\
&  \nonumber
\left. - 1.5 \times10^{-4}\,
\left( \ln \frac{T_\text{rh}}{\text{GeV}}\right)^3 
\right]
\,\quad\quad\quad\quad\quad\quad\quad \textcolor{red}{\text{(BBN)}}\,
\end{tcolorbox}
{\setlength{\parindent}{0 ex} for $T_\text{rh}  < T_\text{bbn}^\text{(MD)}$ and,}
\begin{tcolorbox}[enhanced, ams align, drop fuzzy shadow,
  colback=yellow!10!white,colframe=yellow!50!black]
  \label{spinCMB}
\left.   \sigma(2.5\times 10^{13}\text{g})\right|_{\text{+spin}}   \, \lesssim \, 
\text{Exp}\left[ \right.
&
-6.67- 0.098\, \ln \frac{T_\text{rh}}{\text{GeV}} 
+ 2.4\times10^{-3}\,\left(\ln\frac{T_\text{rh}}{\text{GeV}}\right)^2 
\\
&  \nonumber
\left. - 4\times10^{-5}\,
\left( \ln \frac{T_\text{rh}}{\text{GeV}}\right)^3 
\right]
\quad\quad\quad\quad\quad\quad\quad\quad \textcolor{red}{\text{(CMB)}}\,
\end{tcolorbox}
{\setlength{\parindent}{0 ex} for $T_\text{rh}  < T_\text{cmb}^\text{(MD)}$.  
The Eq. (\ref{kspacM}) translates them into the $k$-space.}

Compared to the spinless case, these bounds are weaker but they are the effective ones for the matter domination era regardless the reheating temperature. They are depicted in Fig. \ref{sigmabbncmb}. 
Collectively, the upper bounds on $\sigma(M)$ for any PBH mass $M$, written also in the momentum space, considering collapse during RD era and during eMD era with and without spin effects, are presented in Fig.   \ref{sigmaT}.

Now that we derived the expressions (\ref{spinBBN})  and (\ref{spinCMB}) for the temperature dependent  variance we can estimate the $T_\text{bbn}^\text{(MD)}$ and $T_\text{cmb}^\text{(MD)}$ from  Eq. (\ref{Tform}).
Plugging in the the upper bound value for the $\sigma_\text{MD}(M)$ for spinning collapse we find the values,
\begin{align}\label{Tformmax}
 T_\text{bbn}^\text{(MD)} = 4\times10^8\,\text{GeV}\,,  \quad\quad\quad
 T_\text{cmb}^\text{(MD)}= 1.3\times10^7\,\text{GeV}\,.
\end{align}

After rewritting the Eq. (\ref{Tform}), we can  also find the maximum reheating temperature value that a PBH with arbitrary mass $M$ forms during matter domination era. This is found  after solving the equation
\begin{align} \label{Trhmd}
T_\text{rh-max} = \left( \frac25\,{\cal I}\right)^{1/2} \,T(M)  \, \sigma^{1/2}(M, T_\text{rh-max})\,,
\end{align}
 where $T(M)$ is given by Eq. (\ref{TM}). We numerically solve this equation to find the $\sigma$ and the mass of  the transition from eMD collapse  to RD collapse for particular reheating temperatures  and make the plots in Fig. \ref{sigmaT}, as well as in Fig. \ref{sigmaX} and \ref{fullPS} presented in the following sections. 
 In these figures the dashed lines in the region of transition from eMD to RD give the upper bound for $\sigma(M)$ if the collapse had been instantaneous.

Plugging in benchmark values, e.g,  $\gamma_\text{M} =0.1$, and  assuming $T_\text{rh}= T_\text{cmb}^\text{(MD)}\simeq 10^7$ GeV where the bounds become stringent, we get $\sigma(4.6\times 10^{17}\,k_{0.05})<9 \times10^{-4}$ for the BBN and $\sigma(1.4 \times 10^{17}\,k_{0.05})<4 \times10^{-4}$ for the CMB. 
The constraints on the variance of the comoving density contrast can be applied on the power spectrum if there is an explicit form of the ${\cal P_R}$ at hand.
In a matter domination era  the relation between the variance and the ${\cal P_R}$ reads
\begin{equation}
\sigma^2(k)= \left( \frac{2}{5} \right)^2  \int \frac{dq}{q}\,W^2\left(\frac{q}{k}\right)\left(\frac{q}{k}\right)^4{\cal P_R}(q)\,.
\end{equation}
For an order of magnitude estimation we can approximate $\sigma \sim (2/5){\cal P_R}^{1/2}$ and for $\gamma_\text{M} =0.1$ and $T_\text{rh}=10^7$ GeV  the constraints on the power spectrum read in momentum space,
${\cal P_R}(4.6\times 10^{17}\,k_{0.05})\lesssim {\cal O}(6\times 10^{-6})$ and   ${\cal P_R}( 1.4 \times 10^{17}\,k_{0.05})\lesssim {\cal O}(10^{-6})$  for the BBN and CMB respectively.  
In Appendix we derive constraints on the  ${\cal P_R}$ and the reheating temperature assuming a particular but representative enough form for the power spectrum.

In summary, the constraints (\ref{kspacCMB}) and the (\ref{spinBBN}), (\ref{spinCMB}) can be seen as {\it width constraints} for the power spectrum peak, that {\it any} inflationary model has to satisfy.

\section{The maximum value for the ${\cal P_R }(k)$} \label{5}
 
 In this section, the maximum value for the  ${\cal P_R }(k)$ will be determined considering {\it both} the constraints from  nonevaporated PBHs, with mass $M_\bullet$, and evaporated PBHs, with mass $M$. We will see that in general cases with intermediate or low reheating temperature the evaporated PBHs pose the stringent constraints on the ${\cal P_R }(k)$ amplitude.
 
 \subsection{Critical reheating temperatures}

If the universe is reheated right after the formation of a PBH with mass $M_\bullet$ then we call this reheating temperature  
$M_\bullet$-critical and we  label it $T^\text{(MD)}_\bullet$. 
In a matter domination universe, contrary to the radiation dominated case, the black hole formation 
is not instantaneous. After the reentry of the overdensity with wavenumber $k$ a finite time for the collapse is required. It is 
\begin{align}  \label{algT}
T^\text{(MD)}_\bullet =T_{\bullet} \,  \left( \frac25 {\cal I}\, \sigma(M_\bullet)\right)^{1/2} \,,
\end{align}
 where $T_{\bullet}$ the temperature that the scale $k^{-1}_\bullet$ enters the Hubble horizon. 
Let us consider  four interesting and representative mass examples of PBHs (see also  Appendix). The critical temperatures read
 \begin{align}
 & \nonumber
 T^\text{(MD)}_\bullet(M_\bullet =10^{18} \text{g}) = 9.7 \times 10^6 \,\gamma^{1/2}_\text{R} \left( \frac{g_*}{106.75}\right)^{-1/4}\, \left( \frac25 {\cal I}\, \sigma(M_\bullet)\right)^{1/2}\, \text{GeV} \\
 & \nonumber
 T^\text{(MD)}_\bullet(M_\bullet =10^{22} \text{g}) = 9.7 \times 10^4 \,  \gamma^{1/2}_\text{R} \left( \frac{g_*}{106.75}\right)^{-1/4}\, \left( \frac25 {\cal I}\, \sigma(M_\bullet)\right)^{1/2}\, \text{GeV} \\
 & \nonumber
 T^\text{(MD)}_\bullet(M_\bullet =10^{29} \text{g}) = 31\,  \gamma^{1/2}_\text{R} \left( \frac{g_*}{106.75}\right)^{-1/4}\, \left( \frac25 {\cal I}\, \sigma(M_\bullet)\right)^{1/2}\, \text{GeV} \\
 & \nonumber
 T^\text{(MD)}_\bullet(M_\bullet =10^{35} \text{g}) = 3.1 \times 10^{-2} \, \gamma^{1/2}_\text{R}  \left( \frac{g_*}{106.75}\right)^{-1/4}\, \left( \frac25 {\cal I}\, \sigma(M_\bullet)\right)^{1/2}\, \text{GeV}\,.
 \end{align}
The numerical value of the $T^\text{(MD)}_\bullet$ depends on the variance  of the density perturbation at the scale $k_\bullet \equiv k(M_\bullet)$.
Assuming the maximum allowed $\sigma_\text{MD}(M, T_\text{rh})$, we find after solving the algebraic equation (\ref{algT}) for  $\sigma_\text{MD}$ given by Eq. (\ref{Cs3}) -spin effects can be ignored here- that,
 $T^\text{(MD)}_\bullet=2.1\times 10^5$ GeV, $3.2\times 10^3$ GeV, $2.9$ GeV and $6 \times 10^{-3}$ GeV for PBH masses $M_\bullet=10^{18}$g, $10^{22}$g, $10^{29}$g and $10^{35}$g respectively.   

\subsection{The maximum ${\cal P_R }(k)$ amplitude}
For the general  definition  
\begin{align}
{\cal P_R }(k \geq k_\text{peak}) \equiv{\cal P_R}_\text{max}\,f(k)
\end{align}
 the  maximum amplitude of the power spectrum of the comoving curvature perturbation 
is ${\cal P_R}_\text{max}={\cal P_R }(k)\left[ f(k) \right]^{-1}$. The $f(k)$ is a function that describes the ${\cal P_R}(k)$ shape for $k>k_\text{peak}$.  We  consider that DM PBHs from at $k_\text{peak}=k_\bullet$.
For the variance of  the comoving density contrast $\sigma^2(k)=\theta^2\, {\cal P_R}(k)$  the general constraints on $\sigma_\text{max}(k)$ obtained in the previous sections can be applied on the power spectrum maximum amplitude,
 \begin{align}
{\cal P_R}_\text{max} \leq \frac{\sigma^2_\text{max}(k)}{\theta^2} \, \left[ f(k) \right]^{-1} ,
\end{align} 
where we assumed that the PBHs mass distribution  is mainly monochromatic.
For a power spectrum ${\cal P_R}(k)$ designed to trigger a sizable PBH formation the ${\cal P_R}_\text{max}$ is bounded by the dynamical constraints on the nonevaporated PBHs, i.e. the fractional abundance $f_\text{PBH}$ must not violate  the bounds depicted in the right panel of  Fig. \ref{YokoR}.  Nevertheless, the evaporating PBH puts additional bounds on the  ${\cal P_R}_\text{max}$ for a  fixed form of the power spectrum tail, that is a fixed $f(k)$. 
In particular, the maximum amplitude of the power spectrum has to satisfy the constraints
\begin{tcolorbox}[enhanced, ams align, drop fuzzy shadow,
  colback=yellow!2!white,colframe=yellow!50!black]
 {\cal P_R}_\text{max} \,  \leq \, \text{Min}\left\{ \frac{\sigma^2_\text{max}(k_\bullet)}{\theta^2} \, \,,   \,
 \frac{\sigma^2_\text{max}(k_\text{bbn})}{\theta^2} \, \left[ f(k_\text{bbn}) \right]^{-1} \right\} \,
 \end{tcolorbox}
 for $T^\text{(MD)}_\text{cmb}  \,  \, \lesssim T_\text{rh} \,\,  \lesssim \,T^\text{(MD)}_\text{bbn}\,$ and
\begin{tcolorbox}[enhanced, ams align, drop fuzzy shadow,
  colback=yellow!2!white,colframe=yellow!50!black]
 {\cal P_R}_\text{max} \,  \leq \, \text{Min}\left\{ \frac{\sigma^2_\text{max}(k_\bullet)}{\theta^2} \,\,,  \, 
  \frac{\sigma^2_\text{max}(k_\text{cmb})}{\theta^2} \, \left[ f(k_\text{cmb}) \right]^{-1} \right\} \,
 \end{tcolorbox}
 for  $T_\text{rh} \lesssim \, T^\text{(MD)}_\text{cmb}$.

\begin{figure}[!htbp]
\begin{subfigure}{.5\textwidth}
  \centering
  \includegraphics[width=1.\linewidth]{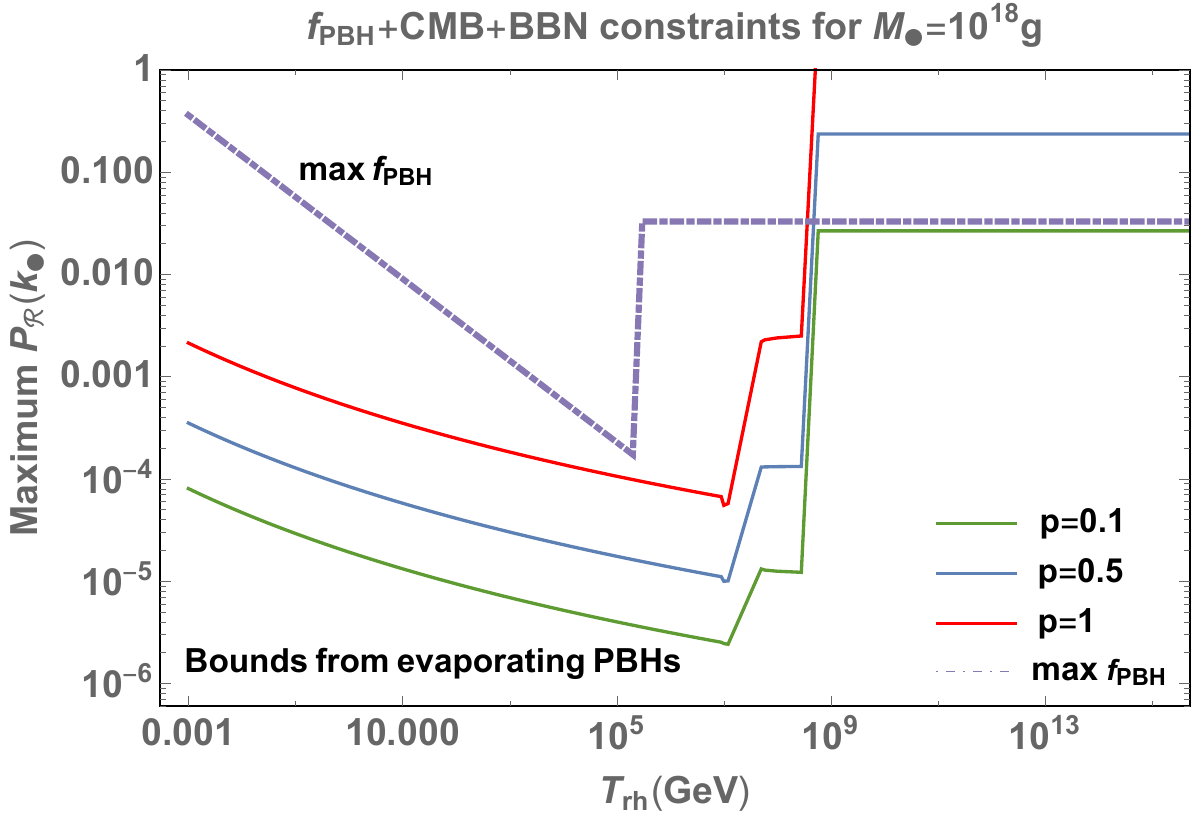}
\end{subfigure} \;
\begin{subfigure}{.5\textwidth}
  \centering
  \includegraphics[width=1.\linewidth]{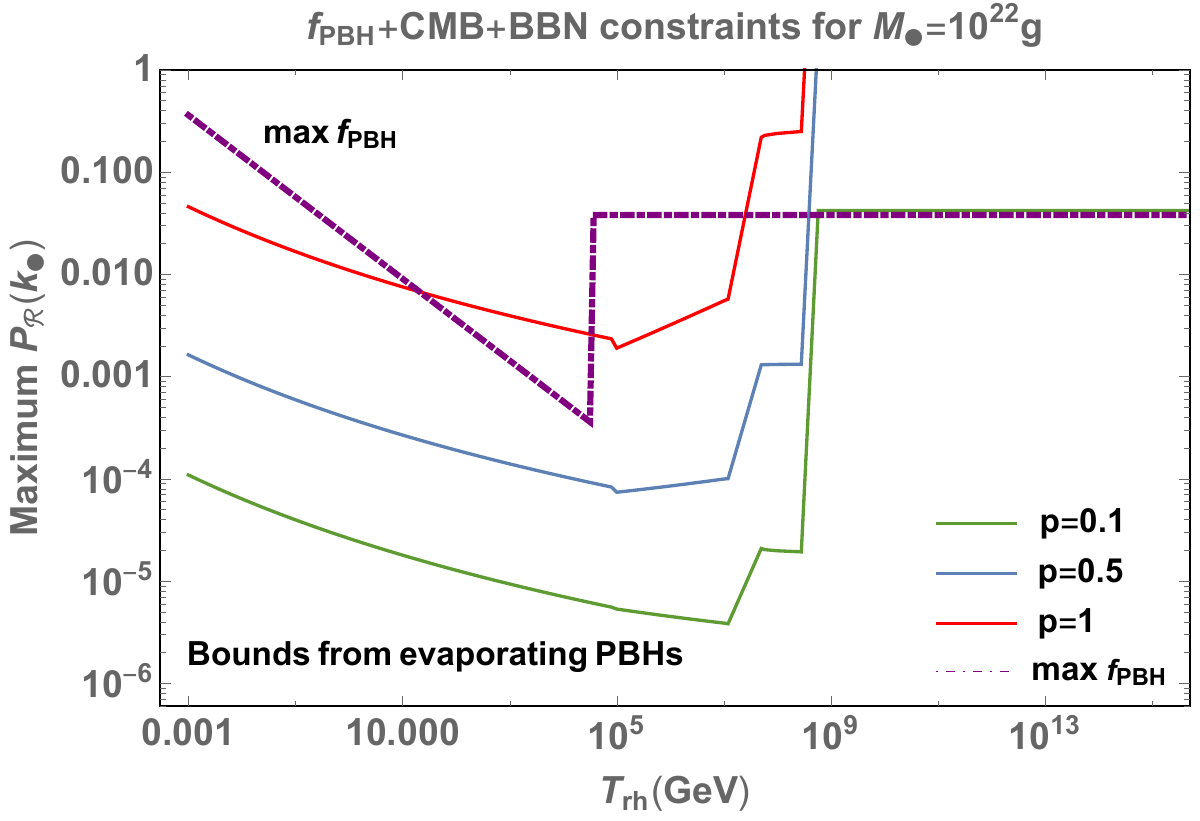}
\end{subfigure}
\\
\\
\begin{subfigure}{.5\textwidth}
  \centering
  \includegraphics[width=1.\linewidth]{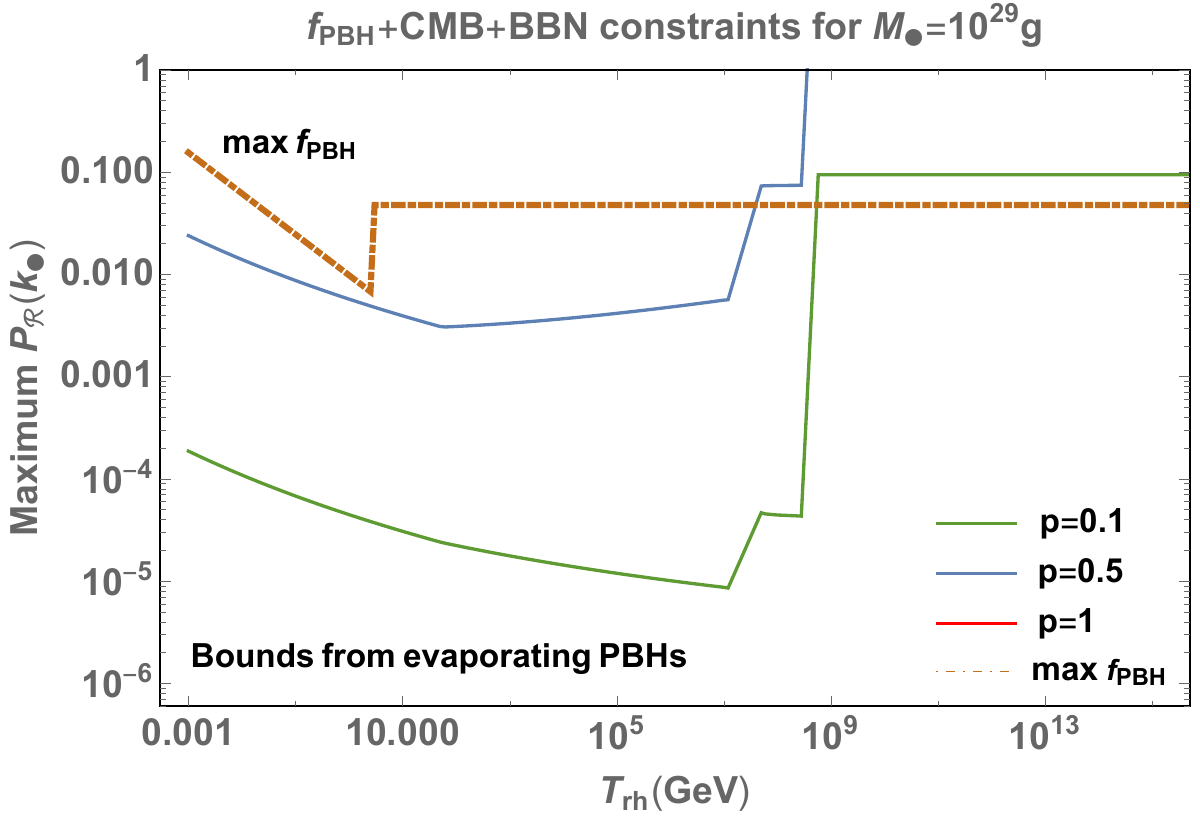}
\end{subfigure} \;
\begin{subfigure}{.5\textwidth}
  \centering
  \includegraphics[width=1.\linewidth]{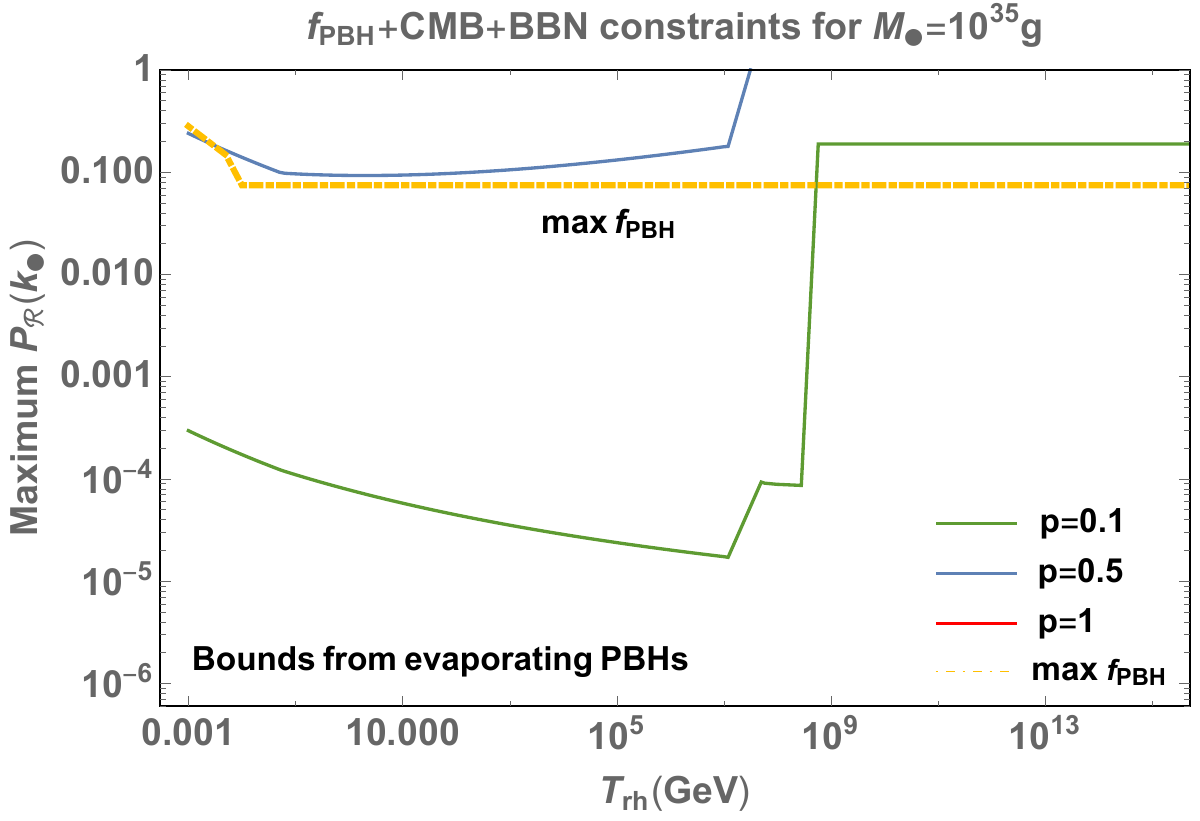}
\end{subfigure}
\caption{\label{AT}~ The plots depict  {\it upper bounds} on the power spectrum amplitude, ${\cal P_R}(k_\bullet)\equiv {\cal P_R}(k_\text{peak})$, from the dark matter PBH abundance (dot-dashed line) and from the CMB and BBN 
constraints from the  evaporating PBH (solid lines)
for three different slopes for the power spectrum tail as described in  Appendix: green (p=0.1), blue (p=0.5) and red (p=1).  
The step-like changes take place at $T^{\text{(MD)}}_\bullet$, $T^{\text{(MD)}}_\text{cmb}$ and $T^{\text{(MD)}}_\text{bbn}$ from left to right.  Benchmark values $\delta_c=0.41$, $\gamma_\text{M}=0.1$ and $\gamma_\text{R}=0.2$ have been used. 
The figure demonstrates that the constraints from the evaporating PBH are often the stringent ones suppressing severely the  PBH abundance.
}
\end{figure}

For $f(k)=(k/k_\bullet)^{-p}$, that we exemplify in Appendix,  it is $\theta^2= 1/2\,(2/5)^2  \,\Gamma \left(2 - \frac{p}{2} \right)$.  The ratio $k/k_\bullet$ depends on the reheating temperature.  According to the expressions (\ref{kmdM}) and (\ref{kRDm})  it is
\begin{equation} \label{kkrelic}
\frac{k}{k_\bullet}  \, =
  \begin{cases} 
 \,   \frac{k_\text{RD}(M)}{k_\text{RD}(M_\bullet)} = \left(\frac{g_*}{g_*(T_\bullet)} \right)^{-1/12}   \, \left(\frac{M}{M_\bullet} \right)^{-1/2}
       \,,  \, \quad\quad \quad\quad\quad\quad\quad\quad \text{for} \quad\quad
        T_\text{rh} > T_k\,  \\      \\ 
       
    \,  \frac{k_\text{MD}(M)}{k_\text{RD}(M_\bullet)} =\eta
   \left( \frac{M}{10^{10}\, \text{g}} \right)^{-1/3} \left( \frac{M_\bullet}{10^{20}\, \text{g}} \right)^{1/2} \left( \frac{T_\text{rh}}{10^{10}\, \text{GeV}} \right)^{1/3}
       
      \; \quad\quad  \text{for} \quad\quad  T_{\bullet} <T_\text{rh} < T_k\,
\\  \\        
         \,  \frac{k_\text{MD}(M)}{k_\text{MD}(M_\bullet)} =  \left(\frac{M}{M_\bullet} \right)^{-1/3}
        \quad \quad\quad \quad\quad\quad\quad\quad\quad\quad\quad\quad\quad\quad\quad  \text{for} \quad\quad T_\text{rh} < T_{\bullet}\,
  \end{cases}
   \end{equation}
where 
$ \eta \equiv 5 \times 10^4 \gamma^{-1/2}_\text{R} \gamma^{1/3}_\text{M} (g_\bullet/106.75)^{1/12}$ and $k^{-1}$ identified either as the CMB or the BBN scale.
$T_k$ is the temperature that the $k^{-1}$ scales enters the horizon.
In the Fig. \ref{AT} the 
 ${\sigma^2_\text{max}(k_\bullet (T_\text{rh}))}/{\theta^2} \, \left[ f(k_\bullet(T_\text{rh})) \right]^{-1}$ bound is depicted with dot-dashed lines and the combined  CMB and BBN bound, ${\sigma^2_\text{max}(k(T_\text{rh}))}/{\theta^2} \, \left[ f(k(T_\text{rh})) \right]^{-1}$, with solid lines. The CMB+BBN bound depends both on the maximum value of the  power spectrum and on the form of the power spectrum tail. We depict in Fig. \ref{AT} three different ${\cal P_R}(k)$ slopes with green, blue and black color respectively, see also Appendix where constraints on the this sort of ${\cal P_R}(k)$ are  further exemplified.

\section{Two-reheating stages} \label{6}

A generic prediction of beyond the Standard Model physics is the existence of additional scalar fields. These scalars  under general initial conditions predict an epoch of early matter domination following inflation.  The mass and decay rate of these scalars vary. For example, in the stringy and supersymmetric frameworks there are scalars, collectively called moduli,  that decay gravitationally and their mass is determined by the scale of the symmetry breaking. A gravitationally decaying scalar $X$ with mass $m_X$ reheats the universe at $T^\text{dec}_X \sim 4 \text{MeV} (m_X/10^5 \text{GeV})^{3/2}$. 
The production of entropy by the modulus dilutes the thermal plasma $\Delta_X$ times, 
\begin{align}
\Delta_X \simeq \frac{T_X^\text{dom}}{T_X^\text{dec}}
\end{align}
where $T_X^\text{dom}$ the temperature that the scalars dominated the energy density of the universe and $T_X^\text{dec}$ the late reheating temperature. 

\begin{figure}[!htbp]
\begin{subfigure}{.5\textwidth}
  \centering
  \includegraphics[width=1.\linewidth]{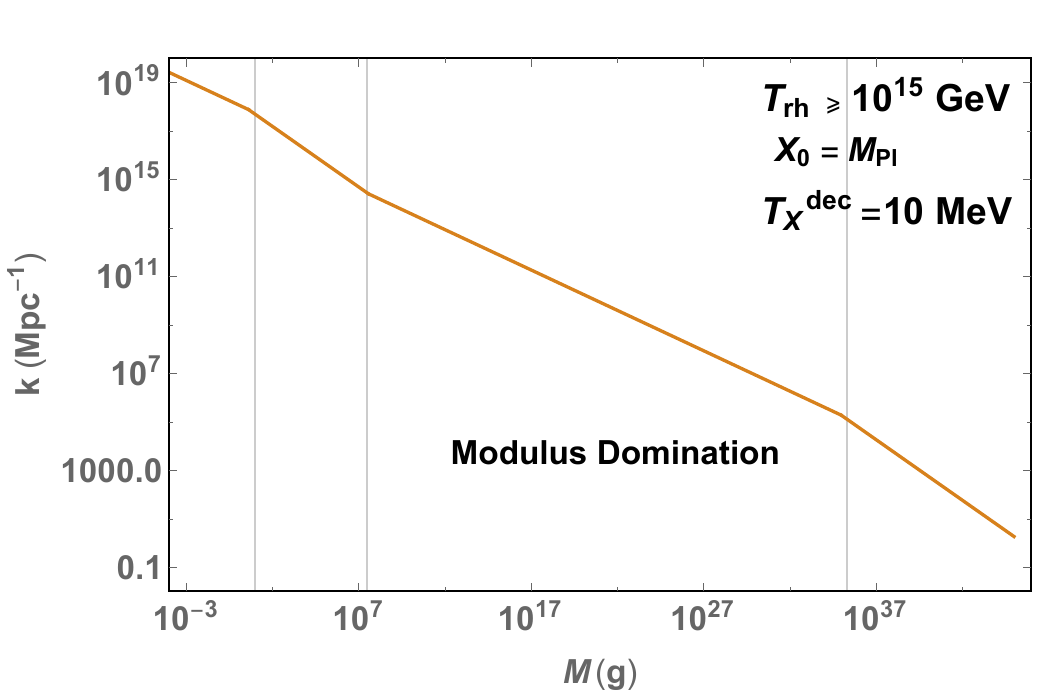}
\end{subfigure} \;
\begin{subfigure}{.5\textwidth}
  \centering
  \includegraphics[width=1.\linewidth]{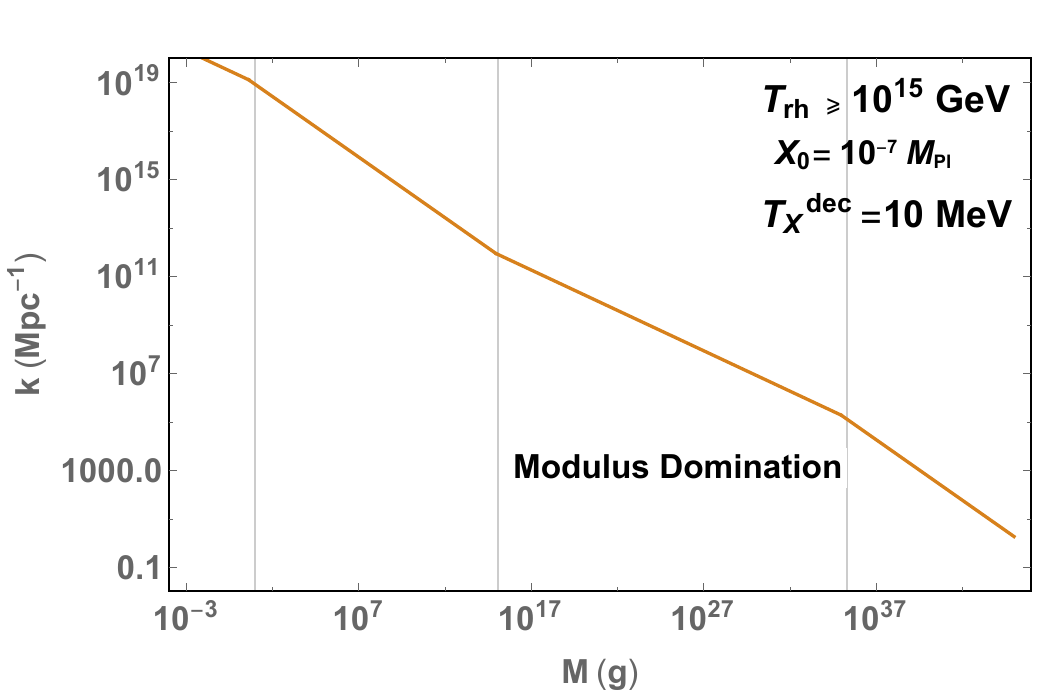}
\end{subfigure}
\caption{\label{kMX}~ The plots depict the $k=k(M)$ relation for the two modulus domination scenarios considered in the text for $\gamma_\text{R}=\gamma_\text{M}=1$. The gridlines indicate, from left to right, the transition from the reheating to the thermal phase,  the modulus domination and the final thermal phase.
}
\end{figure}
The fractional PBH abundance formed during a modulus domination era reads
\begin{align}\label{fpbhX} 
  f^{(\text{MD})}_\text{PBH}(M)  = \,\left(\frac{\beta_\text{MD}(M)}{1.2 \times 10^{-9}}\right) \, 
 \Big(\frac{\gamma_\text{M}}{0.1}\Big)^{\frac{3}{2}} 
\left(\frac{g(T_X^\text{dec})}{10.75}\right)^{-\frac{1}{4}}  
\left(\frac{M}{10^{30}\text{g}}\right)^{-1/2}\,
         \left(  \frac{k}{k_X^\text{dec}} \right)^{-3/2} \,,
 \end{align}
where
\begin{align}\label{kXdec}  
k^{\text{dec}}_X (T_X^\text{dec}, g_*) \, \simeq  10^{4}  \, \text{Mpc}^{-1}
   \left(\frac{T_X^\text{dec}}{ \text{MeV}} \right)^{}
    \left(\frac{g_*(T_X^\text{dec})}{10.75} \right)^{1/4} \,,
\end{align}
and 
\begin{align}\label{kX}  
k_X (M,T_X^\text{dec}) \, \simeq    10^{6}  \, \text{Mpc}^{-1}
\gamma^{1/3}_\text{M} \,
 \left(\frac{M}{10^{30}\, \text{g}} \right)^{-1/3} 
   \left(\frac{T_X^\text{dec}}{ \text{MeV}} \right)^{1/3}
\end{align}
for $k_X^\text{dec}< k_X<k_X^\text{dom}$. 
The $\tilde{N}_X$ are the efolds of modulus-condensate domination, $\tilde{N}_X=\frac43\ln (\Delta_X \,g^{1/4}_\text{dom}/g^{1/4}_\text{dec})$, and it is $k_X^\text{dom}= e^{\tilde{N}_X/2}k_X^\text{dec}$ hence  $k_X^\text{dom}\simeq \Delta_X^{2/3}\,k_X^\text{dec}$.  

Also, we can define the $k$-dependent "dilution" size\footnote{Strictly speaking this is not a dilution. It accounts for the absence of expansion effects on the PBH abundance during the modulus domination.}, 
$\Delta_k\, \equiv  \, \left(  {k}/{k_X^\text{dec}} \right)^{3/2}$ . 
Due to the modulus domination the abundance of PBH formed at scales $k>k_X^\text{dom}$ are diluted $\Delta_X$ times while the abundance of  PBH formed at scales $k$ during the modulus domination era are partially "diluted" $\Delta_k$ times. The horizon masses at the scales  are $k_X^\text{dec}$ and $k_X^\text{dom}$ are respectively,
\begin{align}
&
M_\text{hor}(T_X^\text{dec}, g_*)\, \simeq \, 3\times 10^{38} \,\text{g}  \left(\frac{T_X^\text{dec}}{ \text{MeV}} \right)^{-2}
    \left(\frac{g_*}{10.75} \right)^{-1/2} \\
&
M_\text{hor}(T_X^\text{dom}, g_*)\, =\,  M_\text{hor}(T_X^\text{dec}, g_*) \left(\frac{k_X^\text{dom}}{k_X^\text{dec}}  \right)^{-3}\, =\, 
{ M_\text{hor}(T_X^\text{dec}, g_*)}\, {\Delta_X^{-2}}    
\end{align}

For the entire postinflationary phase the PBH mass $M$ and the horizon scale $k^{-1}$ are related as
\begin{equation} \label{genX}
  k(M) \, =
  \begin{cases} 
 \,   k_\text{MD}(M, T_\text{rh}) =   \left(\text{Eq.} (\ref{kmdM})\right)\times \,e^{-\tilde{N}_X/4} \,, 
 \quad\quad\quad\quad\quad \quad\quad \text{for} \quad\quad
 k_\text{rh}<k<k_\text{end}\,  \\      \\ 
    \,  k_\text{RD}(M, T_\text{rh}, g_*)=\, k_\text{rh}\, \gamma_\text{R}^{1/2}\, \left(\frac{M}{M_\text{rh}}\right)^{-1/2} \left(\frac{g_*}{g_\text{rh}}\right)^{-1/12} 
\;\;\; \quad  \text{for} \quad\quad k_X^\text{dom}<k <k_\text{rh}\, 
\\  \\
   \,  k_X(M,T_X^\text{dec})\, =\, \left(\text{Eq.} (\ref{kX})\right)
   \,
 \quad\quad\quad\quad\quad\quad\quad\quad\quad\quad\quad\quad  \text{for} \quad\quad k_X^\text{dec}< k<k_X^\text{dom}\,       
       \\ \\
    \,   k_\text{RD}(M,T_X^\text{dec}, g_*)=k_X^\text{dec} \gamma_\text{R}^{1/2}  \left(\frac{M}{M_\text{dec}}\right)^{-1/2}\left(\frac{g_*}{g_\text{dec}}\right)^{-1/12} 
  \quad  \text{for} \quad\quad k<k_X^\text{dec} \,
  \end{cases}
   \end{equation}
where $M_\text{dec}\equiv M_\text{hor}(T_X^\text{dec}, g_*)$ and $g_\text{dec}\equiv g_*(T_X^\text{dec})$.

We assume that the scalar $X$ decays just before the BBN nucleosynthesis, thus we assume $M_X \sim 10^5$ GeV and gravitational interactions.  We let the free parameter to be the $\Delta_X$, or equivalently, the $T_X^\text{dom}$. To make the distinction clear, we also assume that the inflaton field decays fast with $T_\text{rh} \sim 10^{15}$ GeV, though a late decaying inflaton together with a modulus domination era is also an interesting possibility.
The corresponding bounds on the $\beta'(M)$ read
\begin{align} \label{betaXbound}
\beta'_\text{MD}(M)\equiv\gamma_\text{M} \beta_\text{MD}(M) < \, 9\times 10^3 \left( \frac{T^\text{dec}_X}{\text{MeV}} \right)^{-1}
 \left( \frac{M_\text{}}{10^{30}\,\text{g}} \right)^{-1/2} \boldsymbol{{C}_M} \,.
\end{align}
The bound on $\beta(M)$ can be translated into a bound on the variance $\sigma$. 
For $M_\text{hor}(T_X^\text{dom})<M/\gamma_\text{M}<M_\text{hor}(T_X^\text{dec})$ the $\sigma$ is given by the theory of pressureless gravitational collapse. For $M/\gamma_\text{R}<M_\text{hor}(T_X^\text{dom})$ the universe is radiation dominated and the bound on $\beta'_\text{RD}(M)$, given by Eq. (\ref{gencon}), is relaxed $\Delta_X$ times. This relaxation is passed to the corresponding $\sigma_\text{RD}(M)$  bound according to the Eq. (\ref{cbr1}).
 For a late decaying modulus field  dominates the energey density for $T_X^\text{dom}<T_\text{cmb}$ the maximum value for $\sigma$, during the pressureless modulus-condensate domination era, is above the threshold value $\sigma_\text{thr}=0.005$. 
Hence one can employ the results of Ref. \cite{Harada:2016mhb} that give the formation rate for spinless gravitational collapse, $\beta_\text{MD}(M_\text{})   \, \simeq  \, 0.056\,  \sigma^5(M_\text{})\,$, and
the bound on the  variance during the modulus domination era reads, 
\begin{align} \label{CsX2}
\sigma_\text{MD}(M)\, < \, 1.1  \,\, \gamma^{-1/5}_\text{M} \left( \frac{T^\text{dec}_X}{ \text{MeV}} \right)^{-1/5} 
\left( \frac{M}{10^{30} \text{g}} \right)^{-1/10} 
 \,\boldsymbol{{C}_M}^{1/5} \,.
\end{align}

\begin{figure}[!htbp]
\begin{subfigure}{.5\textwidth}
  \centering
  \includegraphics[width=1.\linewidth]{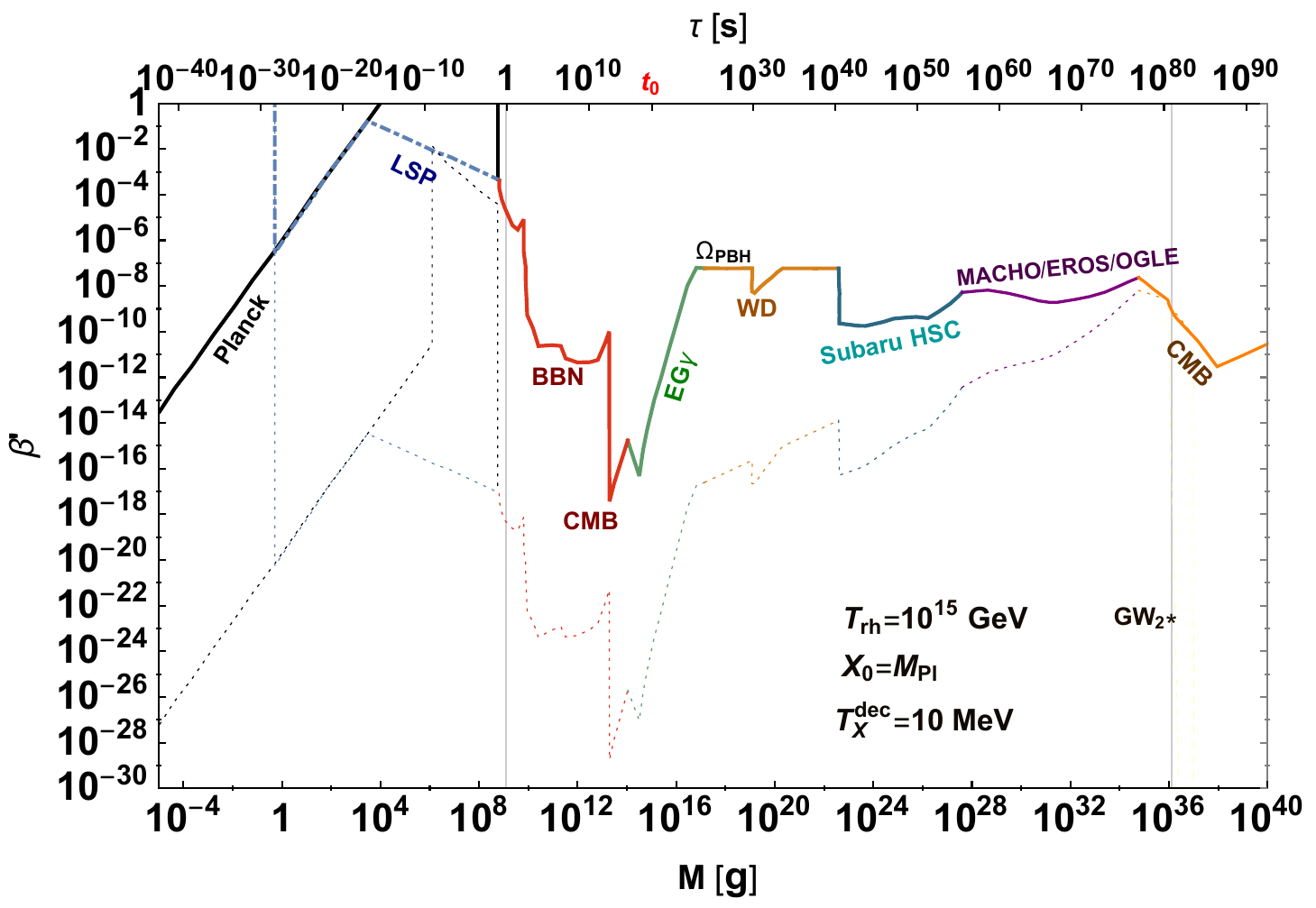}
\end{subfigure}
\begin{subfigure}{.5\textwidth}
  \centering
  \includegraphics[width=1.\linewidth]{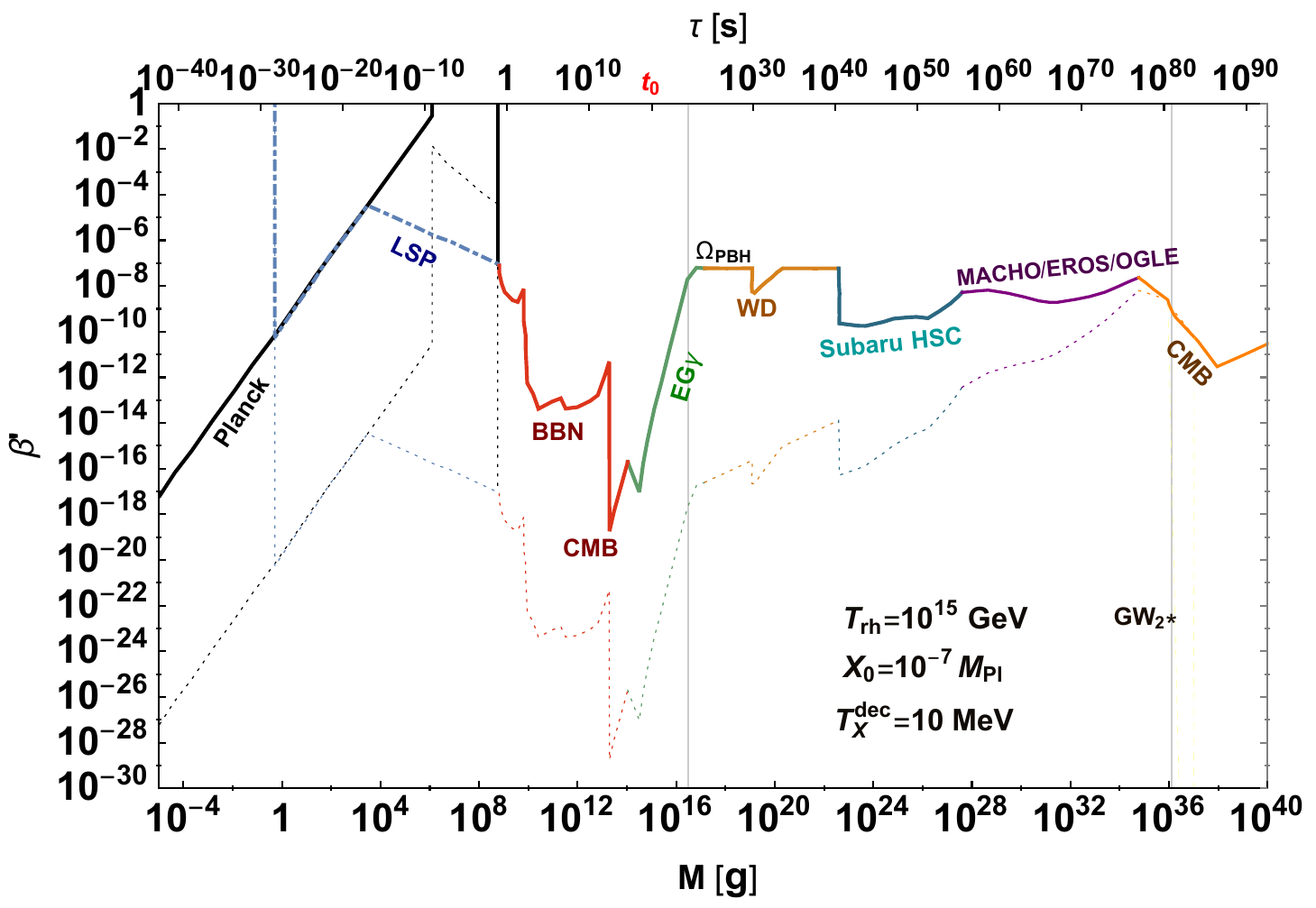}
\end{subfigure}
\caption{\label{betaX}~   {\it Left panel}:  
The combined upper bounds on $\beta'(M)$ for the cosmological scenario with  reheating temperatures $T_\text{rh}=10^{15}$ GeV and a modulus field that dominates the energy density at  $T_X^\text{dom}=5\times 10^{11}$ GeV (left gridline) and decays at $T_X^\text{dec}=10$  MeV (right gridline). 
 The dotted lines depict the $\beta'(M)$ constraints for arbitrarily large reheating  temperature.
  {\it Right panel}:  
As in the right panel with the difference that the modulus field  dominates the energy density at  $T_X^\text{dom}=10^{8}$ GeV.
 }
\end{figure}

\begin{figure} [hbt!]  
\begin{subfigure}{.5\textwidth}
  \centering
  \includegraphics[width=1.\linewidth]{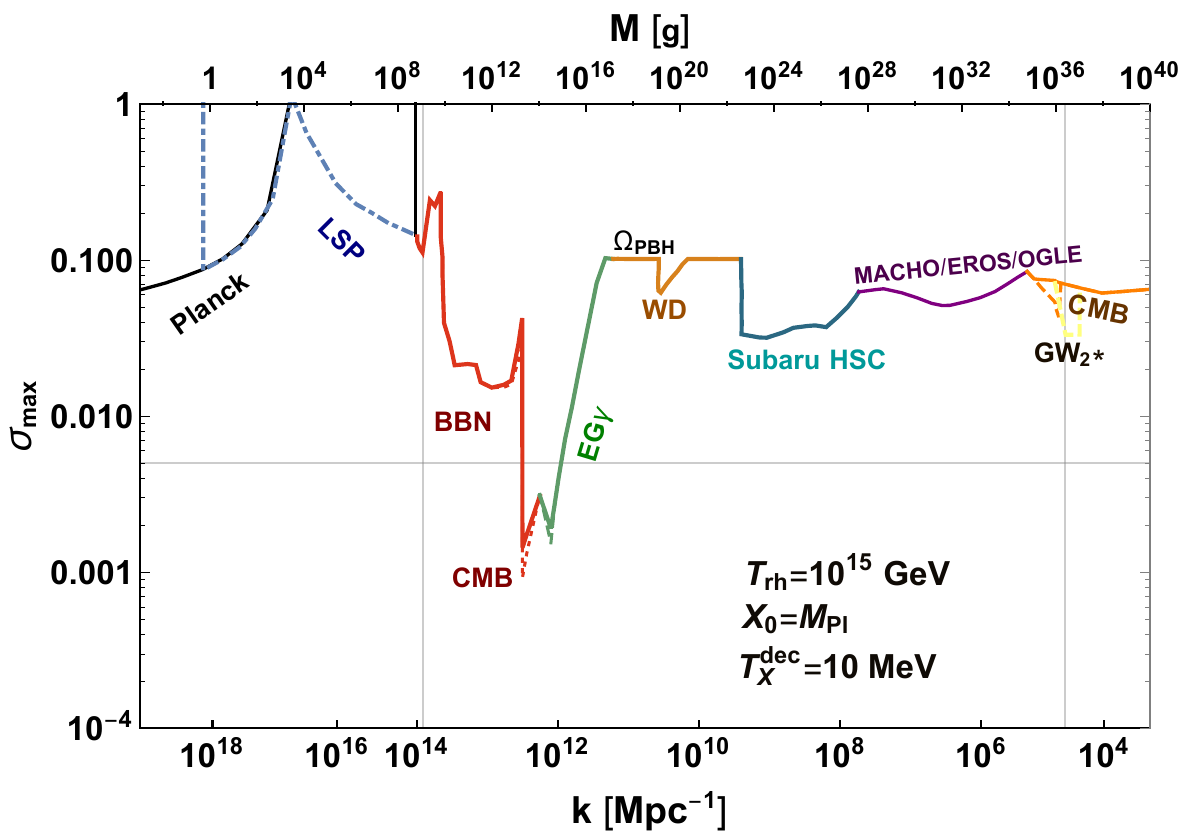}
\end{subfigure}
\begin{subfigure}{.5\textwidth}
  \centering
  \includegraphics[width=1.\linewidth]{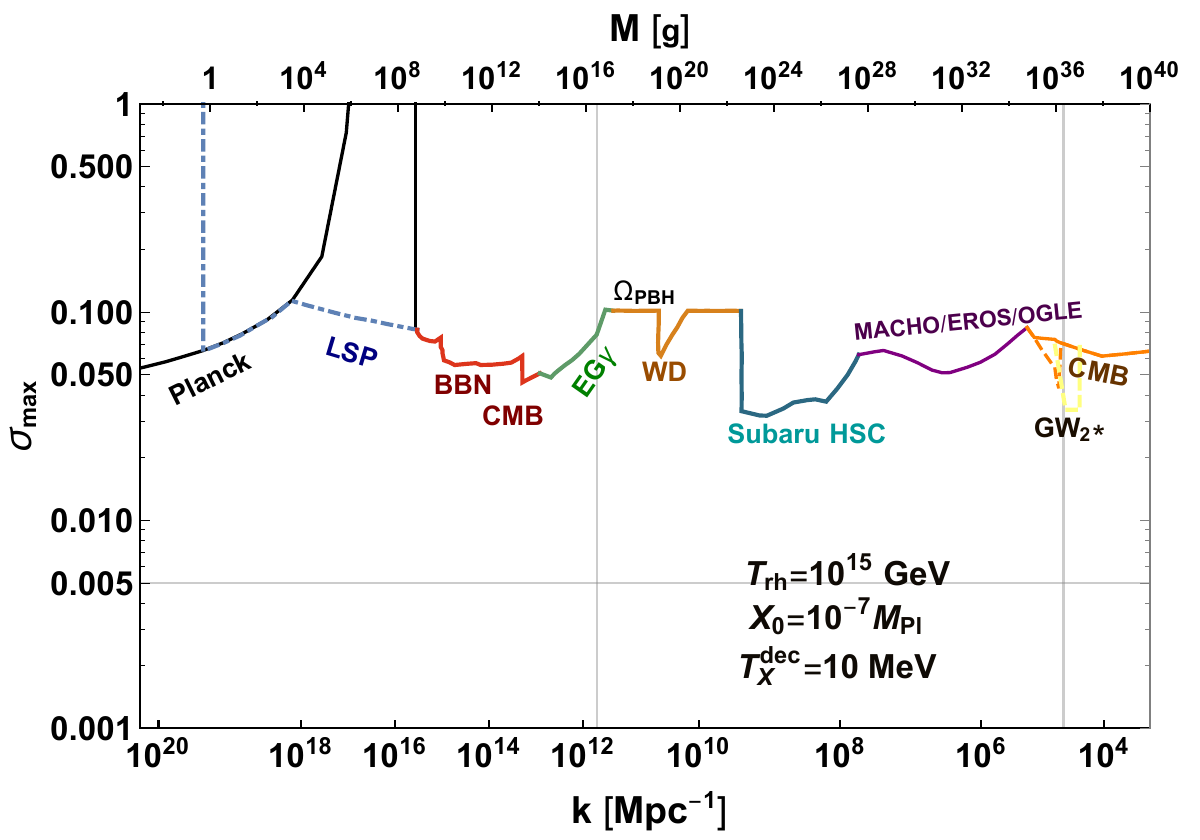}
\end{subfigure}
\caption{\label{sigmaX}~   {\it Left panel}:  
The combined upper bounds on $\sigma(M)$ for the cosmological scenario with  reheating temperatures $T_\text{rh}=10^{15}$ GeV and a modulus field that dominates the energy density at  $T_X^\text{dom}=5\times 10^{11}$ GeV (left gridline) and decays at $T_X^\text{dec}=10$  MeV (right gridline). 
The dotted lines,  for the CMB and extra-galactic gamma rays constraints, give the $\sigma$ bound for spinless gravitational collapse. The dashed lines next to $k_\text{rh}$ give the maximum variance for instantaneous gravitational collapse. 
{\it Right panel}:  
As in the left panel with the modulus field  dominating the energy density at  $T_X^\text{dom}=10^{8}$ GeV. 
 }
\end{figure}

Benchmark values for the $T_X^\text{dom}$ are determined by the initial value of the modulus potential. 
Assuming the simple quadratic potential for the modulus field $V(X)=m^2_X X^2/2 $ then, for  $m_X \sim 10^5$ GeV, it is the initial displacement $X_0$ from the zero temperature minimum that determines the $T_X^\text{dom}$.  
The initial displacement of the modulus field is model dependent.  
For our analysis we assume two distinct cases, a fist with the maximum possible  initial displacement (to avoid a late inflationary phase)  $X_0 \sim M_\text{Pl}$, and a second with an intermediate initial displacement, $X_0 \sim 10^{-7}M_\text{Pl}$. 
 In both cases we assume that the {\it effective} mass of the modulus during inflation is larger than the Hubble scale so that the spectrum of the de-Sitter fluctuations is not transferred to the modulus field, that could otherwise act as a curvaton field.  
For these $X_0$ values we get respectively,
\begin{enumerate}
\item $T_X^\text{dom} \simeq 5 \times 10^{11} \,\text{ GeV}$ and $ \Delta_X\simeq 5\times 10^{13}$ for $X_0\sim M_\text{Pl}$
\item $T_X^\text{dom} \simeq 10^{8} \,\text{GeV} $ and $\Delta_X=10^{10} $  for  $ X_0  \sim 10^{-7}M_\text{Pl}$.
\end{enumerate}

For these two benchmark cases the horizon mass at the time of the modulus decay is $M_\text{hor}(T_X^\text{dec}=10 \text{MeV})\simeq 3\times 10^{36}$ g, and at the time the modulus dominates the energy density is $M_\text{hor}(T_X^\text{dom}=5\times 10^{11} \text{GeV})\simeq 4\times 10^8$g and
$M_\text{hor}(T_X^\text{dom}=10^{8} \text{GeV})\simeq 10^{20}$g, respectively.

\section{Constraints on the primordial power spectrum on all scales} \label{7}

The bounds obtained from evaporating and unevaporating PBH constrain the power spectrum over 45 decades of mass, whereas the CMB direct measurements span only 5 decades. 
The analysis method followed here  to obtain the bounds  is illustrated in Fig. \ref{graphP}. 
 In Fig. {\ref{fullPS}} the upper bounds for the power spectrum of the comoving curvature perturbation are depicted after the assumption that  ${\cal P_R}(k) \simeq (9/4)^2\sigma^2(k)$  for a RD era and ${\cal P_R}(k) \simeq (5/2)^2\sigma^2(k)$ for a eMD era.  
Each panel corresponds to a different cosmic history. Scenarios with reheating temperatures, $T_\text{rh} >10^{15}$ and $T_\text{rh}= 10^7, 10^2, 10^{-2}$ GeV as well as scenarios with an intermediate non-thermal phase  due to a scalar condensate domination have been examined.
For  $T_\text{rh}\simeq T^\text{(MD)}_\text{cmb} \simeq 10^7$ GeV  the constraint on the power spectrum is the stringent one  after the direct $\Delta T/T$ observational constraint at $k \sim 0.05$ Mpc$^{-1}$ \cite{Akrami:2018odb}. 
 At that scale, that corresponds to the horizon mass $2.5 \times 10^{13}\text{g}/\gamma_\text{M}$, the power spectrum of the comoving curvature perturbation has to be ${\cal P_R}(k)\lesssim 9 \times 10^{-7}$. 

In Fig. \ref{fullPS}  the observational constraints, e.g. the CMB constraint, is located in different position in the $k$-space, albeit the position of the constraints on the axis of mass remains the same.  This is either due to the different reheating temperatures or due to a postinflationary non-thermal phase.
In the Fig. \ref{fullPS} we also included the Planck 2018 constraints on the power spectrum \cite{Akrami:2018odb}. 
As in the previous figures, the constraint associated with GWs is depicted with dotted-dashed lines because it is partial and included only as reference.
Moreover, in the right upper panel of Fig. {\ref{fullPS}}  we have added the $k$-range where future observational probes, that we briefly outline below, can  constrain further and significantly the power spectrum amplitude.

\subsection{Additional constraints on the primordial power spectrum} 

 Apart from the direct constraints coming from nonevaporated PBH the power spectrum at large scales can be constrained by other effects.
 PBH generation requires large density perturbations that in turn source the generation of gravitational waves, see Ref. \cite{Sasaki:2018dmp} for  a review.
 Scalar perturbations and tensor  perturbations  are coupled beyond the linear order \cite{Ananda:2006af, Baumann:2007zm} and hence the induced gravitational waves are also stochastic. 
These gravitational waves are produced at the horizon crossing of the scalar perturbations, hence simultaneously with the potential PBH generation, and their frequency is related with the PBH mass as $f_\text{GW}\simeq 7\times 10^{-9} \gamma^{1/2}_\text{} (M_\bullet/M_\odot)^{-1/2}$ Hz. The amount of the gravitational waves depends on the type and the
amplitude of the curvature power spectrum. 
Low frequency gravitational waves are severely constrained by the pulsar timing experiments \cite{Saito:2009jt, Saito:2008jc} whereas, higher frequencies will be subject to future observational probes, see e.g. \cite{Moore:2014lga, Bartolo:2018qqn}. An interesting scenario is that PBHs with mass $10^{-12} M_\odot$ can comprise most of the dark matter in the universe and in such a case their production is associated with a mHz gravitational wave signal that can be tested by LISA \cite{Cai:2018dig, Unal:2018yaa, Bartolo:2018rku}.

In addition due to the Silk damping, that is the erase of acoustic oscillations of $k^{-1}$ that falls within the  photon diffusion scale,  energy is transferred to the background homogeneous plasma \cite{Chluba:2015bqa}. 
Depending on the redshift $z>10^3$  that the damping occurs there exist two types of CMB distortions, the  $\mu$-distortion at scales $50$ Mpc$^{-1}  \lesssim k  \lesssim 10^4$ Mpc$^{-1}$ and $y$-distortion on larger scales.
So far $\mu$-type spectral distortion of the CMB has not be detected and for Gaussian primordial density perturbations  PBHs in the mass range $ 2\times 10^{4} M_\odot \lesssim M_\bullet \lesssim 2\times 10^{13}M_\odot$ are excluded. Smaller scales, that correspond to PBH masses $M_\bullet \lesssim 2\times 10^4 M_\odot$ are still possible to be probed by the measurement of the baryon-to-photon ratio and put constraints on the power spectrum amplitude \cite{Nakama:2014vla, Inomata:2016uip}.
 We comment that if the scalar perturbations are non-Gaussian then the constraints on the ${\cal P_R}(k)$ change \cite{Nakama:2016gzw, Garcia-Bellido:2017aan}.
 In such a case, the $\mu$-distortion as well the bounds from the stochastic gravitational wave background can weaken  depending on the degree of the non-Gaussianity of the primordial perturbations.

\begin{figure}[!htbp]
\begin{subfigure}{.5\textwidth}
  \centering
  \includegraphics[width=1.\linewidth]{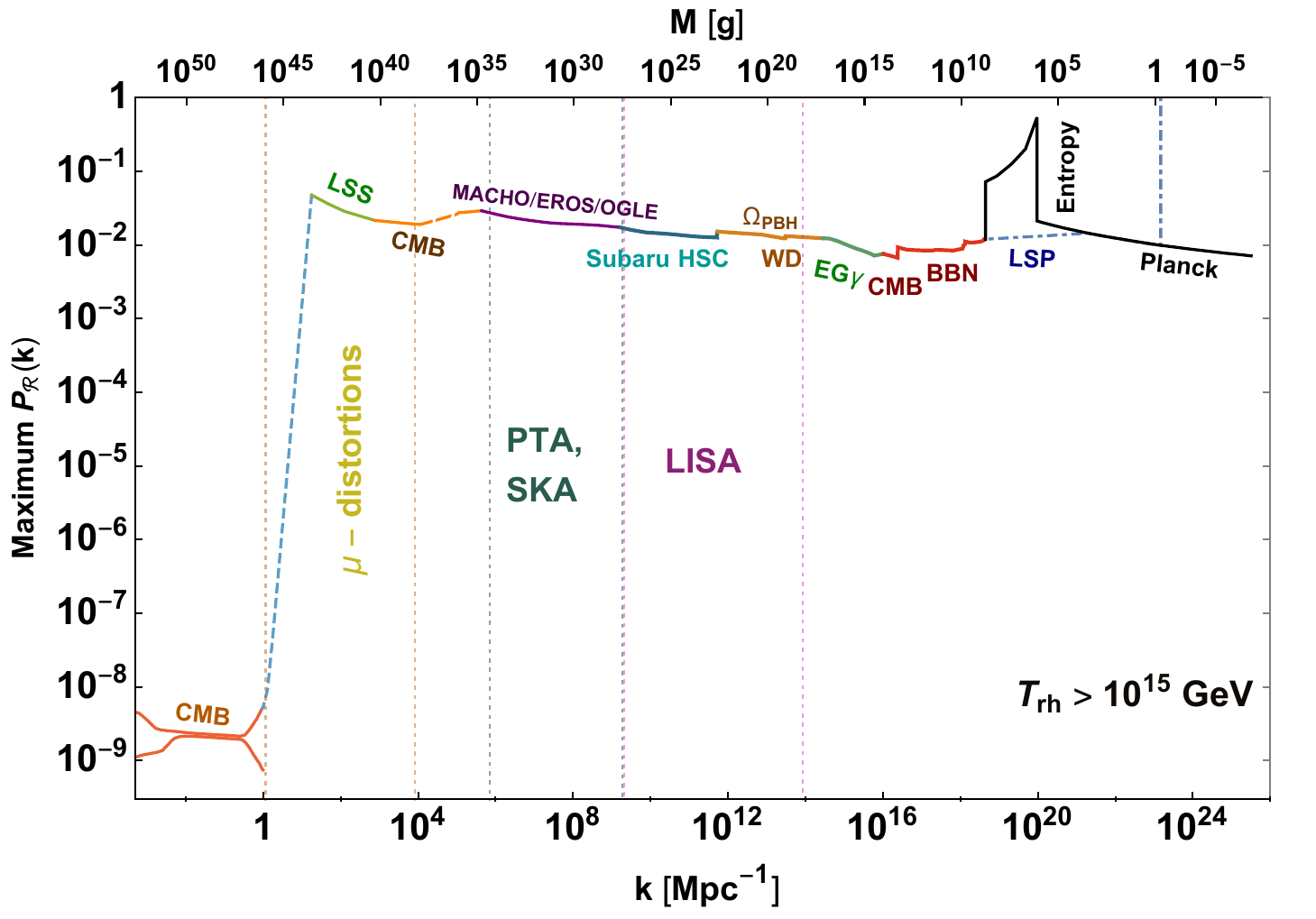}
\end{subfigure} \;
\begin{subfigure}{.5\textwidth}
  \centering
  \includegraphics[width=1.\linewidth]{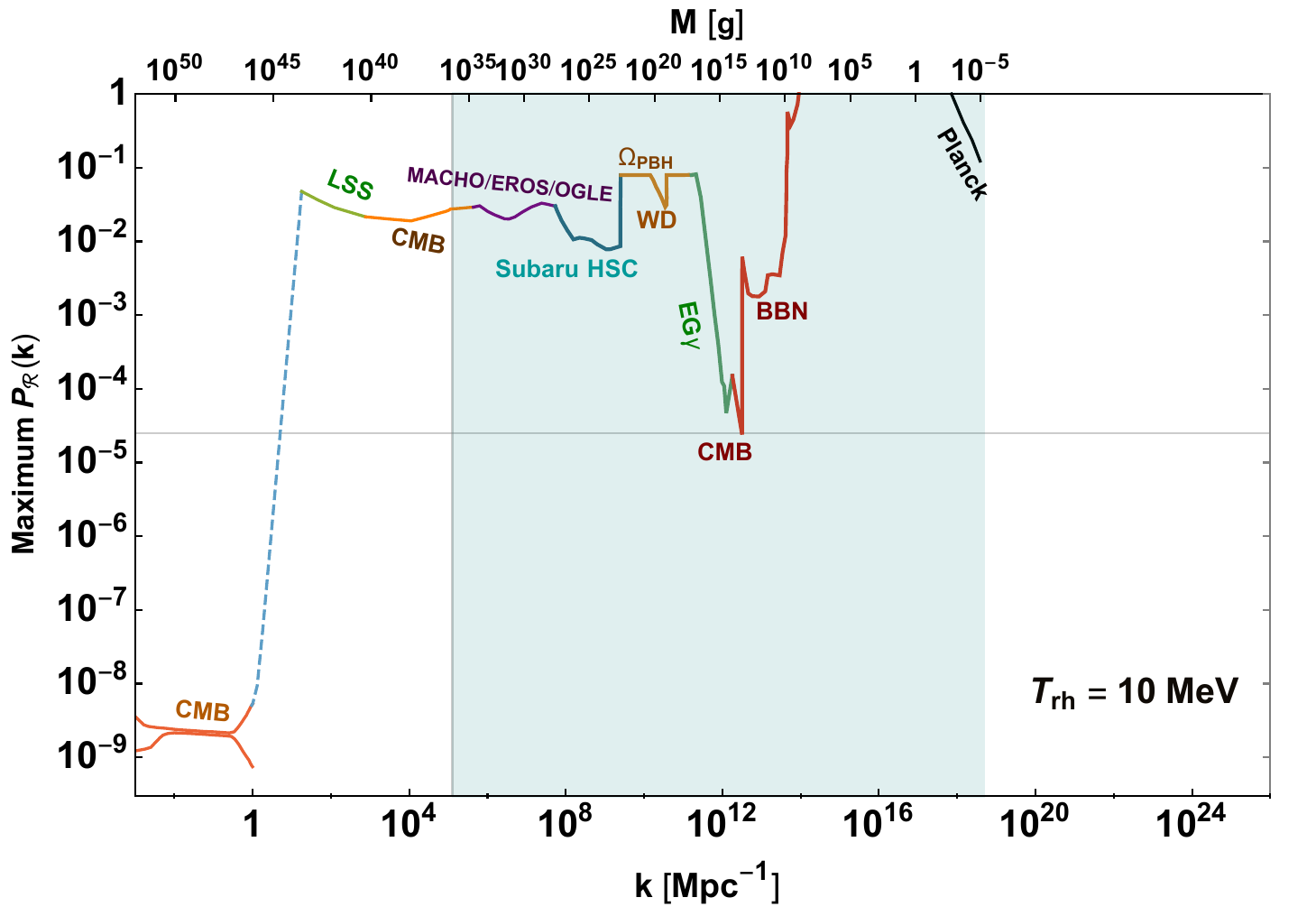}
\end{subfigure}
\begin{subfigure}{.5\textwidth}
  \centering
  \includegraphics[width=1.\linewidth]{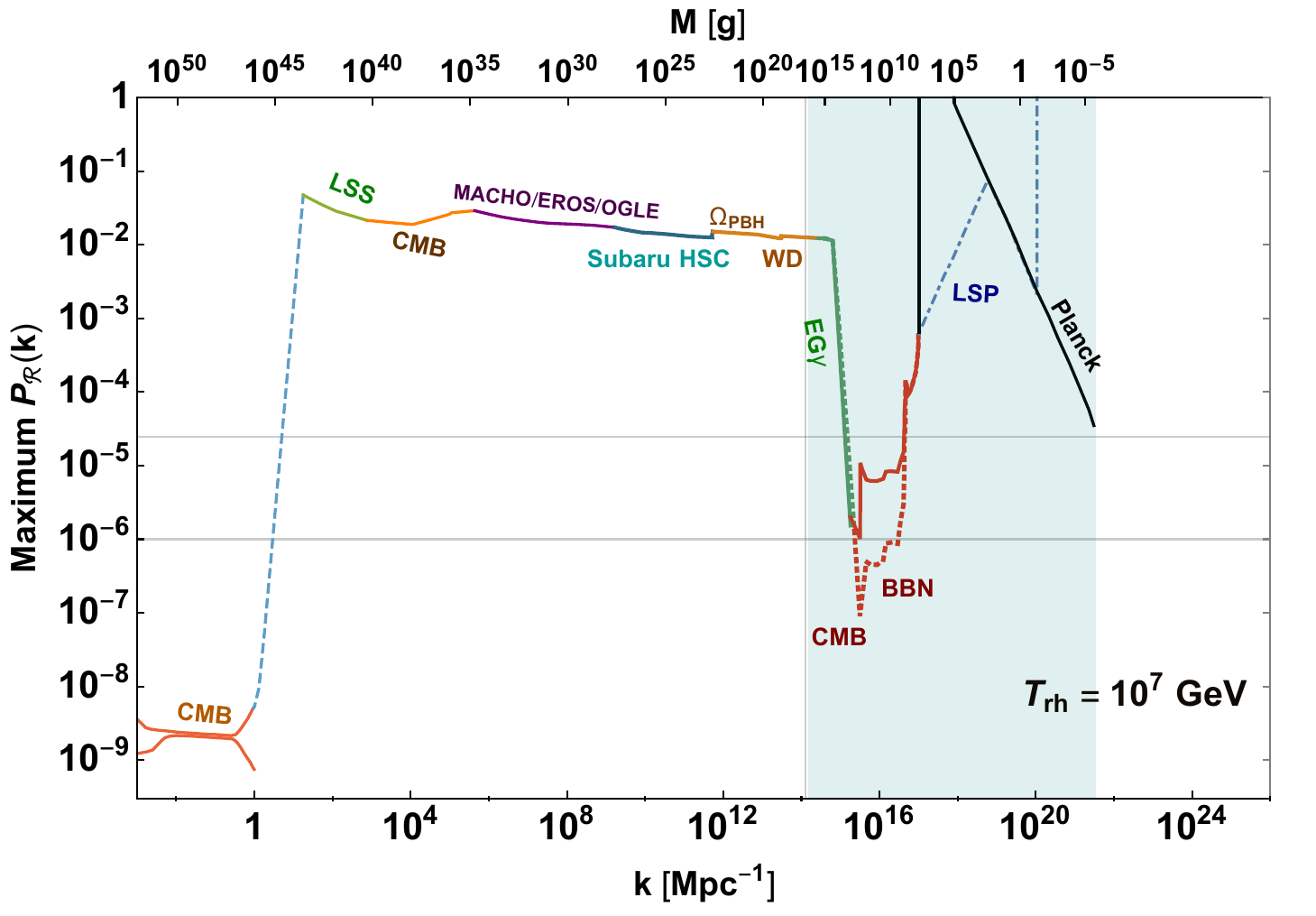}
\end{subfigure} \;
\begin{subfigure}{.5\textwidth}
  \centering
  \includegraphics[width=1.\linewidth]{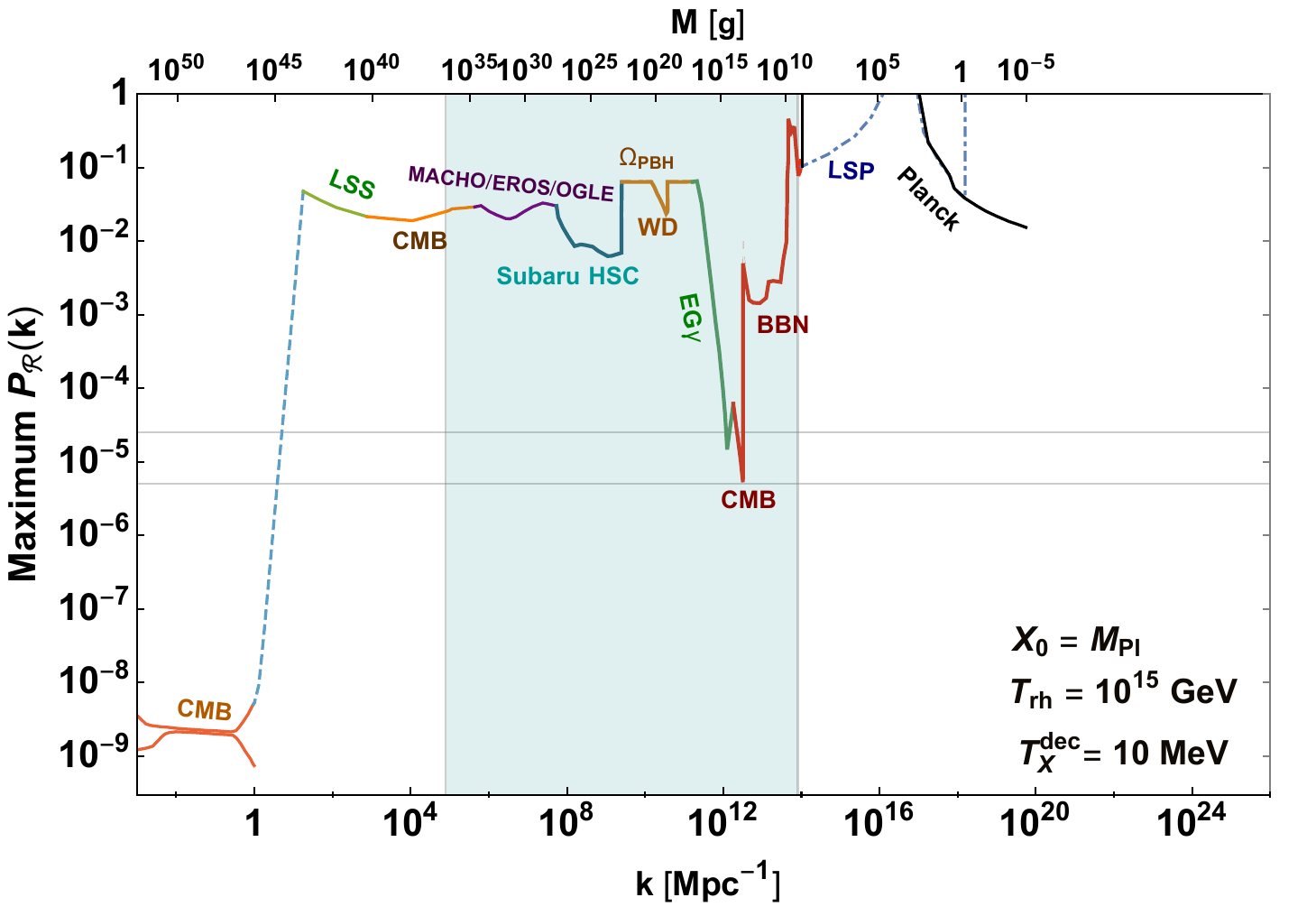}
\end{subfigure}
\begin{subfigure}{.5\textwidth}
  \centering
  \includegraphics[width=1.\linewidth]{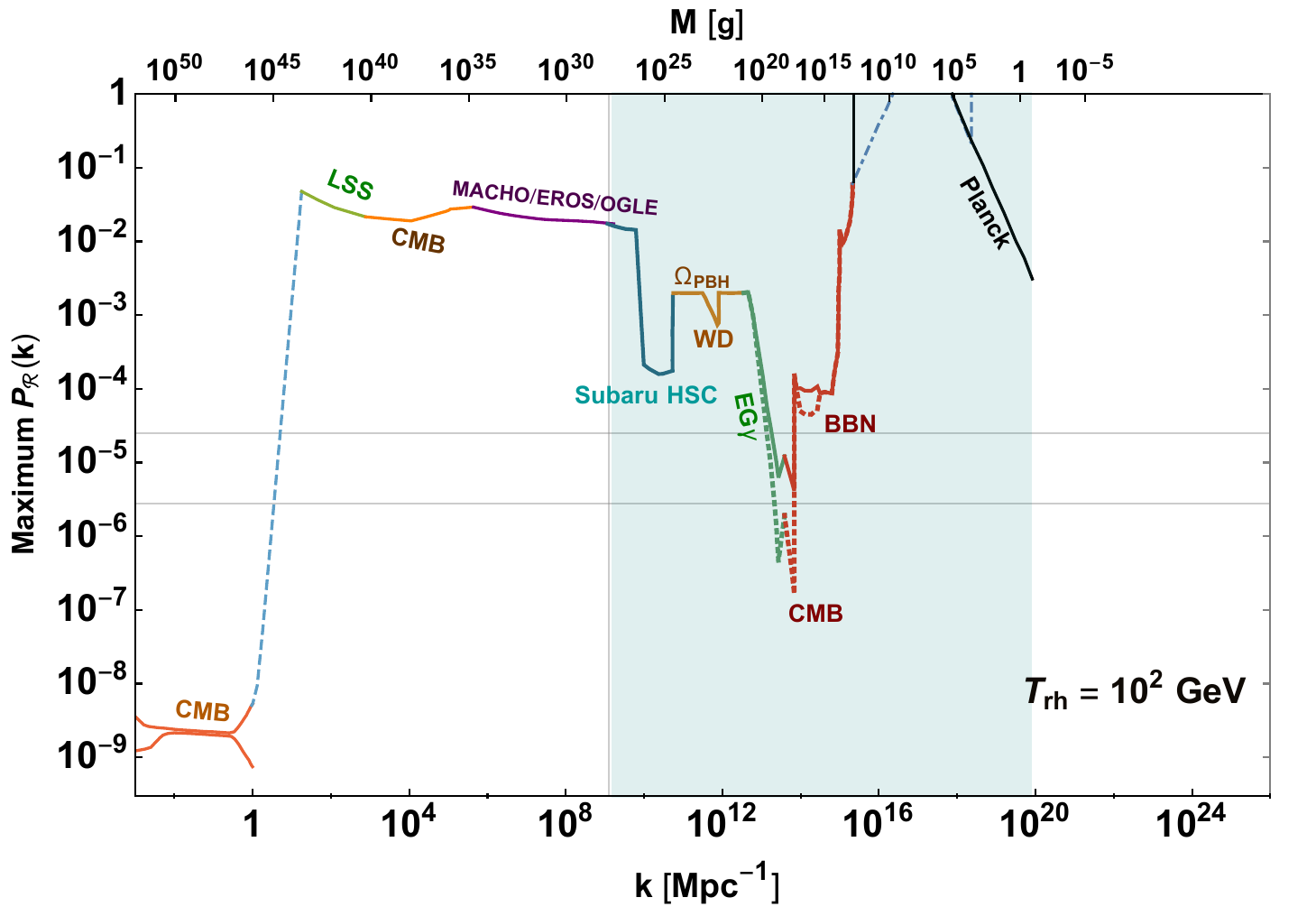}
\end{subfigure} \;
\begin{subfigure}{.5\textwidth}
  \centering
  \includegraphics[width=1.\linewidth]{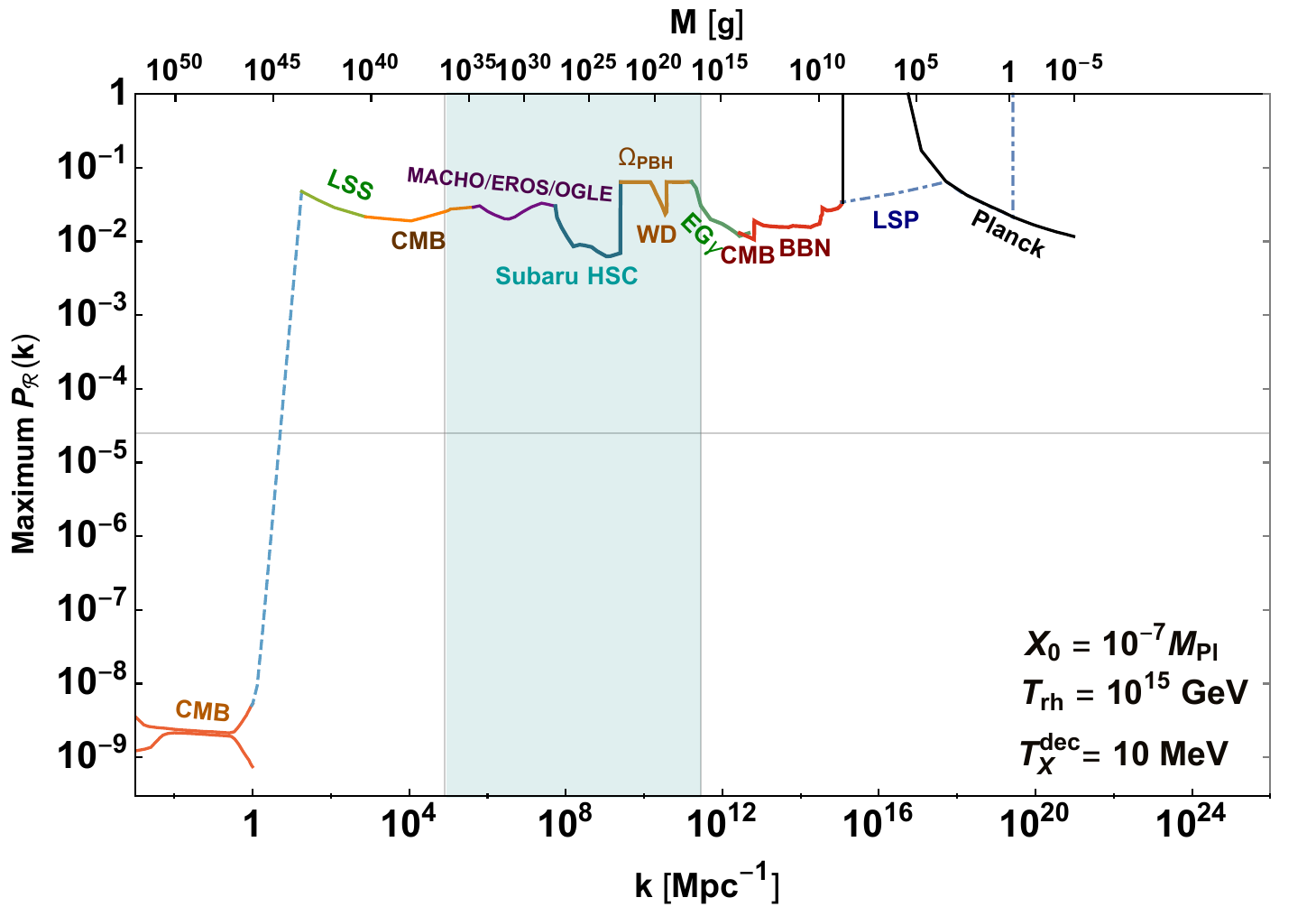}
\end{subfigure}
\caption{\label{fullPS}~ The plots depict upper bounds on the (full) power spectrum of the comoving curvature perturbation coming from constraints on evaporated and nonevaporated PBHs for different early universe cosmic histories  after the assumption that ${\cal P_R}(k) = \theta^{-2} \sigma^2_\text{max}(k)$, where $\theta=2/5$ and $4/9$ for eMD and RD eras respectively. The blue shaded areas correspond to a scalar-condensate dominated non-thermal  phase, caused either by the inflaton or a modulus field. The dashed lines give the upper bounds for spinless gravitational collapse. The upper horizontal gridline is the $\sigma=0.005$ theoretical threshold.  
For reference, in the left upper panel the probing $k$-range of GW antennas and $\mu$-distortions is also depicted.
}
\end{figure}

The $\mu$ and $y$ type distortions of the CMB black body spectrum as well as the secondary stochastic gravitational waves
 usually correspond to very large PBH masses and the interest of this work is mainly on PBH with short lifetime, $M_\bullet \ll M_\odot$,  and the ${\cal P_R}(k)$  features at large wavenumbers.
  We neither examine here the implications  for the ${\cal P_R}(k)$ due to the $\mu$ and $y$ type distortions nor due to stochastic gravitational waves, see e.g. \cite{Byrnes:2018txb} for a recent work. We do not examine implications on the power spectrum of ultracompact minihalos of dark matter \cite{Bringmann:2011ut}, or from the decay of metastable vacua \cite{Tetradis:2016vqb, Gorbunov:2017fhq, Canko:2017ebb} either.  
Following Ref. \cite{Carr:2009jm} we included only the GW  constraint, $\beta(M) \lesssim 10^{-52}$, on the mass band $10^2 -10^4 M_\odot$ that comes from pulsar timing data  
due to the generation second-order tensor perturbations \cite{Saito:2008jc}. 
Also for comparison,  in the upper right panel of Fig. \ref{fullPS}, where a net radiation dominated phase is presented, we add with dotted lines  
  the $k$-range where the $\mu$-distortion constraints apply and the range that will be probed by gravitational wave antennas  LISA and Square Kilometre Array (SKA), as well as pulsar time arrays (PTA) that can search for secondary gravitational waves.

\section{Conclusions} \label{8}

 The gravitational observation of black hole mergers by LIGO offers us an unprecedented piece of information about the dark sector of the universe. 
 This direct observation of black holes motivated the cosmologists to explain the LIGO events by PBHs \cite{kam} as well as to investigate the scenario that lighter PBHs may comprise a significant fraction of the dark matter in the universe.
 Since this sort of dark matter candidates originate from the primordial density perturbations, the presence or the absence of PBHs provides us with an indirect insight into the spectrum of the primordial density perturbation far beyond the scales directly accessible in the CMB.
 
 In this work we focused on PBH scenarios with masses smaller than $10^{15}$ g
 which, if ever generated, will have evaporated by now. These PBHs, although absent from our galaxies, are expected to have interesting cosmological implications for the mechanisms that generate the ${\cal P_R}(k)$
as well  for the details of the early cosmic history.
We have shown that models designed to produce PBH dark matter with mass $M_\bullet > 10^{17}$g have to pass strict constraints in scales significant smaller where ephemeral PBHs form with mass $M\ll M_\bullet$. 
In addition, these constraints are much sensitive to the reheating temperature of the universe.
In particular, we investigated the implication of the evaporating PBHs on the variance of the density perturbations for different reheating temperatures and in scenarios where the early universe has been dominated by a modulus scalar field, see Fig. \ref{sigmaT} and Fig. \ref{sigmaX}.
We explicitly examined the reheating temperature constraints in scenarios with  PBH dark matter  in four different mass scales:  asteroid mass range, $M_\bullet \sim 10^{18}$g,  lunar mass range, $M_\bullet \sim 10^{22}$g,  planet mass range $M_\bullet \sim 10^{29}$g, and the LIGO mass  range (also called intermediate black hole mass range ) $M_\bullet \sim 10^{35}$g, see Fig. \ref{PST} in Appendix and Fig. \ref{AT}.

The main result of this work is that the variance of the density perturbations generated by any inflationary model has to satisfy a specific constraint in the large $k$ limit of the spectrum. 
 Additionally to the dynamical  constraints for the nonevaporated PBH relics,
we found the $\sigma(k)$ constraint given by the Eq. (\ref{kspacCMB}) for reheating temperatures
$T_\text{rh}  \, \gtrsim  T_\text{bbn}^\text{(MD)}$, 
the constraint given by Eq. (\ref{spinBBN}) for $T_\text{cmb}^\text{(eMD)} \,\,\, \lesssim T_\text{rh}  < T_\text{bbn}^\text{(MD)},$ and the constraint given by Eq. (\ref{spinCMB}) for $T_\text{rh} \lesssim  T_\text{cmb}^\text{(MD)}$, where $ T_\text{bbn}^\text{(MD)}\simeq 4\times 10^8$ GeV and 
$ T_\text{cmb}^\text{(MD)}\simeq 10^7$ GeV.  The combined constraints with respect to the reheating temperature are depicted in Fig. \ref{sigmabbncmb}.

We conclude that significant power in the large $k$-limit might be 
 in conflict with the observations.  
 Mechanisms that generate PBH relics with asteroid or lunar mass scale  are required, by the derived bounds on evaporating PBHs, to have a very narrow ${\cal P_R}(k)$ peak, see Appendix for details.
Remarkably,  these mass windows are rather interesting because the relic PBHs can explain the entire dark matter found in the galaxies. For heavier PBH relics there is more freedom however, since the ${\cal P_R}(k)$  amplitude maximizes closer to the Planck pivot scale $k_{0.05}\equiv 0.05$ Mpc$^{-1}$ where the spectrum  is tightly constrained \cite{Akrami:2018odb},  these scenarios are often in conflict  by the  spectral index value and running.
As expected, the most strict observational constraints on the power spectrum come actually from the CMB anisotropies at the scale $k_{0.05}$. 
Next to this constraint it is usually found to be the CMB bound from PBH evaporation at the $k=k(2.5\times 10^{13} \text{g},T)$. It stringent for $T_\text{rh}= T_\text{cmb}^\text{(MD)} \sim 10^7$ GeV, where the power spectrum of the comoving curvature perturbation has to be ${\cal P_R}(k)\lesssim 10^{-6}$.

The derived constraints on $\sigma(k\gg k_{0.05})$, translated into constraints on the spectral index value of the power spectrum tail  (see  Appendix) constrain the inflation models and have considerable implications for the inflationary model building in general. 
 Models that generate  a blue-tilted spectrum or a broad peak at large $k$
 are severely constrained. 
Running mass inflation models can realize a blue-tilted spectrum
and are required to have  
an appropriately balanced running and running of the running of the spectral index, see Ref. \cite{Kohri:2018qtx} for an investigation of viable ${\cal P_R}(k)$ for varying $T_\text{rh}$.  
Double inflation, inflaton-curvaton models  or inflection point models that yield a significant power at  large wavenumbers of the ${\cal P_R}(k)$ might also be in conflict with the derived constraints, in particular if the reheating temperature is low.
Summarizing, an inflationary model is {\it ruled out} if the power spectrum ${\cal P_R}(k)$ violates the bounds dictated by Eq. 
(\ref{kspacCMB}), (\ref{spinBBN}) and (\ref{spinCMB}), that can be seen as  {\it width constraints} of the power spectrum peak. 

The results of this work show
that the
PBHs 
can be regarded as a powerful tool to probe the primordial fluctuations on much smaller scales and give us insights into the dynamics that generated the seeds of the cosmic structure, even if PBHs do not comprise the observed dark matter in the universe. 
\\
\\
\begin{figure}[hbt!]
\begin{tikzpicture}[->,>=stealth',shorten >=2pt,auto,node distance=4.5cm,
                    semithick]
  \tikzstyle{every state}=[fill=blue!10 ,draw=none,text={black},  
  font = \footnotesize
  \bfseries]

  \node[state]   (A)                    { \scalebox{1.1} {${ \beta_{\rm max}(M)}$}};
 \node[state]   (Y)      [ left of=A] { \scalebox{1.1} {$ Y_{\rm PBH, max}(M)$}}; 
  \node[state]    (A2) [below right of=A] {$\beta_{\rm MD, max}(M, T_{\rm rh})$};
  \node[state]    (A1) [below left of=A] {$\beta_{\rm RD, max}(M)$};
   \node[state]   (B1) [ below right of=A2] {$\sigma_{\rm MD, max}(M,  T_{\rm rh})$};
  \node[state]         (B2) [below left of=A2] {$\sigma_{\rm RD, max}(M)$};
  \node[state]         (C1) [below  of=B1] {$\sigma^{\rm +spin}_{\rm MD, max}(M,  T_{\rm rh})$};
  \node[state]         (C2) [ below left of=B1] {$\sigma^{\rm spinless}_{\rm MD, max}(M,  T_{\rm rh})$};
  \node[state]         (P1) [ below  left of=C2] { \scalebox{1.2} {${\cal P_R}_{,\rm max}(k)$} };

  
  
  {\footnotesize {
  \path 
 (Y) edge              node {} (A)  
  (A) edge              node {$M>\gamma M_{\rm rh}$} (A1)
            edge              node {$M<\gamma M_{\rm rh}$} (A2)
        (A2) edge              node {$\sigma_{\rm max}>\sigma_{\rm cr}(M, T_{\rm rh})$} (B1)
            edge   node {$\sigma_{\rm max}<\sigma_{\rm cr}(M, T_{\rm rh})$} (B2)
            (A1) edge      node {} (B2)
            
            (B1) edge         node {$\sigma_{\rm max}<0.005$} (C1)
            edge   [bend right]                     node {$\sigma_{\rm max} >0.005$} (C2)
            
               (C1) edge              node {} (P1)
             (C2) edge              node {$k(M, T_{\rm rh})$}(P1)
          (B2) edge              node {$k(M)$}(P1)
             ;
}}
\end{tikzpicture}
\caption{\label{graphP}~ The graph illustrates the steps followed to derive the upper bounds for the power spectrum of the comoving curvature perturbations. The $Y_\text{PBH, max}$ refers to the maximum value of the PBH number density-over-entropy with mass  $M$ allowed by the observations. The $Y_\text{PBH, max}$  can specify the maximum value for the $\beta(M)$  only after the reheating temperature is known. 
The size of the $\beta$ depends on the variance $\sigma$ of the density perturbations in a different way if the collapse takes place in a background with or without  pressure. If there is thermal pressure (RD) the collapse is effectively instantaneous whereas, if the there is no pressure (eMD) the collapse has a duration determined by the $\sigma_\text{cr}$ value. Also, in eMD era the size of the variance determines whether the PBH formation-rate is affected by the spin. Taking all these into account, the maximum value for the ${\cal P_R}(k)$ is derived.  
} 
\end{figure}

\section*{Acknowledgments}

\noindent 
I would like to thank Antonio Riotto  and Jun'ichi Yokoyama for their  comments. 
This work is supported by the IKY Scholarship Programs for Strengthening Post Doctoral Research, co-financed by the European Social Fund ESF and the Greek government.

\appendix

\section{Parametrizing the morphology of the $P_{\cal R}(k)$} \label{Ap}

In the main text of this paper, we computed upper bounds for the  variance of the comoving density contrast without specifying the power spectrum -with the exception of the Fig. \ref{AT}.   In Appendix we assume a particular morphology for the power spectrum of the comoving curvature perturbation in order to illustrate the sort of the power spectra that are compatible with the constraints derived in this work. 
For simplicity, we omit possible critical collapse effects and assume  the horizon-mass approximation for the PBHs mass function. In this regard, the bounds that we derive are conservative since critical effects are expected to increase the PBHs abundance in the low mass tail \cite{Yokoyama:1998xd, Carr:2016drx} that we focus on.

We recall that the comoving curvature perturbation, ${\cal R}_k$, is related to the metric perturbations $\Phi(k)$ for modes $k$ outside the horizon via the relation
\begin{equation}
{\cal R}_k=-\frac{(5+3w)\Phi(k)}{3+3w}
\end{equation} 
where the metric in the longitudinal gauge reads, $ds^2=-(1+2\Phi)dt^2+a^2(t)(1-2\Phi)dx^2$. Typical inflationary scalar perturbations are well approximated by the expansion  
\begin{align} \nonumber
\ln {\cal P_R}(k)=\ln A_{0.05}+(n_s-1)\ln(k/k_{0.05}) +\frac12 \frac{d\ln n_s}{d \ln k}\ln(k/k_{0.05})^2+\frac{1}{6}\frac{d^2\ln n_s}{d \ln k^2}\ln(k/k_{0.05})^3+...
\end{align}
where $k_{0.05}=0.05$ Mpc$^{-1}$ is the pivot scale used by the Planck collaboration \cite{Akrami:2018odb} and $n_s$ the scalar spectral index.  If the running of the spectral index $d \ln n_s/d \ln k$ and the running of the running are nonzero the power spectrum amplitude may increase  significantly at smaller scales. 
The enhancement of the power spectrum with amplitude ${\cal P_R}_\text{max}$ can generate PBHs at the scale $k_\bullet$, identified with the $k_\text{peak}$.

We focus on the power spectrum tail since it is this part of the spectrum that is relevant to the PBH evaporation constraints. A general description if ${\cal P_R}(k)={\cal P_R}(k_\text{peak}) \, f(k)$ for $k>k_\text{peak}$.
Let us parametrize the tail of the power spectrum  by the  function 
$f(k)=\left({k}/{k_\text{peak}}\right)^{s(k, p, \alpha)}$ where $s(k, p, \alpha)\equiv{-p\left(\log\frac{k}{k_\text{peak}} \right)^{\alpha-1}}$.
For the values $\alpha=1$ or $\alpha=2$ this parametrization describes well a large class of power spectra designed to generate PBH, e.g inflationary models with inflection point. 
In the following we will examine analytically the simplest $\alpha=1$ case,
\begin{equation} \label{PR1}
{\cal P_R}(k\geq k_\text{peak})= {\cal P_R}_\text{max} \left(\frac{k}{k_\text{peak}}\right)^{-p} \,
\end{equation} 
where ${\cal P_R}_\text{max} \equiv  {\cal P_R}(k_\text{peak})$. 
The power $p$ can be viewed as the spectral index of the power spectrum tail, $p\equiv 1-n_s^\text{(tail)}$.

\subsection{Radiation domination}

In the approximation that the inflation stage is instantaneously followed by a thermal radiation dominated epoch it is $k_\text{rh}=k_\text{end}$. The variance of the perturbation during RD era that instantaneously follows inflation is given by 
\begin{equation}
\sigma^2(k\geq k_\text{peak})= \left( \frac{4}{9} \right)^2  {\cal P_R}(k_\text{peak})  \int_{k_\text{peak}}^{k_\text{end}} \frac{dq}{q}\,W^2\left(\frac{q}{k}\right)\left(\frac{q}{k}\right)^4   \left(\frac{q}{k_\text{peak}}\right)^{-p}    \,.
\end{equation}
The integration gives
\begin{align}
\sigma^2(k\geq k_\text{peak})
&
= \left( \frac{4}{9} \right)^2 \frac{{\cal P_R}_\text{max}}{2} 
\left(     \frac{k}{k_\text{peak}}     \right)^{-p} 
\left[ \Gamma \left(2-\frac{p}{2}, \, \frac{k^2_\text{peak}}{k^2}\right) -  
\Gamma \left( 2-\frac{p}{2}, \, \frac{k^2_\text{end}}{k^2} \right)  \right] \\
&
\simeq \left( \frac{4}{9} \right)^2 \frac{{\cal P_R}_\text{max}}{2}
\left(     \frac{k}{k_\text{peak}}     \right)^{-p} 
 \Gamma \left(2-\frac{p}{2}, \, 
\frac{k^2_\text{peak}}{k^2}\right) \,,
\end{align}
where we took into account that $ \Gamma \left(2-\frac{p}{2}, \, \frac{k^2_\text{peak}}{k^2}\right) \gg 
\Gamma \left( 2-\frac{p}{2}, \, \frac{k^2_\text{end}}{k^2} \right) $.
For $k>k_\text{peak}$  the incomplete Gamma function can be expanded and the variance is approximated as
\begin{equation} \nonumber
\sigma^2(k\geq k_\text{peak}) \simeq \left( \frac{4}{9} \right)^2 \frac{{\cal P_R}_\text{max}}{2}
\left(     \frac{k}{k_\text{peak}}     \right)^{-p} 
\left[
\Gamma \left(2 - \frac{p}{2} \right) + \left( \frac{k}{k_\text{peak}}  \right)^p 
\left(\frac{2}{p-4}\,  \left(\frac{k^2_\text{peak}}{k^2}\right)^2+ {\cal O}\left(\frac{k^2_\text{peak}}{k^2} \right)^3 \right)
\right]
\end{equation} 
After keeping the leading terms the variance squared reads,
\begin{align} \label{PS1Rex} \nonumber
\sigma^2(k) & \simeq \left( \frac{4}{9} \right)^2 \frac{{\cal P_R}_\text{max}}{2}
\left( \frac{k_\text{}}{k_\text{peak}}     \right)^{-p} 
\Gamma \left(2 - \frac{p}{2} \right)\, \\
& \simeq  \, 0.1\, \,\Gamma \left(2 - \frac{p}{2} \right)\, {\cal P_R}(k) \, \equiv\, \theta^2\, {\cal P_R}(k)   \quad\quad\quad\quad \, \text{for} \quad k>k_\text{peak}  \,.
\end{align} 

\vspace*{.5cm}

\begin{figure}[!htbp]
\begin{subfigure}{.5\textwidth}
  \centering
  \includegraphics[width=1.\linewidth]{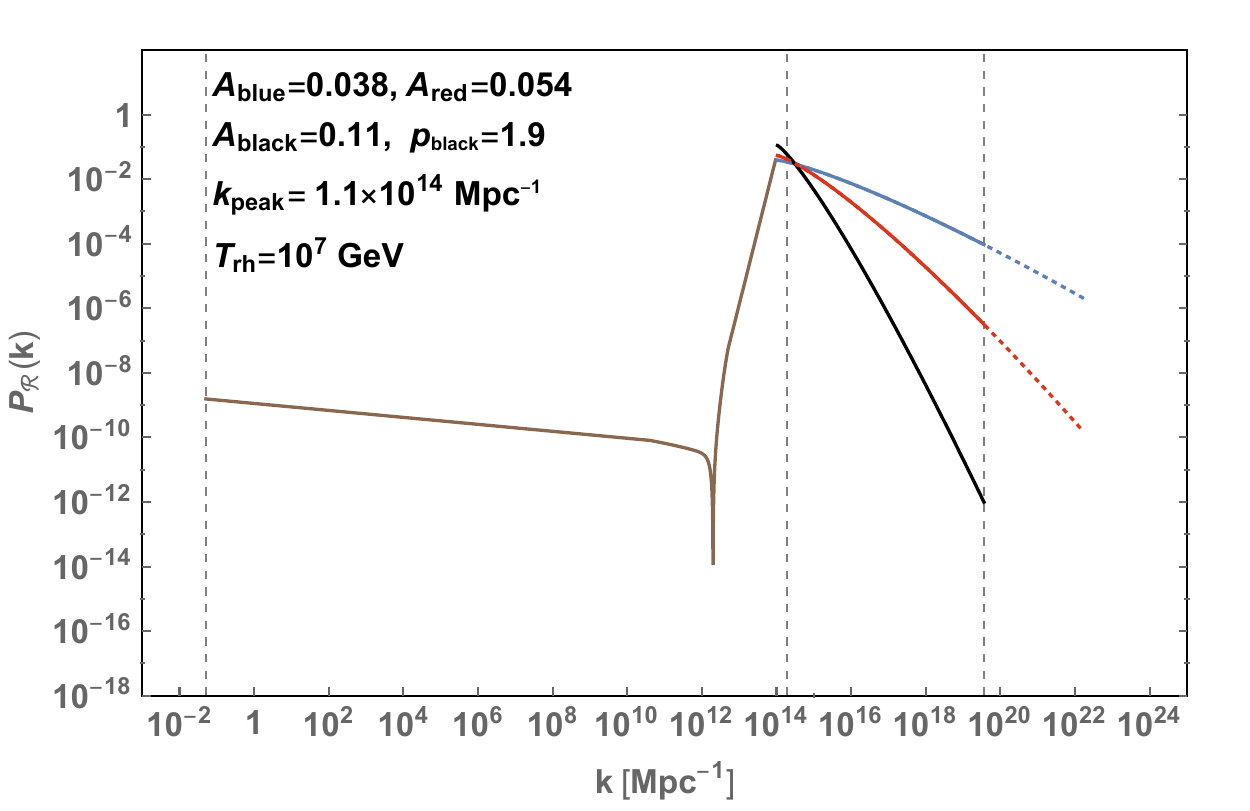}
\end{subfigure}%
\begin{subfigure}{.5\textwidth}
  \centering
  \includegraphics[width=1.\linewidth]{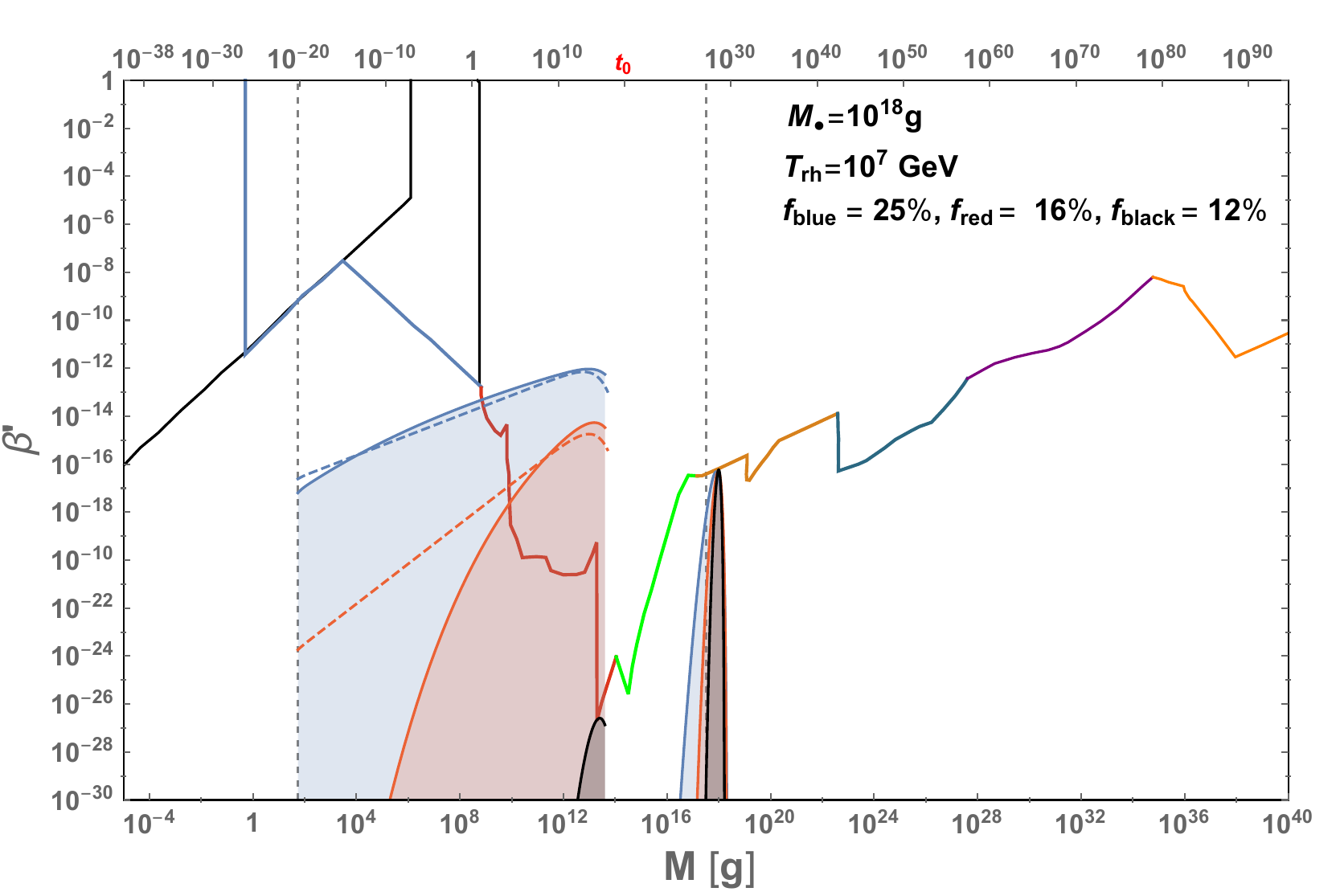}
\end{subfigure}
\begin{subfigure}{.5\textwidth}
  \centering
  \includegraphics[width=1.\linewidth]{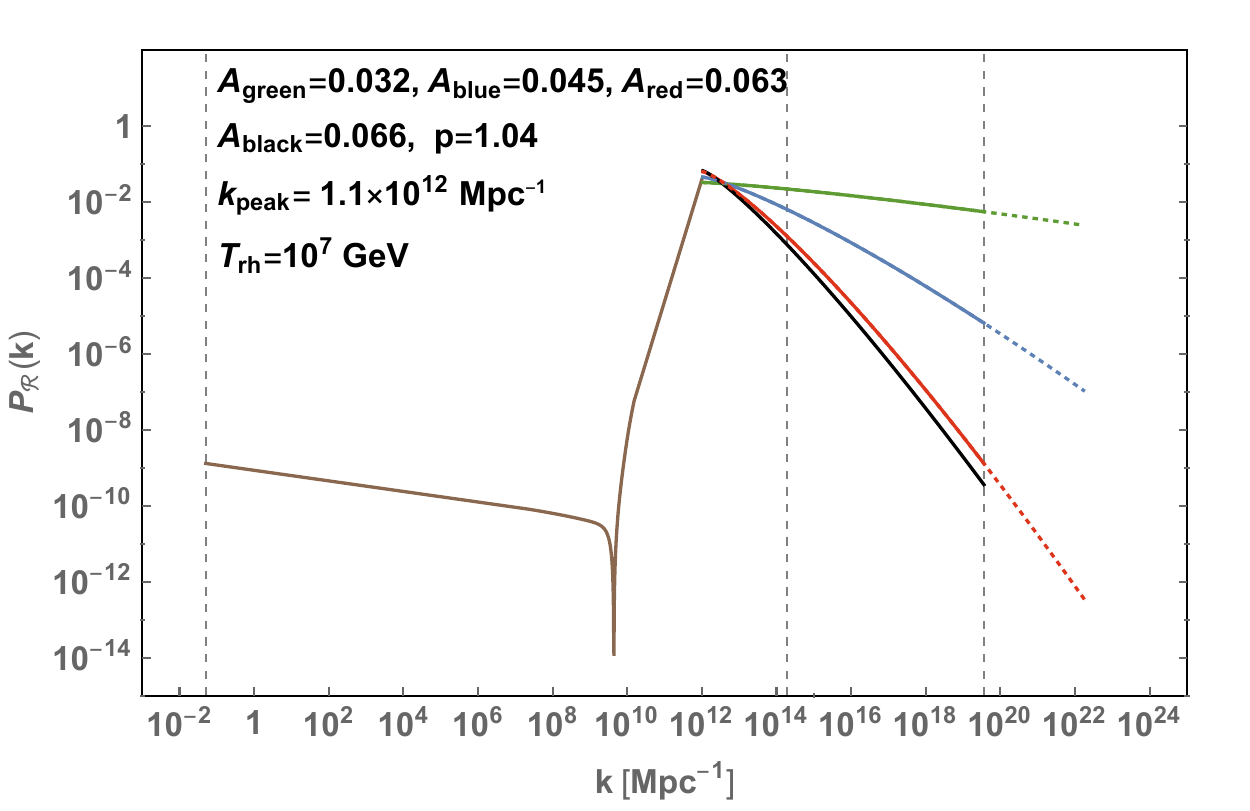}
\end{subfigure}%
\begin{subfigure}{.5\textwidth}
  \centering
  \includegraphics[width=1.\linewidth]{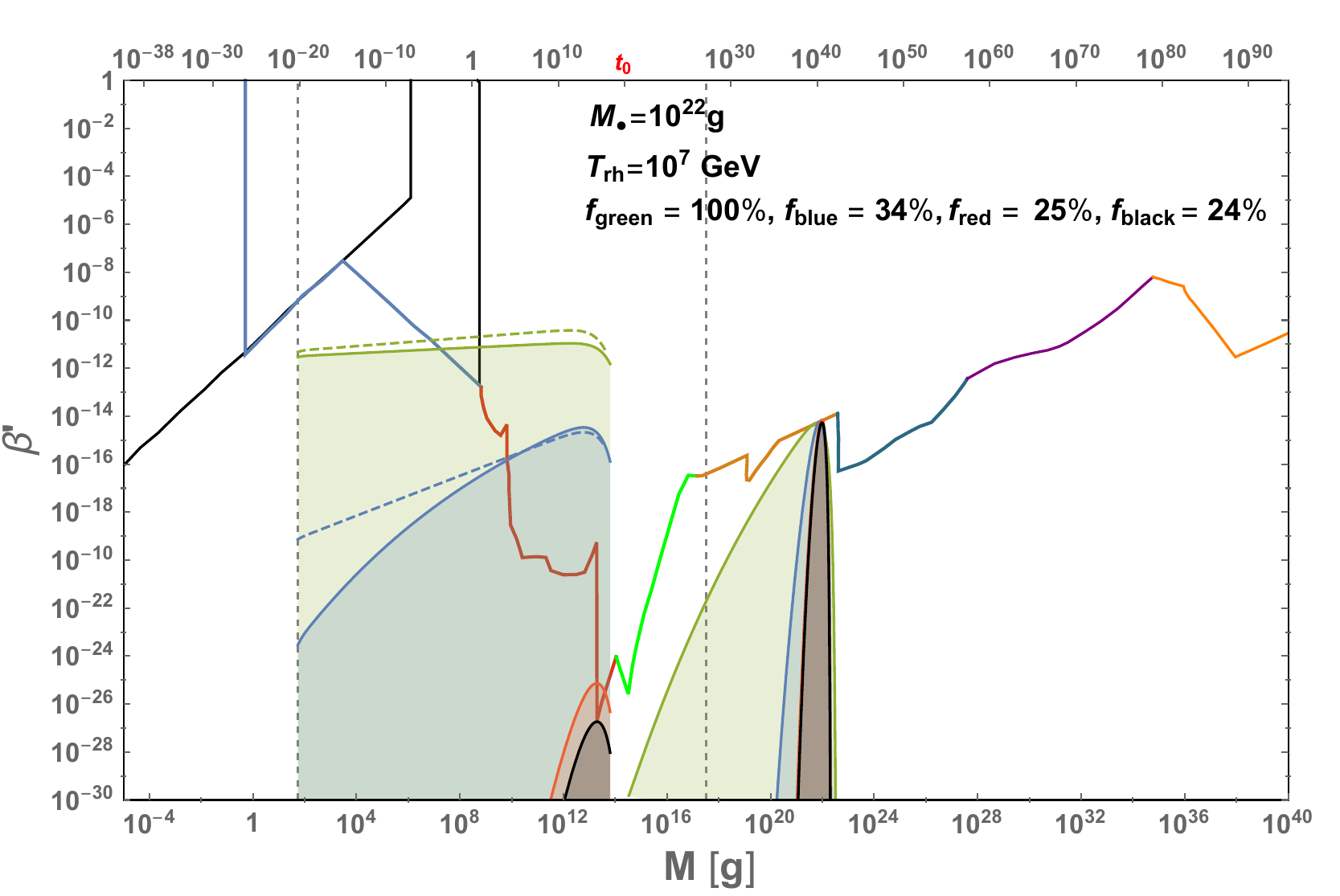}
\end{subfigure}
\begin{subfigure}{.5\textwidth}
  \centering
  \includegraphics[width=1.\linewidth]{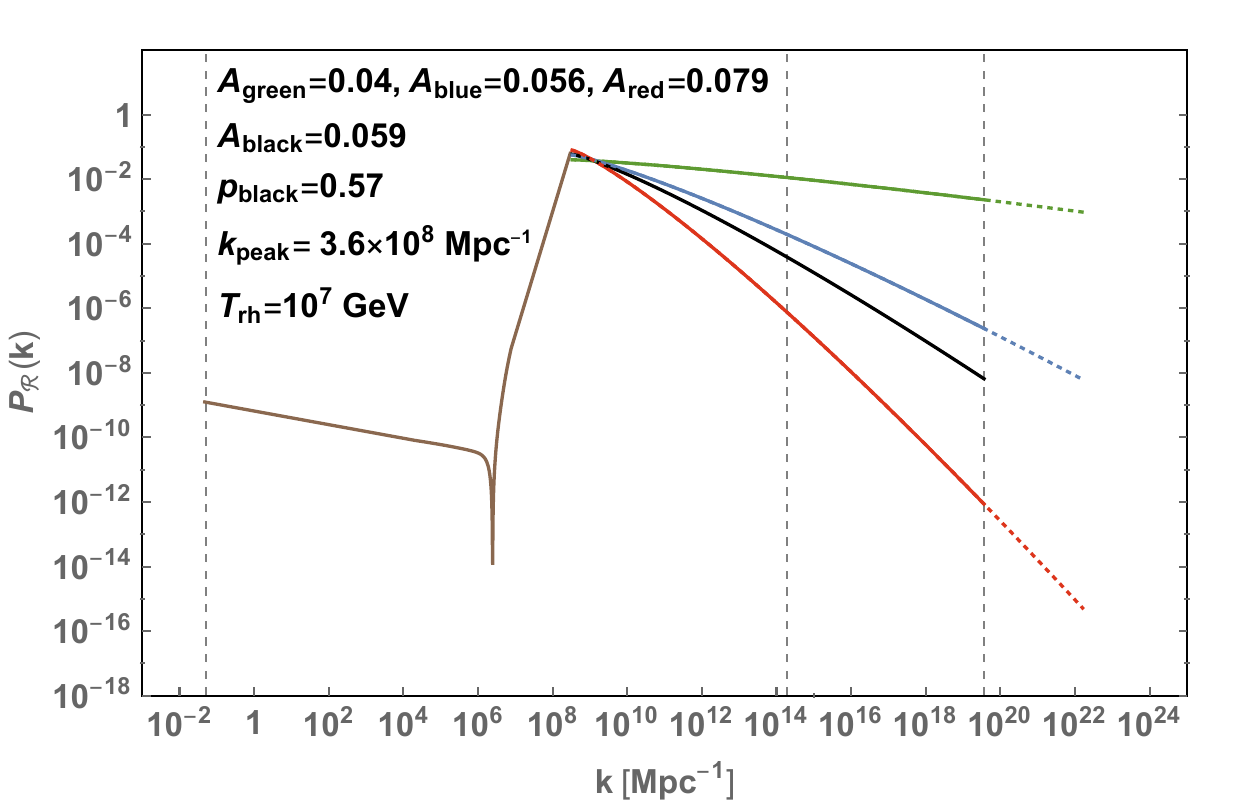}
\end{subfigure}%
\begin{subfigure}{.5\textwidth}
  \centering
  \includegraphics[width=1.\linewidth]{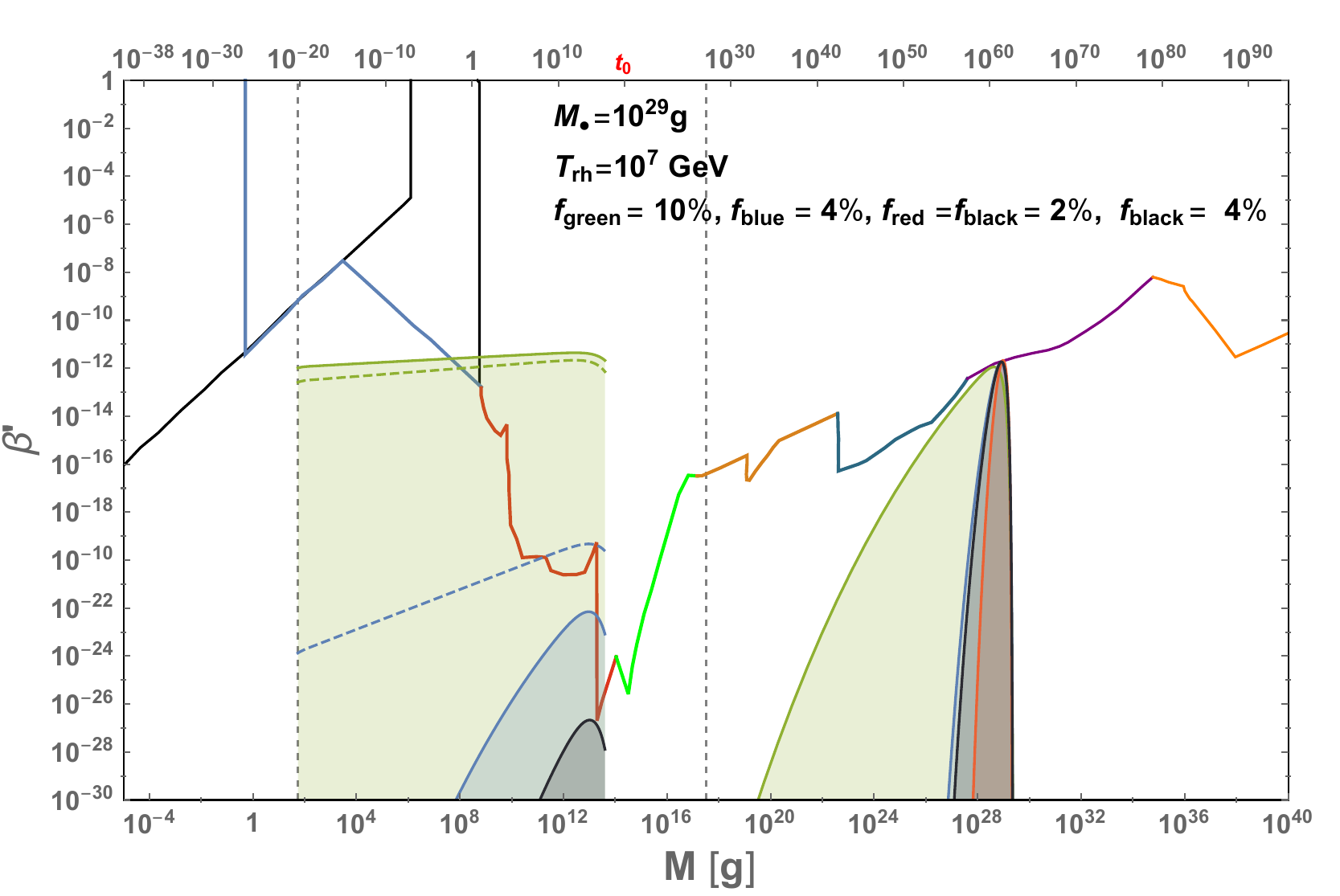}
\end{subfigure}
\begin{subfigure}{.5\textwidth}
  \centering
  \includegraphics[width=1.\linewidth]{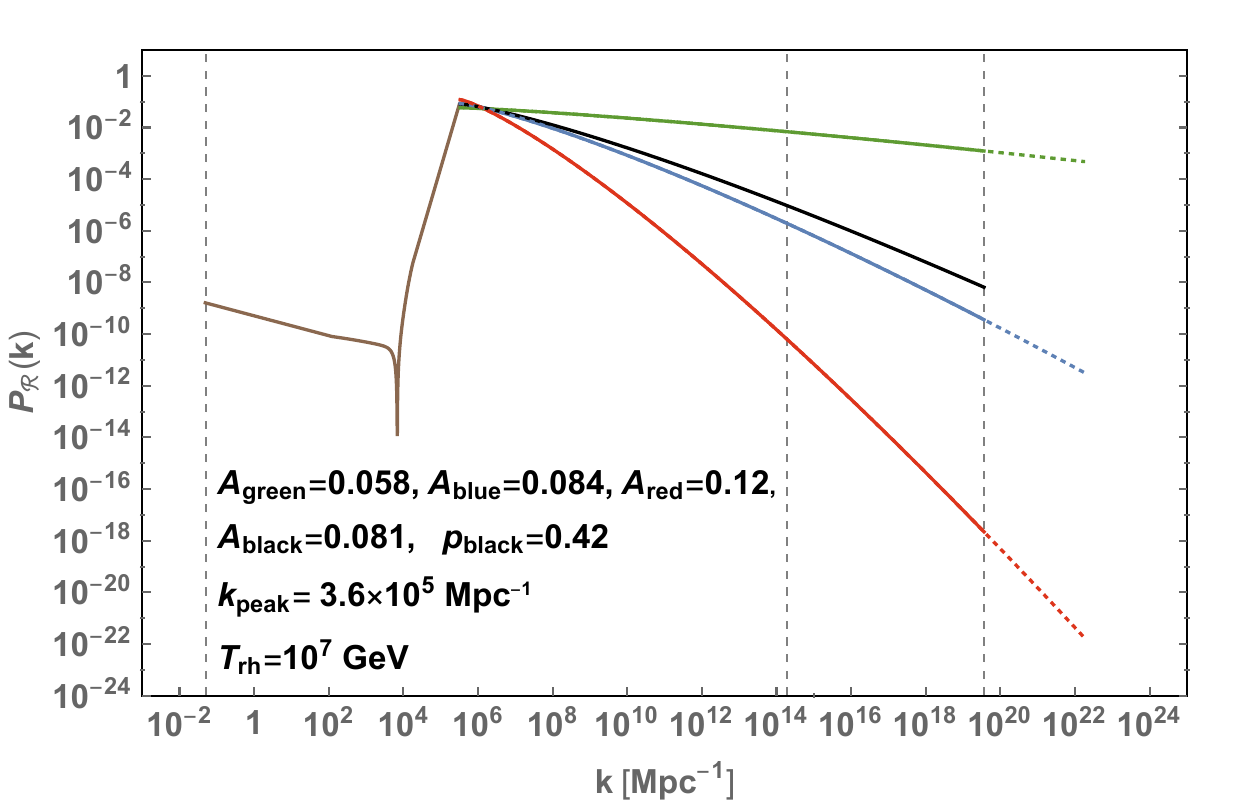}
\end{subfigure}%
\begin{subfigure}{.5\textwidth}
  \centering
  \includegraphics[width=1.\linewidth]{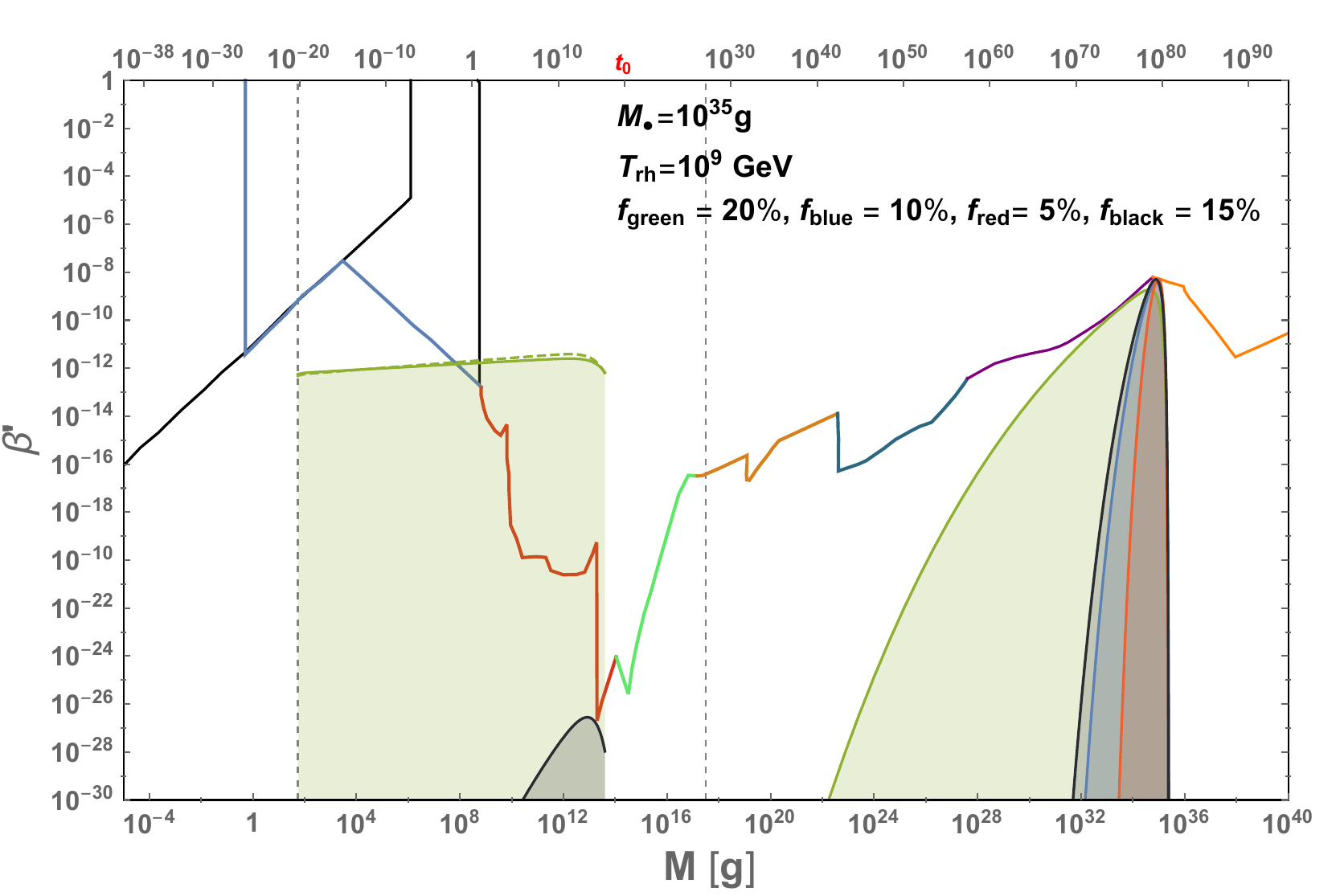}
\end{subfigure}
\caption{\label{PST}~ 
{\it Left panels}: Typical  ${\cal P_R}(k)$ are depicted designed to produce PBH with $M_\bullet =M_\text{peak}=10^{18}, 10^{22}, 10^{29}, 10^{35}$ g  
for an inflaton that decays 
at $T_\text{rh}=10^7$ GeV.  
 The different slopes of the power spectra are fitted by the Eq. (\ref{PR1}) for $p=0.1$ (green),  0.5 (blue),  1 (red). 
For each slope the amplitude of the  ${\cal P_R}(k)$ peak, $A\equiv {\cal P_R}_\text{max}$, differs so that the PBH abundance maximizes. The black line corresponds to the critical slope and
separates the ${\cal P_R}(k)$ lines {\it ruled out} by PBH evaporation (above the black line)  from those ruled in (below). The vertical gridlines from left to right show the scales $k_{0.05},  k_\text{rh}, k_\text{end}$ for $T_\text{rh}=10^7$ GeV. The dotted ends extend to $k_\text{end}$ if it was  $T_\text{rh}=10^{15}$ GeV.
 {\it Right panels}: The mass fraction of the universe that collapses into PBH and the $f_\text{PBH}$ for each color against the observational constraints are depicted. The dashed lines correspond to spinless collapse.
}
\end{figure} 

\subsubsection{Constraints on the power spectrum tail}

The constraint (\ref{sRD}) can be now applied on the power spectrum and reads for the BBN and CMB respectively
\begin{align}
 &{\cal P_R}\left(k(5\times 10^{10} \, \text{g}) \right) \, \lesssim \, \frac{0.014} {\Gamma\left(2-\frac{p}{2}\right)} \,\left(\frac{\delta_c}{0.41}\right)^2 \left[1+ 0.028\ln\left(\frac{\gamma_\text{R}}{0.2} \right)\right]^{-1}\,\,\quad\quad\quad\quad \text{(BBN)} \\ 
 \nonumber\\
 &{\cal P_R}\left(k(2\times10^{13} \, \text{g}) \right) \, \lesssim  \, \frac{0.012} {\Gamma\left(2-\frac{p}{2}\right)} \,\left(\frac{\delta_c}{0.41}\right)^2 \left[1+ 0.023\ln\left(\frac{\gamma_\text{R}}{0.2} \right)\right]^{-1}\,\,\quad\quad\quad\quad \text{(CMB)}
\end{align}
These upper bounds, ${\cal P_R}_\text{bound}$, constrain the power spectrum maximum amplitude ${\cal P_R}_\text{max}$ and the power $p$ at a particular scale. It is ${\cal P_R}(M)={\cal P_R}_\text{max}(M/M_\bullet)^{p/2}$ for $M<M_\bullet$, thus we get
\begin{align} \label{pcmb}
p \, \gtrsim \,2\, \frac{\ln {\cal P_R}_\text{max}-\ln{\cal P_R}_\text{bound}(M)}{\ln M_\bullet-\ln M}\,.
\end{align}  
where $M_\bullet$ is the characteristic {\it relic} PBH mass that an inflationary model predicts.  Scenarios with flat power spectra, $p \sim 0$, that generate a significant amount of PBH relics are  {\it ruled out}.
The stringent constraint is for the CMB, which for benchmark values $\gamma_\text{R}=0.2$, $\delta_c=0.41$, reads
\begin{align} \label{pcmb2}
p \, \gtrsim \, 2\,\frac{\ln ({\cal P_R}_\text{max}/10^{-2})-0.18}{\ln (M_\bullet/10^{20}\,\text{g})+15.4}\,.
\end{align}  

\subsection{Matter domination}
Let us assume that after inflation reheating follows that lasts $N_\text{rh}$  e-folds of expansion, then the (\ref{PR1}) rewrites,  ${\cal P_R}(k\geq k_\text{rh})= {\cal P_R}(k_\text{rh})  \left({k}/{k_\text{rh}}\right)^{-p} \,.$
The variance of the comoving density contrast during early matter domination is 
\begin{equation}
\sigma^2(k\geq k_\text{rh})= \left( \frac{2}{5} \right)^2  {\cal P_R}(k_\text{rh})  \int_{k_\text{rh}}^{k_\text{end}} \frac{dq}{q}\,W^2\left(\frac{q}{k}\right)\left(\frac{q}{k}\right)^4   \left(\frac{q}{k_\text{rh}}\right)^{-p}    \,.
\end{equation}
Also here the integration, after taking into account  that  $ \Gamma \left(2-\frac{p}{2}, \, \frac{k^2_\text{rh}}{k^2}\right) \gg 
\Gamma \left( 2-\frac{p}{2}, \, \frac{k^2_\text{end}}{k^2} \right) $ 
and expanding the  incomplete Gamma function, 
gives the variance squared that at leading order,
\begin{align} \label{PS1ex}
\sigma^2(k) 
 \simeq  \, \frac{1}{2}\left( \frac{2}{5} \right)^2\, \,\Gamma \left(2 - \frac{p}{2} \right)\, {\cal P_R}(k) \,\equiv \, \theta^2\, {\cal P_R}(k)  \quad\quad\quad\quad \, \text{for} \quad k>k_\text{rh}  \,.
\end{align}

In the Fig. \ref{PST} power spectra that generate  four representative  relic PBH masses $M_\bullet=10^{18}$g, $10^{22}$g, $10^{29}$g and $10^{35}$g are depicted for three different tails with steepness, $p=0.1$ (in green), $p=0.5$ (in blue) and $p=1$ (in red). 
In black the less steep {\it allowed} slope is depicted, the critical slope, that separates the allowed from the disallowed power spectra.  
The reheating temperature is chosen to be $T_\text{rh}=10^7$ GeV where the CMB constraints are the stringent ones and the role of the evaporating PBHs on the inflationary model selection is more manifest.
On the plots, the total PBH fractional density $
f_\text{PBH, tot}=\int_M\, d \ln M\, f_\text{PBH} (M)
$ is also computed. We have  assumed a one-to-one correspondence between the wavenumber $k$ of the perturbations and the PBHs masses. 

%

\subsubsection{Constraints on the power spectrum tail and the reheating temperature}

Let us derive here a  bound on the reheating temperature, associated with the slope of the ${\cal P_R}(k)$.

The power spectrum value at the time of reheating, ${\cal P_R}(k_\text{rh})$ depends on the reheating temperature,  
and  we can pursue further the implications of the constraint (\ref{betaMDbound}), 
that we rewrite it here as 
\begin{align}
\sigma(M) \, < \sigma_\text{MD,max}( \boldsymbol{C_M}, M, T_\text{rh}, \gamma_\text{M})\,,
\end{align}
where $\sigma(M)$ is given by Eq. (\ref{PS1ex}) and $\sigma_\text{max}$ given  by the observational constraints (independent of the ${\cal P_R}(k)$ form). It is $\sigma(k)= \theta \, {\cal P_R}^{1/2}(k_\text{rh}) \, (k/k_\text{rh})^{-p/2} $ and the power of the comoving curvature perturbations that reentrer the horizon at the time of reheating is
 \begin{align}
{\cal P_R}(k_\text{rh})\,=\, {\cal P_R}_\text{max} \, \left( 9.6 \times 10^{-5} \right)^p \,\gamma_\text{R}^{p/2} 
\left( \frac{T_\text{rh}}{10^{10}\, \text{GeV}}\right)^{-p}\,
 \left( \frac{M_\bullet}{10^{20}\, \text{g}}\right)^{-p/2}  \left(\frac{g_*}{106.75} \right)^{-p/4}\,. 
\end{align}
Substituting the ratio $k/k_\text{rh} $ from Eq. (\ref{krhM}) and (\ref{gen}) and neglecting the finite time of the gravitational collapse  we get the constraint on the reheating temperature, 
\begin{align} \label{Tcon}
T_\text{rh}^{p/6} \, \geq \, \frac{\xi({\cal P_R}_\text{max},\gamma_\text{M}, \gamma_\text{R} , M_\bullet, M, p)}{\sigma_\text{MD,max}( \boldsymbol{C_M}, M, T_\text{rh}, \gamma_\text{M})} \,,
\end{align}
where $\xi$ is given by the expression, 
\begin{align}
\xi \,\equiv \, \theta \, {\cal P_R}_\text{max}^{1/2} \, (2.1\times 10^{-5})^{p/2} \gamma^{p/4}_\text{R} \gamma^{-p/6}_\text{M} \left( \frac{g_*}{106.75}\right)^{-p/24} \left(\frac{M_\bullet}{10^{20}\, \text{g}}\right)^{-p/4}\left( \frac{M}{10^{10}\, \text{g}}\right)^{p/6}
\end{align}

In the approximation of the spinless collapse  the $ \sigma_\text{MD,max}(M, T_\text{rh}, \gamma_\text{M})$ is explicitly calculated, see Eq. (\ref{Cs3}), and the reheating temperature is constrained to be
 \begin{align} \label{Tconz}
\left(\frac{T_\text{rh}}{10^{10}\, \text{GeV}}\right)^{\frac{p}{6}-\frac{1}{5}} \, > \,  \,
\xi({\cal P_R}_\text{max},\gamma_\text{M}, \gamma_\text{R} , M_\bullet, M, p) \, 2.8 \, \gamma_\text{M}^{{1}/{5}} 
\, \left(  \frac{M}{10^{10} \, \text{g}}  \right)^\frac{1}{10}
\boldsymbol{C_M}^{-1/5}\,.
\end{align}

In fact spin effects cannot be neglected at the mass range relevant to  the CMB observables  and one has to solve numerically the  (\ref{Tcon}).  Numerics show that, qualitatively, a similar behavior to the condition (\ref{Tconz}) is found. 

Flat or not steep power spectrum tails are ruled out for any reheating temperature $T_\text{rh} \lesssim T^\text{(MD)}_\text{cmb} \sim 10^7 $ GeV.  
{\it The constraints are stringent for light PBH dark matter.} As illustrated in  \ref{PST},
PBH with mass $M_\bullet=10^{18}$ g are allowed to be generated in a sizable amount only if the ${\cal P_R}(k)$ slope parameter is particularly large, $p\gtrsim 1.9$  and for $M_\bullet=10^{22}$ g only if $p\gtrsim 1$. 
This means that  the power spectrum must have the shape of a narrow peak in order to  generate PBHs in accordance with the observational constraints.  
For $M_\bullet=10^{29}$ g and $M_\bullet=10^{35}$ g the bounds on the power spectrum tail are weaker but still considerable: unless $p\gtrsim 0.6$ and $p\gtrsim 0.4$ respectively the CMB anisotropy  observables alter.


\end{document}